\documentclass[aps,prb,superscriptaddress,amsmath,amssymb,showpacs,floatfix,reprint,longbibliography]{revtex4-2}
\usepackage{wasysym}

\usepackage{graphicx}
\usepackage{float}

\usepackage{multirow}    
    
\usepackage{color}
\usepackage[colorlinks=true, urlcolor=blue, linkcolor=blue, citecolor=blue, pdftex]{hyperref}

\usepackage{tikz}

\DeclareRobustCommand{\DavidStarOut}{\tikz[baseline={([yshift=-.35em]current bounding box.center)},x=0.16em,y=0.16em]{
  \draw (0, 2.000)--(-0.5774, 1.000)--(-1.732, 1.000)--(-1.155, 0)--(-1.732, -1.000)--(-0.5774, -1.000)--(0, -2.000)--(0.5774, -1.000)--(1.732, -1.000)--(1.155, 0)--(1.732, 1.000)--(0.5774, 1.000)--(0, 2.000)--(-0.5774, 1.000)}}
    
\DeclareMathOperator{\Tr}{Tr}

\definecolor{karlocolor}{rgb}{0.9,0.2,0}
\definecolor{oldcolor}{rgb}{0.3,0.3,1.0}
\definecolor{newcolor}{rgb}{0,0.0,0.5}
\definecolor{danicolor}{rgb}{0,0.6,0.6}

\begin{document}

\title{The algebraic spin liquid in the SU(6) Heisenberg model on the kagome lattice}

\author{D\'aniel V\"or\"os}
\affiliation{Department of Theoretical Physics, Institute of Physics,
Budapest University of Technology and Economics, Műegyetem rakpart 3, H-1111 Budapest, Hungary}
\affiliation{Institute for Solid State Physics and Optics, Wigner Research Centre for Physics, H-1525 Budapest, P.O. Box 49, Hungary}

\author{P\'eter Kr\'anitz}
\affiliation{Department of Theoretical Physics, Institute of Physics,
Budapest University of Technology and Economics, Műegyetem rakpart 3, H-1111 Budapest, Hungary}
\affiliation{Institute for Solid State Physics and Optics, Wigner Research Centre for Physics, H-1525 Budapest, P.O. Box 49, Hungary}

\author{Karlo Penc}
\affiliation{Institute for Solid State Physics and Optics, Wigner Research Centre for Physics, H-1525 Budapest, P.O. Box 49, Hungary}
\affiliation{Department of Physics, Indian Institute of Technology Madras, Chennai 600036, India}

\date{\today}

\begin{abstract}
We explore the Dirac spin liquid (DSL) as a candidate for the ground state of the Mott insulating phase of fermions with six flavors on the Kagome lattice, particularly focusing on realizations using $^{173}$Yb atoms in optical lattices.
Using mean-field theory and variational Monte Carlo simulations, we demonstrate that the Dirac spin liquid (DSL) has the lowest variational energy among SU(6) symmetry-preserving trial wave functions with a periodicity of a 12-site unit cell, as well as uniform chiral states with larger unit cells. 
It remains a local minimum even when small second-nearest neighbor and ring exchange interactions are introduced.
To characterize the DSL, we calculate the static and dynamic structure factor of the Gutzwiller projected wavefunction and compare it with mean-field calculations. 
The static structure factor shows triangular-shaped plateaus around the $\mathrm{K}$ points in the extended Brillouin zone, with small peaks at the corners of these plateaus.
The dynamical structure factor consists of a gapless continuum of fractionalized excitations.
Our study also presents several complementary results, including bounds for the ground state energy, methods for calculating three-site ring exchange expectations in the projective mean field, the boundary of ferromagnetic states, and the non-topological nature of flat bands in the DSL band structure.
\end{abstract}


\maketitle


\section{Introduction}

Competing interactions realized by geometrical frustration in spin-1/2 Mott insulators with SU(2) symmetry can give rise to quantum spin liquid ground states in quasi-two-dimensional systems. 
Alternatively, increasing the local Hilbert space from two to $N$ and enhancing the symmetry from SU(2) to SU($N$) induces quantum fluctuations that may favor a quantum spin liquid ground state.
  While such symmetry enhancements can unlock exotic quantum phases, their experimental realization is challenging. 
 Notable examples include optically trapped ultracold alkaline earth metal atoms, which achieve high symmetry due to the near-perfect SU($N = 2I + 1$) symmetry originating from the nuclear degree of freedom.
\cite{Argument_that_Yb_is_SU6_New_Journal_of_Physics_2009, First_two_orbital_SUN_realization_with_cold_atoms_Nature_2010}. 
Here, $I$ represents the nuclear spin, and $N$ can be as high as 10, as 
demonstrated with $^{87}$Sr isotopes exhibiting SU(10) symmetry \cite{Degenerate_Fermi_Gas_of_Stroncium_PRL_2010, Double_degenerate_Bose_Fermi_Mixture_of_Strontium_PRA_2010}. 
 Recently, an SU(6) symmetric Mott insulator was realized using $^{173}$Yb isotopes \cite{Degenerate_Fermi_Gas_of_Ytterbium_PRL_2007, SU2_x_SU6_PhysRevLett_2010, SU6_Nature_2012}, and further studied experimentally in \cite{SUN_one_dimension_experiment_Nature_2014, SUN_Equation_of_State_in_3D_Fermi_Hubbard_model_optical_lattice_experiment_PRX_2016, SpinOrbit_SU6_PhysRevA_2016, SU2_x_SU6_interisotope_inerorbital_interactions_PRA_2021, SUN_bosonization_experiment_PRX_2020, SU6_antifferomagnetic_correlations_Nature_2022}. 

 An SU($N = 2I + 1$) symmetric $N$-component Fermi gas in an optical lattice is well described by the SU($N$) Hubbard model \cite{SU6_Nature_2012, Argument_that_Yb_is_SU6_New_Journal_of_Physics_2009}
\begin{align}
    \mathcal{\tilde H} = &-t \sum_{\langle i,j \rangle, \sigma} (f^{\dagger}_{i,\sigma} f^{\phantom{\dagger}}_{j,\sigma} + \text{H.c.}) + U \sum_{i, \sigma' > \sigma} n_{i, \sigma} n_{i, \sigma'} \nonumber \\
    &+ \sum_{i,\sigma} \varepsilon_i n_{i, \sigma}, 
\label{eq:Hubbard_Hamiltonian}
\end{align}
where $f^{\phantom{\dagger}}_{i, \sigma}$ annihilates a fermion with spin (flavor) $\sigma \in \lbrace -I, \dots ,I  \rbrace $ on site $i$, $n_{i,\sigma} = f^{\dagger}_{i,\sigma} f^{\phantom{\dagger}}_{i,\sigma}$ is the fermion number operator, and $\varepsilon_i$ stands for the strength of the harmonic confinement potential. 
The Hamiltonian remains SU($N$) symmetric due to the independence of $t$, $U$, and $\varepsilon_i$ from $\sigma$. 
By tuning the scattering lengths, one can adjust the strength and sign of the interaction $U$ \cite{Kitagawa_Two_color_photoassociation_spectroscopy_of_ytterbium_atoms_and_the_precise_determinations_of_s_wave_scattering_lengths_PRA_2008,Escobar_Two-photon_photoassociative_spectroscopy_of_ultracold_Strontium_scatterin_length_determination_PRA_2008}. 
For strong enough on-site repulsion, the leading order perturbation theory in $t/U$ of the Hamiltonian (\ref{eq:Hubbard_Hamiltonian}) results in the SU($N$) symmetric antiferromagnetic Heisenberg model
\begin{equation}
\label{eq:Heisenberg_Hamiltonian}
    \mathcal{H} = J \sum_{\langle i, j\rangle} \mathbf{T}_{i} \cdot \mathbf{T}_{j} = J \sum_{\langle i, j\rangle} \sum_{a=1}^{N^2-1} T_{i}^{a} T_{j}^{a},
\end{equation}
where the summation runs over the $\langle i, j\rangle$ nearest neighbor sites, and $T_{i}^{a}$ are the SU($N$) spin operators, with $a=1,2,\dots N^2-1$, satisfying the commutation relations of the su($N$) Lie algebra. 
The filling of the Hubbard model determines the irreducible representation of the SU($N$) spins at each site and the dimension of the local Hilbert space.
Specifically, the $1/N$ filled Hubbard model with one fermion per site corresponds to an SU($N$) Heisenberg model describing spins in the $N$-dimensional fundamental representation.

While the SU(2) spins need the high frustration on the kagome lattice to stabilize a spin liquid phase, the SU(6) is a large enough symmetry to stabilize a liquid phase even on bipartite lattices. Chiral spin liquids for SU(6) Hubbard models were found for large enough filling (i.e., beyond the fundamental representation) on the square \cite{SUN_square_Heisenberg_large_N_limit_chiral_spin_liquid_with_semiclassical_argument_against_magnetic_order_PRL_2009,NOT_SU6_fundamental_SUN_square_lattice_chiral_spin_liquids_and_valence_cluster_states_and_self_consistent_minimalization_algorithm_PRB_2011}, cubic \cite{SUN_Heisenberg_3D_cubic_lattice_large_N_spin_liquids_PhysRevB_2013} and kagome \cite{Marston-Zeng_chiral_kagome_1991,hastings2000_gap_opening_with_anisotropy} lattices.
There were indications of chiral spin liquid ground states at 1/6 filling of the SU(6) Hubbard model on the triangular \cite{SUN_triangular_lattice_chiral_spin_liquid_Karlo_ED_VMC_PRL_2016,SUN_triangular_lattice_Gang_Chen_PRR_2021}, on the honeycomb \cite{Szirmai_SU6_honeycomb_chiral_spin_liquid_with_Gutzwiler_projection_PhysRevA_2011, SU6_and_SU4_honeycomb_Szirmai_PhysRevA_2013, SUN_honeycomb_PRB_using_J_2_to_induce_transition_from_chiral_spin_liquid_to_valence_cluster_solid_Gang_Chen_2022}, and the square \cite{SUN_square_Hubbard_has_a_chiral_spin_liquid_PRA_2016} lattices. The 1/6 filling corresponds to the fundamental representation of the SU(6) Heisenberg model. However, the honeycomb and square lattices were also argued to develop 6-site singlet plaquette orders \cite{SU6_honeycomb_plaquette_order_Karlo_PRB_2016, SU6_square_as_SU4_bilinear_biquadratic_Heisenberg_model_with_PEPS_ED_showing_plaquette_and_dimer_orders_PRB_2020, SU6_square_plaquette_order_found_with_ED_on_small_clusters_arxive_2023}. In contrast, the square lattice was further predicted to have a charge conjugation symmetry breaking ground state \cite{paramekanti_bilinear_biquadratic_2007}.

In this study, we investigate the SU(6) symmetric antiferromagnetic Heisenberg model in the fundamental representation on the kagome lattice using the variational Monte Carlo (VMC) method. 
Similar to the SU(2) Heisenberg model, which is not magnetically ordered \cite{SU2_Heisenberg_model_on_the_kagome_lattice_is_NOT_ordered}, we expect the SU(6) Heisenberg model to also exhibit disorder due to increased quantum fluctuations.

 We propose that the ground state of this model may be well described by the Dirac spin-liquid (DSL) state, previously suggested for the SU(2) case \cite{SU2_Dirac_spin_liquid_Ran_PRL_2007}. 
 To support our claim, we search for possible perturbations of the Dirac spin liquid in the subspace of magnetically disordered SU(6) singlet states (valence bond patterns assuming a 12-site unit cell), as discussed in Refs.~\onlinecite{hastings2000_gap_opening_with_anisotropy,SU2_Dirac_spin_liquid_Hermele_Ran_PRB_2008,Song_Wang_Vishwanath_monopole_and_chiral_instabilities_of_the_Dirac_spin_liquid_Nature_2019,Urban_Spin_Peierls_NatComm2024}. We found that the DSL is energetically the most favorable state among the ones considered here.
Furthermore, we also find the chiral spin liquid states to be higher in energy than the Dirac spin liquid, a matter of debate in the SU(2) case  \cite{Chalker_kagome_1992,Waldtmann_1998,SU2_Dirac_spin_liquid_Ran_PRL_2007,Sun_Kagome_chiral_QM_2024}. 
Gapless spin liquids have already been proposed as ground state candidates on various lattices, both for $N = 2$ \cite{Dirac_cone_staibility_with_Hermele_SUN_self_conjugate_large_N_limit_on_the_square_lattice_and_QED_PRB_2004, PhysRevLett.95.247203, SU2_Dirac_spin_liquid_Ran_PRL_2007, SU2_Dirac_spin_liquid_Hermele_Ran_PRB_2008, SU2_Dirac_spin_liquid_Hermele_Ran_PRB_2008, PhysRevB.87.060405, PhysRevB.89.020407, PhysRevB.93.144411, PhysRevX.7.031020, PhysRevX.8.011012, Becca, 2019PhRvX...9c1026F, Ferrari_kagome} and $N = 4$ \cite{SU4_Honeycomb, Tetramerization_of_SU4_honeycomb_Karlo, Mi_SU4_cikkunk}.

We note that two forms of stability can be distinguished for a fermionic variational wave function: quantum field-theoretical stability and energetic stability. The former requires that interactions between fermions become irrelevant in the renormalization group sense, leading to deconfined fermions. For Dirac fermions, this is expected to occur for $N > N_c$, although the precise value of $N_c$ remains an open question
\cite{Song_Wang_Vishwanath_monopole_and_chiral_instabilities_of_the_Dirac_spin_liquid_Nature_2019, Dirac_cone_staibility_Appelquist_PRL_1988, Nash_Higher_order_corrections_to_N_C_for_Dirac_cone_stability_PRL, Dirac_cone_staibility_Appelquist_PRL_1990, Dirac_cone_staibility_with_Hermele_SUN_self_conjugate_large_N_limit_on_the_square_lattice_and_QED_PRB_2004, Lu_honeycomb_SU_N_Dirac_cone_stability_PRB_2011}. 
While we have not explicitly checked quantum field-theoretical stability for the SU(6) case, the large $N$ value is certainly beneficial. For the remainder of this paper, 'stability' will refer exclusively to energetic stability.

We organize the paper as follows. 
In Sec.~\ref{sec:DSL}, we introduce the mean-field theory of the antiferromagnetic Heisenberg model, using the parton decomposition of the spin operators. We also introduce the Dirac spin liquid, which is the Gutzwiller projected wavefunction of the Fermi sea ground state of a hopping Hamiltonian having $\pi$-fluxes in every elementary plaquette of the Kagome lattice. 
We classify the perturbations of the DSL within the 12-site unit cell in Sec.~\ref{sec:possible_instabilities_of_the_Dirac_spin_liquid}, based on the irreducible representations of the $O_h$ group (which is isomorphic to the symmetry group of the 12-site unit cell, as explained in Appendix.~\ref{App:isomorphy_of_the_12_site_unit_cell_and_the_O_h_group}). 
In Sec.~\ref{sec:second_nearest_neighbor_and_ring_exchange}, we calculate the energies of the different perturbations of the DSL and show that the DSL has the lowest energy using a variational Monte Carlo method to sample the Gutzwiller projected wave functions.  
In Sec.~\ref{sec:structure_factor}, we calculate the flavor-flavor correlations and the structure factor; in Sec.~\ref{sec:dynamical_structure_factor}, we calculate the dynamical structure factor. 
We summarize our results in Sec.~\ref{sec:conclusion}. 
The appendices contain supplementary results.
Appendix~\ref{sec:gauge_trans} presents the gauge transformations of the DSL, which ensure the projective symmetry of the $\pi$-flux mean-field Hamiltonian.
Appendix~\ref{App:isomorphy_of_the_12_site_unit_cell_and_the_O_h_group} discusses the isomorphy of the symmetry group of the 12-site unit cell to the $O_h$ group. 
Appendix~\ref{sec:perms_diag} details the VMC method and the calculation of the ring exchange operators using diagonal expectation values. 
The projected mean field is presented in Appendix~\ref{sec:proj_mean_field}. 
In Appendix~\ref{sec:lower_bound}, we calculate the lower bound to the ground state energy. The stability of the ferromagnetic phase is discussed in Appendix~\ref{sec:ferro}. 
Finally, in Appendix~\ref{sec:flatBands}, we discuss the localized states related to the flat bands of the DSL ansatz. 

\section{Description of main results}

To assist the reader, we summarize our main findings below, along with the corresponding sections, figures, and tables where they are detailed.
  
\begin{itemize}

\item Perturbations of the DSL corresponding to symmetry-broken ansätze with real hopping amplitudes are classified according to their irreducible representations in Table~\ref{tab:real_perturbations}. The VMC energies for small perturbations are shown in Figs.~\ref{fig:Local_stability_of_1D_real_ansatze} and \ref{fig:Local_stability_of_2D_real_ansatze}, demonstrating the stability of the DSL.

\item Perturbations describing chiral trial wave functions with complex hopping amplitudes are classified based on their irreducible representations in Table~\ref{tab:complex_perturbations}. Their variational energies and expectation values are plotted in Fig.~\ref{fig:VMC_chiral_staggered}.

\item Chiral states with flux values of $\Phi=2\pi/3$ and $\pi/2$ through the elementary unit cell are discussed in Sec.~\ref{sec:chirals}, and their energies are provided in Table~\ref{tab:chirals}. 

\item The phase diagram in the parameter space of second-neighbor and ring exchanges is shown in Figures~\ref{fig:Local_stability_of_1D_real_ansatze}, \ref{fig:Local_stability_of_2D_real_ansatze},
\ref{fig:Global_stability_of_David_star}, and \ref{fig:global_stability_chiral_staggered}.

\item In Sec.\ref{sec:proj_mean_field}, we extended the projective mean-field method—an analytical approximation incorporating the occupancy constraint on a bond—to include three sites and SU($N$) fermions. The results are compared to VMC calculations in Fig.~\ref{fig:VMC_chiral_staggered}.

\item Figure~\ref{fig:structure_factors} illustrates the structure factor,  showing that the key features are well captured by the mean-field approximation once the corrections from Eq.~(\ref{eq:TdTMFcorr}) are included.

\item Figure~\ref{fig:dynamical_structure_factor_48} compares the dynamical structure factor obtained from the mean-field approximation and VMC, highlighting their qualitative and quantitative agreement for a relatively small 48-site cluster. Building on this, Fig.~\ref{fig:dynamical_structure_factor_continuum} displays the gapless continuum of fermionic excitations, the flavorons.

\item We provide a lower bound for the energy in Eqs.~(\ref{eq:P1_bound}) and (\ref{eq:T1T2_bound}).

\item The region of the stability of the fully polarized state against one- and two-magnon excitations is presented in Fig.~\ref{fig:FM_instab}. 

\item The localized states of the flat bands in the $\pi$-flux state are depicted in Fig.~\ref{fig:localized_states}. Unlike the conventional Kagome flat band, the flat band for the DSL is non-topological.

\end{itemize}

\section{Mean-field ground state and the Dirac spin liquid}
\label{sec:DSL}

\begin{figure}[bt]
\centering
\includegraphics[width=0.7\columnwidth]{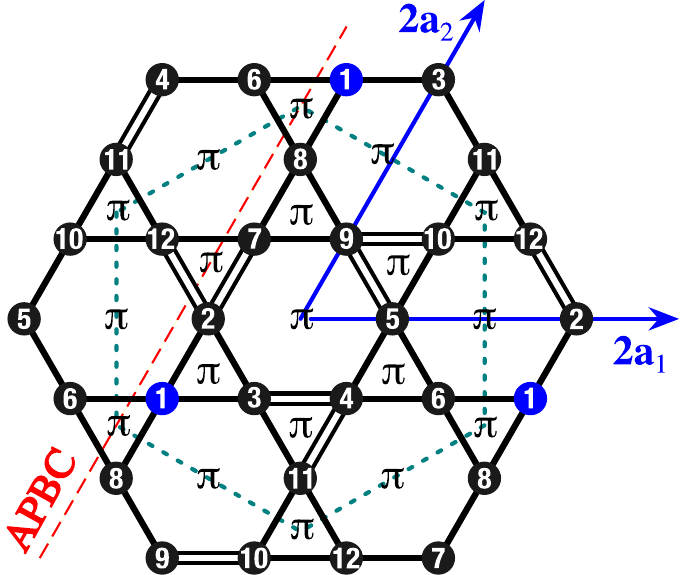}
\caption{The 12-site quadrupled unit cell of the mean-field Hamiltonian Eq.~(\ref{eq:mean_field_Hamiltonian}) in real space and Eq.~(\ref{eq:mean-field_Hamiltonian_in_k_space}) in reciprocal space, describing the Dirac spin liquid ansatz. The black solid bonds represent $t^{\text{DSL}}_{i,j} = - 1$ and the white (inverted) $t^{\text{DSL}}_{i,j} = 1$.
The product of the hoppings around every elementary triangular and hexagonal plaquette is $-1$, corresponding to a $\pi$ flux threading through each plaquette. The $2 \mathbf{a}_1$ and $2 \mathbf{a}_2$ are the primitive vectors of the quadrupled unit cell, where $\mathbf{a_1}= \left(1,0 \right)$ and $\mathbf{a}_2=\left(\frac{1}{2},\frac{\sqrt{3}}{2}\right)$ are the primitive vectors of the unit cell of the kagome lattice. The red dashed line shows the antiperiodic boundary condition along the edge of the cluster parallel to $\mathbf{a}_2$, which we impose to make the mean-field ground state non-degenerate.
\label{fig:DSL_with_APBC}
}
\end{figure}

In the fermionic parton mean-field approach \cite{Affleck_pi_flux,NOT_SU6_fundamental_Martson_Affleck_bilinear_biquadratic_PRL_1989}, the SU(6) spin operators in Eq.~(\ref{eq:Heisenberg_Hamiltonian}) are decomposed into fermionic degrees of freedom (partons):
\begin{equation}
  T_i^{a} = \frac{1}{2} \sum_{\sigma, \sigma' = A}^{F} f^{\dagger}_{i, \sigma} \lambda^{a}_{\sigma, \sigma'} f^{\phantom{\dagger}}_{i, \sigma'} \;
  \label{eq:spin_operator_with_partons_SU_4}
\end{equation}
where the  $6\times6$ matrices $\lambda^{a}$ (with $a \in \lbrace 1, \dots ,35 \rbrace$) generalize the eight SU(3) Gell-Mann matrices \cite{Gell_Mann_matrices}
to 35 $6\times 6$ matrices of the SU(6). In the fundamental representation, a singly occupied site with a fermion of flavor $\sigma \in \{A, B,\dots, F\}$ represents a spin basis state $\sigma$, and the system is one-sixth filled.

To obtain the mean-field approximation, we substitute Eq.~(\ref{eq:spin_operator_with_partons_SU_4}) into Eq.~(\ref{eq:Heisenberg_Hamiltonian}), replacing pairs of fermionic operators with their mean-field expectation values:
\begin{equation}
\label{eq:mean_field_Hamiltonian}
\mathcal{H}_{\text{MF}} = 
  \sum_{\sigma = A}^{F} 
  \sum_{\langle i,j \rangle} t_{i,j} 
    f^{\dagger}_{i,\sigma} 
    f^{\phantom{\dagger}}_{j,\sigma}.
\end{equation}
Here, the hopping amplitudes $t_{i,j}$ are determined by the self-consistency equation
\begin{equation}
t_{i,j} = J \sum_{\sigma = A}^{F} \langle \text{FS}|  f^{\dagger}_{i,\sigma} f^{\phantom{\dagger}}_{j,\sigma} | \text{FS} \rangle ,
\label{eq:self_consist}
\end{equation}
where $| \text{FS} \rangle$ is the Fermi sea ground state of $\mathcal{H}_{\text{MF}}$.

 However, instead of determining the $t_{i,j}$ self-consistently, we will treat them as independent variational parameters, minimizing the variational energy $\langle \psi | \mathcal{H} | \psi \rangle / \langle \psi | \psi \rangle$ for the Heisenberg Hamiltonian (\ref{eq:Heisenberg_Hamiltonian}). 
 The variational wavefunction $| \psi \rangle$, which approximates the ground state of the Heisenberg model, is constructed by applying the Gutzwiller projector $P_{\text{G}}$ to the Fermi sea, such that $| \psi \rangle = P_{\text{G}} | \text{FS} \rangle$. The role of $P_{\text{G}}$ is to enforce single occupancy on each lattice site, thereby correctly mapping the fermionic Hilbert space to the Hilbert space of the SU(6) spin operators. Both $|\text{FS}\rangle$ and $P_{\text{G}}|\text{FS}\rangle$ are SU(6) singlets and are independent of the overall magnitude of the hopping terms, depending only on their relative ratios.
 Different choices of $t_{i,j}$ lead to different mean-field Hamiltonians $\mathcal{H}_{\text{MF}}$, Fermi sea states $|\text{FS}\rangle$, and corresponding variational wavefunctions $|\psi \rangle = P_{\text{G}}|\text{FS}\rangle$, allowing us to compare their variational energies. We refer to the set of hopping parameters $t_{i,j}$ and its associated $\mathcal{H}_{\text{MF}}$ as an ansatz.

In the mean-field Hamiltonian (\ref{eq:mean_field_Hamiltonian}) we neglect the pairing terms like $f^{\dagger}_{i,\sigma} f^{\dagger}_{j, \sigma'}$, which naturally appears in the context of the SU(2) symmetric Heisenberg model as a singlet operator $f^{\dagger}_{i,\uparrow} f^{\dagger}_{j, \downarrow}$ and describe a $Z_2$ spin 
liquid \cite{Projective_Symmetry_Group_Wen}. Namely, in the fundamental representation of SU(6), six fermions combine to form a singlet, and this is beyond the mean-field description. 
However, already in the case of the SU(2) Heisenberg model on the kagome lattice, the U(1) Dirac spin liquid ansatz was shown to give a good approximation of the actual ground state \cite{SU2_Dirac_spin_liquid_Ran_PRL_2007, SU2_Dirac_spin_liquid_Hermele_Ran_PRB_2008, PhysRevB.84.020407, PhysRevB.87.060405, PhysRevB.89.020407, PhysRevB.91.020402, PhysRevX.7.031020, Ferrari_kagome}. 

The DSL corresponds to real uniform hopping amplitudes with $t^{\text{DSL}}_{i,j}= \pm 1$, where we choose the signs such that their directed product around every elementary triangular and hexagonal plaquette is negative, as shown in Fig.~(\ref{fig:DSL_with_APBC}). The flux $\phi$ of a plaquette is defined as 
\begin{equation}
e^{i \phi} = \prod_{\langle i, j\rangle \in \text{plaquette}} \frac{t_{i,j}}{|t_{i,j}|}.
\end{equation}
The negative product around every plaquette corresponds to $\phi^{\text{DSL}} = \pi$, so we refer to the Fermi sea of the DSL as $|\pi \text{FS} \rangle$, with its associated flux structure denoted as $\pi_{\hexagon} \pi_{\triangle} \pi_{\bigtriangledown}$. 
As we will see later, this flux structure also allows different hopping amplitudes, while the Dirac spin liquid (DSL) specifically requires all hopping amplitudes to have equal magnitudes.
Let us also remark that using this convention, the regular kagome lattice with uniform negative hopping amplitudes $-t$ and a flat band at the top with energy $2t$ corresponds to having a $\pi$ flux through the triangles and $0$ through the hexagon, i.e., $0_{\hexagon} \pi_{\triangle} \pi_{\bigtriangledown}$.

The Hamiltonian introduced in \cite{SU2_Dirac_spin_liquid_Ran_PRL_2007} requires a doubled unit cell with a 6-site basis to accommodate the hoppings generating the $\pi$ fluxes.
Here, we will use a quadrupled unit cell with a 12-site basis, with a choice of hopping amplitudes shown in Fig.~\ref{fig:DSL_with_APBC}. The connection between the two ansätze is discussed in Appendix \ref{App:connection_between_the_DSL_ansatze}.
Even though the hopping structures of both Hamiltonians break the wallpaper group symmetries $\mathsf{g}$ of the kagome lattice, one can restore these symmetries by combining the symmetry operations $\mathsf{g}$ with suitable site-dependent but flavor-independent gauge transformations $G:f^{{\dagger}}_{j,\sigma} \rightarrow e^{i \phi(j)}f^{{\dagger}}_{j,\sigma}$, so that
\begin{equation}
\mathcal{H}_{\text{MF}} = G_{\mathsf{g}} \mathsf{g} \mathcal{H}_{\text{MF}} \mathsf{g}^{-1} G^{-1}_{\mathsf{g}},
\label{eq:projective_symmetry_of_H_MF} 
\end{equation}
which is called projective symmetry \cite{First_Parton_construction_article_Jain, *Second_Parton_construction_article, Projective_construction_Wen, Projective_Symmetry_Group_Wen} (see Appendix \ref{App:projective_symmetries}). Since the spin operators are insensitive to $G_{\mathsf{g}}$, their expectation values taken in states $|\text{FS}\rangle$ and $P_{\text{G}} |\text{FS}\rangle$ will possess all the symmetries of the kagome lattice, justifying the name spin liquid.
To make $|\text{FS}\rangle$ non-degenerate for all system sizes, we have applied antiperiodic boundary condition (APBC) on one boundary and periodic on the other, as shown in Fig.~\ref{fig:DSL_with_APBC}. The APBC can break the $C_6$ and $\sigma$ projective symmetries, which are then restored in the thermodynamic limit (see \cite{Mi_SU4_cikkunk} for details).

\begin{figure}[t]
\centering
\includegraphics[width=0.9\columnwidth]{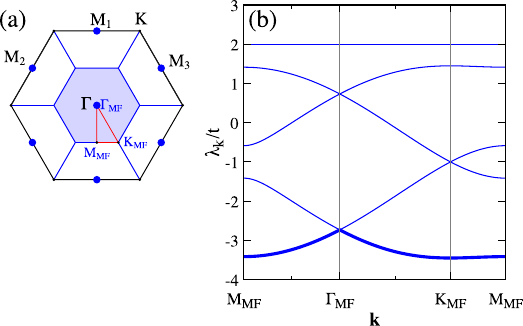}
\caption{
(a) The Brillouin zones of the kagome lattice (black hexagon) and of the quadrupled unit cell having 12 sites (blue shaded hexagon). The blue points at $\Gamma$ and $\mathrm{M}$ momenta denote the Fermi points of the DSL (the $\mathrm{M}$'s are also $\Gamma_{\text{MF}}$).
(b) The bands of the DSL mean-field Hamiltonian $\mathcal{H}_{\text{DSL}}(\mathbf{k})$ (\ref{eq:mean-field_Hamiltonian_in_k_space}) along the path $\mathrm{M}_{\text{MF}}-\Gamma_{\text{MF}}-\mathrm{K}_{\text{MF}}-\mathrm{M}_{\text{MF}}$ in the Brillouin zone of the 12-site unit cell. 
Partons describing SU(2) spins occupy the lowest three bands up to the Dirac point at the $\Gamma$ point, while for the SU(6), they occupy only the lowest band (drawn by the thick line). Each dispersive band is two-fold degenerate, while the flat band is four-fold degenerate. 
\label{fig:DSL_bands}
}
\end{figure}

In the reciprocal space, assuming the quadrupled unit cell with 12 sites and periodic boundary conditions, the mean-field Hamiltonian is the following $12\times12$ matrix
\begin{widetext}
\begin{equation}
\mathcal{H}_{\text{DSL}}(\mathbf{k}) = 
\left(
\begin{array}{cccccccccccc}
0 & -1 & -1 & 0 & 0 & -\bar r_1 & 0 & -\bar r_2 & 0 & 0 & 0 & 0 \\
-1 & 0 & -1 & 0 & 0 & 0 & 1 & 0 & 0 & 0 & 0 & \bar r_1\\
-1 & -1 & 0 & 1 & 0 & 0 & 0 & 0 & 0 & 0 & -\bar r_2& 0 \\
0 & 0 & 1 & 0 & -1 & -1 & 0 & 0 & 0 & 0 & \bar r_2& 0 \\
0 & 0 & 0 & -1 & 0 & -1 & 0 & 0 & 1 & -1 & 0 & 0 \\
-r_1& 0 & 0 & -1 & -1 & 0 & 0 & -r_1 \bar r_2& 0 & 0 & 0 & 0 \\
0 & 1 & 0 & 0 & 0 & 0 & 0 & -1 & -1 & 0 & 0 & -\bar r_1\\
-r_2& 0 & 0 & 0 & 0 & -r_2 \bar r_1& -1 & 0 & -1 & 0 & 0 & 0 \\
0 & 0 & 0 & 0 & 1 & 0 & -1 & -1 & 0 & 1 & 0 & 0 \\
0 & 0 & 0 & 0 & -1 & 0 & 0 & 0 & 1 & 0 & -1 & -1 \\
0 & 0 & -r_2& r_2& 0 & 0 & 0 & 0 & 0 & -1 & 0 & -1 \\
0 & r_1& 0 & 0 & 0 & 0 & -r_1& 0 & 0 & -1 & -1 & 0 \\
\end{array}
\right),
\label{eq:mean-field_Hamiltonian_in_k_space}
\end{equation}
\end{widetext}
where the $i$-th row and $j$-th column contains $t^{\text{DSL}}_{i,j}=\pm 1$, 
multiplied by $r_1 = e^{i\mathbf{k} \cdot 2\mathbf{a}_1}$ or $r_2 = e^{i\mathbf{k} \cdot 2\mathbf{a}_2}$ when $t^{\text{DSL}}_{i,j}$ crosses the unit cell boundary (the $\bar r_1 = e^{-i\mathbf{k} \cdot 2 \mathbf{a}_1}$ and $\bar r_2 = e^{-i\mathbf{k} \cdot 2\mathbf{a}_2}$ are the complex conjugates). 
The effect of the antiperiodic boundary condition (shown in Fig.~\ref{fig:DSL_with_APBC}) is to multiply the hoppings crossing the antiperiodic boundary by $-1$, resulting in $r_1 \to e^{i \pi/L_1} r_1$, where $L_1$ is the number of times the quadrupled unit cell is repeated in the direction $2\mathbf{a}_1$ in the finite cluster. Alternatively, one can shift the $\mathbf{k}$ momentum values to absorb the minus sign. 
The characteristic polynomial of the $\mathcal{H}_{\text{DSL}}(\mathbf{k})$, providing the eigenvalues $\lambda_{\mathbf{k}}$,  is
\begin{equation}
0=  \left( \lambda_{\mathbf{k}}^4 + 4 t \lambda_{\mathbf{k}}^3- 8 t^3 \lambda_{\mathbf{k}} + 2 t^4 \gamma_{\mathbf{k}} \right)^2 \left( \lambda_{\mathbf{k}} - 2 t \right)^4,
\label{eq:characteristic_polynomial}
\end{equation}
where 
\begin{equation}
\gamma_{\mathbf{k}} =  \cos 2\mathbf{k} \cdot \mathbf{a_1}  + 
\cos 2\mathbf{k} \cdot \mathbf{a_2} + \cos 2\mathbf{k} \cdot [\mathbf{a_1} - \mathbf{a_2} ] -1 .
\end{equation} 
The band structure consists of a four-fold degenerate flat band at $\lambda_{\mathbf{k}} = 2 t$ and four two-fold degenerate dispersive bands, shown in Fig.~\ref{fig:DSL_bands}. Unlike in the zero-flux case, the dispersive bands are separated by a gap from the flat bands, and the dispersive bands touch at Dirac points (see Appendix \ref{sec:flatBands} for more details about flat bands).

Reversing the sign of every hopping in the DSL ansatz in Fig.~\ref{fig:DSL_with_APBC} changes the flux structure from $\pi_{\hexagon}\pi_{\triangle}\pi_{\bigtriangledown}$ to $\pi_{\hexagon}0_{\triangle}0_{\bigtriangledown}$, therefore we denote this ansatz as the uniform $\pi_{\hexagon}0_{\triangle}0_{\bigtriangledown}$. In the case of SU(2), these two states are equivalent due to a symmetry under a combination of a spin-rotation and a time-reversal transformation (as explained in section 9.2.7 in Ref.~\onlinecite{Wens_book}, also in \cite{Projective_Symmetry_Group_Wen}). However, they are inequivalent in the case of SU(6). The mean-field band structure of the uniform $\pi_{\hexagon}0_{\triangle}0_{\bigtriangledown}$ ansatz is that of the DSL in Fig.~\ref{fig:DSL_bands} turned upside-down, so that the four-fold degenerate high energy flat band becomes the lowest energy band in the uniform $\pi_{\hexagon}0_{\triangle}0_{\bigtriangledown}$ ansatz, which makes the Fermi sea highly degenerate.

The Hamiltonian discussed in \cite{SU2_Dirac_spin_liquid_Ran_PRL_2007} with a doubled unit cell has the same band structure as the quadrupled unit cell ansatz, with the difference that the degeneracy of each band is halved and the size of the Brillouin zone gets doubled.
In the case of SU(2), the mean-field ground state has a half-filled band structure, filling the lowest three bands up to a Dirac-Fermi point. For SU(3), the third filling also gives a Dirac point, but it is unstable and the ground state trimerizes instead \cite{Arovas_simplex_sun,*Corboz_simplex_sun}.

In the case of SU(6), to fulfill the one fermion per site constraint in average, i.e., $\langle \text{FS} | n_{i} | \text{FS} \rangle = \sum_{\sigma} \langle \text{FS} | f^{\dagger}_{i,\sigma} f^{\phantom{\dagger}}_{i,\sigma} | \text{FS} \rangle = 1$ $\forall i$, the system has to be $1/6$ filled. Consequently, the mean-field ground state fills only the lowest energy band up to a Dirac Fermi point.
For the fermionic correlations, we get
\begin{equation}
  |\langle \pi \text{FS}|  f^{\dagger}_{i,\sigma} f^{\phantom{\dagger}}_{j,\sigma} | \pi \text{FS} \rangle| = 0.138051\,,
\label{eq:fdf}
\end{equation}
independently of the flavor $\sigma$. From Eq.~(\ref{eq:self_consist}), the self-consistency gives $t_{i,j} = \pm 0.828306 J$ in the thermodynamic limit, with signs chosen to satisfy the DSL ansatz.

Using variational Monte Carlo calculations to evaluate the Gutzwiller projected Fermi sea of the DSL ansatz (see Appendix~\ref{sec:VMC} for details), we got 
\begin{equation}
\langle\mathbf{T}_{i} \cdot \mathbf{T}_{j} \rangle  = -0.4276 \pm 0.0001
\end{equation}
 for the energy of the nearest neighbor bonds. We also determined an exact lower bound for the energy in Sec.~\ref{sec:lower_bound}, $\langle \mathbf{T}_1\cdot \mathbf{T}_2 \rangle \geq -0.472623 $. Since VMC is a variational treatment that provides an upper bound, the energy shall be between these values. Experimentally, one can use lattice modulation spectroscopy \cite{Greif_PRL_2011} to measure the nearest-neighbor correlations, see Refs.~\onlinecite{SU6_Nature_2012,SU6_antifferomagnetic_correlations_Nature_2022} for the SU(6) symmetric $^{173}$Yb cold atoms. 

\section{Symmetry classification of the perturbations of the Dirac spin liquid in the quadrupled unit cell}
\label{sec:possible_instabilities_of_the_Dirac_spin_liquid}

\begin{widetext}
\begin{table*}[bt]
\caption{
The symmetry classification of the real-valued perturbations of the DSL according to the irreducible representations of a 12-site unit cell.
The first row lists the (directed) bonds $\langle i,j \rangle$ of the upward pointing triangles, the second the bonds of the downward pointing triangles $C_2 \langle i,j \rangle$ (the $C_2$ rotates the upward pointing triangles into downward-pointing ones, and vice-versa). The third and fourth rows display the unperturbed hoppings in the $\mathcal{H}_{\text{DSL}}(\mathbf{k})$ given by Eq.~(\ref{eq:mean-field_Hamiltonian_in_k_space}). 
The $A_1$, $E$, $T_1$, and $T_2$ in the first column are the irreducible representations (irreps). Subsequent columns are amplitudes of the perturbations $\tilde t_{i,j}$, where the $\langle i,j \rangle$ nearest neighbor pairs are given in the first two rows. From each irreducible representation, there is a gerade and an ungerade one. For the gerade irreps, $\tilde t_{C_2(i,j)} = \tilde t_{i,j}$, while for the ungerade ones $\tilde t_{C_2(i,j)} = -\tilde t_{i,j}$. Every row of an irrep defines an ansatz belonging to the irrep, as does any linear combination of the rows of a multidimensional irrep. The hopping structures corresponding to these ansätze are drawn in Figs.~\ref{fig:Local_stability_of_1D_real_ansatze} and \ref{fig:Local_stability_of_2D_real_ansatze}.  The last column indicates the degeneracy of the bands for the gerade and ungerade cases, respectively.
\label{tab:real_perturbations}
}
\begin{ruledtabular}
\begin{tabular}{cccccccccccccc}
$\langle i,j \rangle$	
&$\langle 2, 1 \rangle$	&$\langle 1, 3 \rangle$	&$\langle 3, 2 \rangle$	&$\langle 5, 4 \rangle$	&$\langle 4, 6 \rangle$	&$\langle 6, 5 \rangle$	&$\langle 8, 7 \rangle$	&$\langle 7, 9 \rangle$	&$\langle 9, 8 \rangle$	&$\langle 11, 10 \rangle$	&$\langle 10, 12 \rangle$	&$\langle 12, 11 \rangle$\\
$C_2\langle i,j \rangle$	
&$\langle 5, 10 \rangle$	&$\langle 10,9 \rangle$	&$\langle 9, 5 \rangle$	&$\langle 2, 7 \rangle$	&$\langle 7, 12 \rangle$	&$\langle 12, 2 \rangle$	&$\langle 11, 4 \rangle$	&$\langle 4, 3 \rangle$	&$\langle 3, 11 \rangle$	&$\langle 8, 1 \rangle$	&$\langle 1, 6 \rangle$	&$\langle 6, 8 \rangle$\\
$t^{\text{DSL}}_{i,j}$ &$-1$	&$-1$	&$-1$	&$-1$	&$-1$	&$-1$	&$-1$	&$-1$	&$-1$	&$-1$	&$-1$	&$-1$\\
$t^{\text{DSL}}_{C_2(i,j)}$ &$-1$	&$1$	&$1$	&$1$	&$-\bar r_1$	&$r_1$	&$r_2$	&$1$	&$-\bar r_2$	&$-r_2$	&$-\bar r_1$	&$-r_1 \bar r_2$\\
\hline
irrep. &\multicolumn{12}{c}{$\tilde t_{i,j}$}	&deg.\\
\hline
$A_1$	&$1$	&$1$	&$1$	&$1$	&$1$	&$1$	&$1$	&$1$	&$1$	&$1$	&$1$	&$1$	&$\times2,\times2$\\
\\
\multirow{2}{*}{$E$}	&$0$	&$1$	&$-1$	&$0$	&$1$	&$-1$	&$0$	&$1$	&$-1$	&$0$	&$1$	&$-1$	&\multirow{2}{*}{$\times2,\times2$}\\
&$\frac{2}{\sqrt{3}}$	&$\frac{-1}{\sqrt{3}}$	&$\frac{-1}{\sqrt{3}}$	&$\frac{2}{\sqrt{3}}$	&$\frac{-1}{\sqrt{3}}$	&$\frac{-1}{\sqrt{3}}$	&$\frac{2}{\sqrt{3}}$	&$\frac{-1}{\sqrt{3}}$	&$\frac{-1}{\sqrt{3}}$	&$\frac{2}{\sqrt{3}}$	&$\frac{-1}{\sqrt{3}}$	&$\frac{-1}{\sqrt{3}}$	&\\
\\
\multirow{3}{*}{$T_1$}	
&$\beta$	&$\beta$	&$\alpha$	&$-\beta$	&$-\beta$	&$-\alpha$	&$-\beta$	&$-\beta$	&$-\alpha$	&$\beta$	&$\beta$	&$\alpha$	&\multirow{3}{*}{$\times2,\times1$}\\
&$\beta$	&$\alpha$	&$\beta$	&$\beta$	&$\alpha$	&$\beta$	&$-\beta$	&$-\alpha$	&$-\beta$	&$-\beta$	&$-\alpha$	&$-\beta$	&\\
&$\alpha$	&$\beta$	&$\beta$	&$-\alpha$	&$-\beta$	&$-\beta$	&$\alpha$	&$\beta$	&$\beta$	&$-\alpha$	&$-\beta$	&$-\beta$	&\\
\\
\multirow{3}{*}{$T_2$}	&$0$	&$1$	&$-1$	&$0$	&$-1$	&$1$	&$0$	&$1$	&$-1$	&$0$	&$-1$	&$1$	&\multirow{3}{*}{$\times2,\times1$}\\
&$1$	&$0$	&$-1$	&$1$	&$0$	&$-1$	&$-1$	&$0$	&$1$	&$-1$	&$0$	&$1$	&\\
&$1$	&$-1$	&$0$	&$-1$	&$1$	&$0$	&$-1$	&$1$	&$0$	&$1$	&$-1$	&$0$	&\\
\end{tabular}
\end{ruledtabular}
\end{table*}%
\end{widetext}

Since we are interested in the energetical stability of DSL against perturbations of the mean-field ansatz, it is useful to classify the possible perturbations as irreducible representations of the point group symmetries of the 12-site unit cell. To do so, we modify the hoppings $t^{\text{DSL}}_{i,j}$ of the DSL mean-field Hamiltonian, $\mathcal{H}_{\text{DSL}}(\mathbf{k})$ in Eqs.~(\ref{eq:mean-field_Hamiltonian_in_k_space}) and (\ref{eq:mean_field_Hamiltonian}), as
\begin{equation}
t_{i,j} = t^{\text{DSL}}_{i,j} (1 + \delta \tilde t_{i,j}),
\label{eq:perturbation_of_the_hoppings}
\end{equation}
where $0 < \delta \ll 1$, $|\tilde t_{i,j}| = \mathcal{O}(1)$. The Gutzwiller projected ground state of such a perturbed mean-field Hamiltonian yields a modified wavefunction $|\psi_{\tilde t_{i,j}} \rangle$. We then compute the variational energy $ \langle \psi | \mathcal{H} |\psi \rangle / \langle \psi |\psi \rangle$ for the perturbed wavefunctions associated with different irreducible representations (irreps). 
If the variational energy of any perturbed ansatz is lower than that of the DSL, this indicates that the DSL is no longer the ground state, and we say that the DSL is locally unstable against the perturbations in Eq.~(\ref{eq:perturbation_of_the_hoppings}). For clarity, we neglect on-site chemical potentials, though they can be easily incorporated.

Following Landau's theory of phase transitions, we expand the grand canonical potential
\begin{equation}
  \Omega(T,\mu) = -T \sum_{\mathbf{k}} \Tr \ln \left( 1+ e^{\beta(\mu \mathcal{N} -  \mathcal{H}_{\mathrm{MF}}(\mathbf{k}))}\right)
\end{equation}
in powers of the perturbation $\delta$,
 where $T=\beta^{-1}$ is the temperature, $\mu$ is the chemical potential ensuring the 1/6-th filling, and the summation is over the momenta $\mathbf{k}$ in the Brillouin zone.
To achieve this, we first expand $\Omega(T,\mu)$ in powers of $\beta$,
\begin{equation}
  \Omega(T,\mu) = T N_s \ln 2 + \sum_{l=0}^{\infty} o_l \beta^l \sum_{\mathbf{k}}\Tr \left\{ \left[\mu \mathcal{N} -  \mathcal{H}_{\mathrm{MF}}(\mathbf{k})\right]^{l+1} \right\} \;,
 \end{equation} 
where $N_s$ is the number of sites.
We expand the traces of the matrix powers in $\delta$ as
\begin{align} 
\Tr \left[ (\mu \mathcal{N} -  \mathcal{H}_{\mathrm{MF}}(q))^l \right] &= \Tr \left[ (\mu \mathcal{N} -  \mathcal{H}_{\mathrm{DSL}}(q))^l \right]  \bigg| 
\nonumber\\
&+ \delta^2 \sum_{\langle i,j \rangle , \langle i',j' \rangle} \tilde t_{i,j} Q^{(l)}_{\langle i,j \rangle , \langle i',j' \rangle} \tilde t_{i',j'} 
\nonumber\\ 
&+ \mathcal{O}(\delta^3).
\label{eq:expansion_of_the_traces_as_a_function_of_delta}
\end{align}
Only gauge-invariant terms appear in the expansion, so the first-order terms linear in $\tilde t_{i',j'}'$ $\tilde t_{i,' j'}'$ are absent.
Block diagonalizing the matrices $Q^{(l)}_{\langle i,j \rangle,  \langle i',j' \rangle}$ for all $l$ simultaneously, we can identify the different irreducible representations of the $\tilde t_{i,j}$ which transform separately under the elements of the symmetry group of the 12-site unit cell. %
Table~\ref{tab:real_perturbations} lists the resulting linear combinations of the $\tilde t_{i,j}$ which can
perturb the Dirac spin liquid and introduce phases that break symmetries. 
 Since the symmetry group of the 12-site unit cell is isomorphic to the $O_h$ group (see Appendix.~\ref{App:isomorphy_of_the_12_site_unit_cell_and_the_O_h_group}), we classify the irreducible representations of the $\tilde t_{i,j}$ using the character table of the $O_h$ group. 
We also take advantage of the fact that the center of the $O_h$ group is the identity and the inversion that constitute a normal subgroup. Consequently, the irreducible representations of the $O_h$ are either even (gerade) and odd (ungerade) under the inversion. The inversion in the $O_h$ corresponds to the twofold rotation $C_2$ around the center of a hexagon in the wallpaper group, so $\tilde t_{C_2(i,j)} = +\tilde t_{i,j}$ in an even (gerade, $g$) irreducible representation, and $\tilde t_{C_2(i,j)} = -\tilde t_{i,j}$ in an odd (ungerade, $u$) one. For each irreducible representation, there is a gerade and an ungerade one. During the process, we need to take into account that $\tilde t_{i,j} = \tilde t_{j, i}^*$, so we classified the real and complex-valued $\tilde t_{i,j}$-s separately. Let us also remark that during the diagonalization of the terms in the expansion (\ref{eq:expansion_of_the_traces_as_a_function_of_delta}), we encountered higher dimensional (e.g. five-dimensional) manifolds that are combinations of the irreducible representations of the $O_h$ group, suggesting a higher symmetry group, but we were not able to identify it.

Our approach is related to the classification of charge- and bond-density wave order parameters in Ref.~\onlinecite{Feng_PhysRevB.104.165136}. There, triple-$\mathbf{Q}$ order parameters were considered, $\mathbf{Q}$ momenta being the $\mathrm{M}_1$, $\mathrm{M}_2$, and $\mathrm{M}_3$ points in the Brillouin zone of the Kagome lattice where van-Hove singularities occur in the band structure. The periodicity of these density waves is compatible with the 12-site unit cell.  

\subsection{Real-valued $\tilde t_{i,j}$ perturbations of the DSL} \label{sec:real_instabilities}

Here, we consider time reversal invariant perturbations, which modify the strength of the hoppings, leaving them real and their signs identical to those of the DSL, keeping the $\pi_{\hexagon} \pi_{\triangle} \pi_{\bigtriangledown}$ flux structure unchanged. There are 24 nearest-neighbor bonds, each having a $\tilde t_{j, i}$. 
They transform according to the one-dimensional $A_{1g}$ and $A_{1u}$, the two-dimensional $E_g$ and $E_u$, the three-dimensional $T_{2g}$ and $T_{2u}$, and two copies (with parameters $\alpha$ and $\beta$) of the three-dimensional $T_{1g}$ and $T_{1u}$ irreducible representation, as shown in Tab.~\ref{tab:real_perturbations}. 
The hopping patterns corresponding to these ansätze are shown in the first column of Figs.~\ref{fig:Local_stability_of_2D_real_ansatze} and \ref{fig:Local_stability_of_1D_real_ansatze}.

\begin{figure}[t]
\centering
\includegraphics[width=0.6\columnwidth]{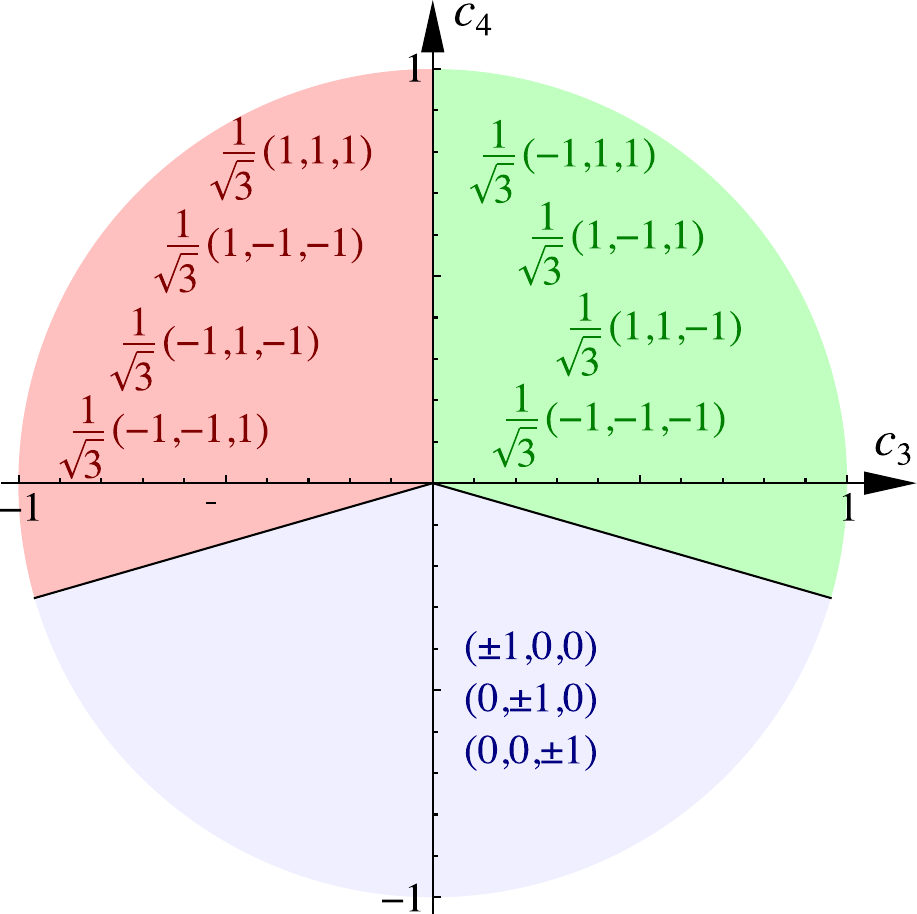}
\caption{\label{fig:c3c4}
  The plot of the $(v_1,v_2,v_3)$ values that minimize the symmetry invariant form of the energy Eq.~(\ref{eq:c2c3c4}) of a three-dimensional irreducible representation ($T_1$ or $T_2$).}
\end{figure}

In the case of higher dimensional irreducible representations, the actual perturbation is a linear combination of the basis. For example, $\sum_{i=1,2,3} v_i T^i_{1g}$ in the case of the three dimensional $T_{1g}$, where we denote the rows of an irrep by a number in the superscript. Assuming that the norm of the basis states is the same, the perturbation will contribute to the ground state energy (or the grand canonical potential) as
\begin{multline}
E_\delta = E_0 + 
c_2 (v_1^2+v_2^2+v_3^2) \delta^2 
+ c_3 v_1 v_2 v_3 \delta^3 \\
+ c_4  (v_1^4+v_2^4+v_3^4) \delta^4,
\label{eq:c2c3c4}
\end{multline}
where $c_2$, $c_3$, and $c_4$ are the coefficients of the $v_1^2+v_2^2+v_3^2$, $v_1 v_2 v_3$, and $v_1^4+v_2^4+v_3^4$ invariants. In the case of a two- and three-dimensional irreducible representation, the normalized linear combination of the rows of an irreducible representation will have the same contribution in $\delta^2$ in the grand canonical potential. 
The perturbation is unstable when $c_2<0$, and the coefficients $c_3$ and $c_4$ of the higher order invariant will determine the precise form of the perturbation. Fig.~\ref{fig:c3c4} presents the minimal energy solutions for the $(v_1,v_2,v_3)$ vector as a function of the $c_3$ and $c_4$. For the even (gerade) irreducible representations, we may assume that the cubic term prevails over the quartic one and the optimal perturbations are the linear combinations with $|v_1|=|v_2| = |v_3| = 1/\sqrt{3}$, with signs such that $v_1 v_2 v_3 c_3 <0$. For the odd (ungerade) irreducible representations, the inversion symmetry cancels the cubic term ($c_3=0$), and the optimal linear combinations are $|v_1|=|v_2| = |v_3| = 1/\sqrt{3}$ for $c_4>0$ and $|v_1|=1$, $v_2=v_3=0$ and equivalents for $c_4<0$.  In the case of the two-dimensional irreducible representations, the cubic invariant is $v_1^3-3v_1 v_2^2$ for both $E_g$ and $E_u$.

Consequently, it is enough to check the variational energy for the ansätze with either $|v_1|=|v_2|=|v_3| = 1/\sqrt{3}$ or $|v_1|=1$,$v_2=v_3=0$. This reduces the possibilities significantly because the most general $v_1 T_{1}^1 + v_2 T_{1}^2 + v_3 T_{1}^3$ could have five free parameters (after fixing normalization).

In Fig.~\ref{fig:Local_stability_of_1D_real_ansatze} we show the single-parameter ansätze $ T_{2g}^1 -  T_{2g}^2 -  T_{2g}^3$, $ T_{2g}^1$, $ T_{2u}^1 -  T_{2u}^2 -  T_{2u}^3$, $ T_{2u}^1$, and $A_{1u}$ with the parameter $\delta$ (here we allow negative $\delta$, which is like multiplying the $\tilde t_{i,j}$ by $-1$). The $ T_{2g}^1 -  T_{2g}^2 -  T_{2g}^3$ (first column in figure) coincides with the famous "pinwheel" pattern in deformed kagome material Rb$_2$Cu$_3$SnF$_{12}$ \cite{Matan:2010cl}.
We did not plot the $A_{1g}$ since it changes the hoppings uniformly, leaving the wave function equivalent to the DSL. 

In Fig.~\ref{fig:Local_stability_of_2D_real_ansatze} we show the two-parameter ansätze 
$- T_{1g}^1 +  T_{1g}^2 +  T_{1g}^3$, $ T_{1g}^3$, 
$ -T_{1u}^1 +  T_{1u}^2 +  T_{1u}^3$, $ T_{1u}^3$,
$ v_1 E_{g}^1 + v_2  E_{g}^2$, and $v_1  E_{u}^1 + v_2 E_{u}^2 $. 
To make the comparison of the different ansätze in Fig.~\ref{fig:Local_stability_of_2D_real_ansatze} unambiguous, we have collected the different hopping amplitudes that appear in Tab.~\ref{tab:hoppings_with_colors_for_local_2_parameter_ansatzes}, also listing their colors.

\begin{table}[h]
\begin{tabular}{ccc}
\hline
\hline
ansatz & color & relative hoppings $t_{i,j}/t^{\text{DSL}}_{i,j}$ \\
\\
\multirow{3}{*}{$- T_{1g}^1 +  T_{1g}^2 +  T_{1g}^3$}	& red & $ t_{\DavidStarOut} \approx 1$\\
& green & $ t_{\hexagon} \approx 1 + 2( \beta - \alpha) \delta $\\
& blue & $ t_{\triangle} \approx 1 - 2( \beta + \alpha) \delta $ \\
\\
\multirow{4}{*}{$ T_{1g}^3 $} & dark green & $  1 + \alpha \delta$ \\
&light green & $  1 - \alpha \delta$ \\
& dark red & $  1 + \beta \delta$ \\
& light red & $  1 - \beta \delta$ \\
\\
\multirow{6}{*}{$- T_{1u}^1 +  T_{1u}^2 +  T_{1u}^3$} & dark green &  $ 1 + (2\beta - \alpha) \delta$ \\
& light green &  $ 1 - (2\beta - \alpha) \delta$ \\
& dark red &  $ 1 + \alpha \delta$ \\
& light red &  $ 1 - \alpha \delta$ \\
& light blue &  $ 1 + (2\beta + \alpha) \delta$ \\
& dark blue &  $ 1 - (2\beta + \alpha) \delta$ \\
\\
\multirow{4}{*}{$ T_{1u}^3 $} & dark green &  $ 1 + \alpha \delta$ \\
& light green &  $ 1 - \alpha \delta$ \\
& dark red &  $ 1 + \beta \delta$ \\
& light red &  $ 1 - \alpha \delta$ \\
\\
\multirow{3}{*}{$\alpha |E^1_g \rangle + \beta |E^2_g \rangle$} & green &  $ 1 + (\alpha - \beta) \delta$ \\
& red &  $ 1 + 2 \beta \delta$ \\
& blue &  $ 1 - (\alpha + \beta) \delta$\\
\\
\multirow{6}{*}{$\alpha |E^1_u \rangle + \beta |E^2_u \rangle$} & dark red &  $ 1 + 2 \beta \delta$ \\
& light red &  $ 1 - 2 \beta \delta$ \\
& dark blue &  $ 1 + (\alpha + \beta) \delta$ \\
& light blue &  $ 1 - (\alpha + \beta) \delta$ \\
& light green &  $ 1 + (\alpha - \beta) \delta$ \\
& dark green &  $ 1 - (\alpha - \beta) \delta$\\
\hline
\hline
\end{tabular}
\caption{Hoppings for all ansätze in Fig.~\ref{fig:Local_stability_of_2D_real_ansatze}
}
\label{tab:hoppings_with_colors_for_local_2_parameter_ansatzes}
\end{table}

We can identify the $- T_{1g}^1 +  T_{1g}^2 +  T_{1g}^3$ ansatz with the David star ansatz studied in \cite{hastings2000_gap_opening_with_anisotropy} for the SU(2) case. There, it had a single parameter $\delta$, which is identical to $\alpha = 1$ and $\beta = 0$ in our notation, so that each bond on the edge of a David star is strengthened as $t_{i,j} = t^{\text{DSL}}_{i,j} (1+\delta)$, while all other hoppings are weakened as $t_{i,j} = t^{\text{DSL}}_{i,j} (1-\delta)$ (see Tab.~\ref{tab:hoppings_with_colors_for_local_2_parameter_ansatzes}).  Some linear combinations are equivalent, for example, $ T_{1g}^1 -  T_{1g}^2 +  T_{1g}^3$ is a David star shifted in $\mathbf{a}_2$ direction. 
Allowing for $\beta \neq 0$, we consider a more general David star ansatz, 
\begin{subequations}
\label{eq:DavidStar}
\begin{align}
t^{\DavidStarOut}_{i,j} &= t^{\text{DSL}}_{i,j} (1 + \alpha \delta)\\
t^{\triangle}_{i,j} &= t^{\text{DSL}}_{i,j} \left[1- (2 \beta + \alpha) \delta\right]\\
t^{\hexagon}_{i,j} &= t^{\text{DSL}}_{i,j} \left[1 + (2 \beta - \alpha) \delta\right],
\end{align}
\end{subequations}
shown in the first row of Fig.~\ref{fig:Local_stability_of_2D_real_ansatze}.
The $t^{\DavidStarOut}_{i,j}$ denotes the hoppings amplitudes on the edge of the David star (red bonds in Fig.~\ref{fig:Local_stability_of_2D_real_ansatze}), the $t^{\triangle}_{i,j}$ on the interstar triangles (blue) and $t^{\hexagon}_{i,j}$ on the hexagon within the star (shown by green). 
After Gutzwiller projecting the ground state of the David star mean-field Hamiltonian (\ref{eq:mean_field_Hamiltonian}), only the ratios of the hopping amplitudes remain essential.
Therefore, using the notations $t^{\DavidStarOut}_{i,j} = t^{\text{DSL}}_{i,j} t_{\DavidStarOut}$, $t^{\triangle}_{i,j} = t^{\text{DSL}}_{i,j} t_{\triangle}$, and $t^{\hexagon}_{i,j} = t^{\text{DSL}}_{i,j} t_{\hexagon}$, we can divide with $1 + \alpha \delta$, getting $t_{\DavidStarOut} = 1$, and use $t_{\triangle} \approx 1 - 2( \beta + \alpha) \delta$ and $t_{\hexagon} \approx 1 + 2( \beta - \alpha) \delta  $ as the two free parameters (also listed in Tab.~\ref{tab:hoppings_with_colors_for_local_2_parameter_ansatzes}), instead of $\alpha \delta$ and $\beta \delta$. This will facilitate physical interpretation when we consider much bigger deformations of the Dirac spin liquid in Fig.~\ref{fig:Global_stability_of_David_star}.
The David star is special because, for weak $t_{\triangle}$ hoppings, the nearly decoupled 12 sites tend to host two SU(6) singlet states, see Sec.~\ref{sec:phase_diagram_of_David_star}. This can be contrasted with the $A_{1u}$ configuration, which results in decoupled triangles -— an ideal starting point for the trimerized phase of the SU(3) Heisenberg model \cite{Arovas_simplex_sun, Corboz_simplex_sun}. Another interesting case is $|t_{\DavidStarOut}| < |t_{\triangle}|$ and $|t_{\DavidStarOut}| < |t_{\hexagon}|$ when the six spins in the hexagon tend to form a singlet and the spins in the decoupled triangles the 20-dimensional self-conjugate irreducible representation, see Sec.~\ref{sec:phase_diagram_of_David_star}.

Refs.~\onlinecite{hastings2000_gap_opening_with_anisotropy,SU2_Dirac_spin_liquid_Hermele_Ran_PRB_2008} also considered the dimerized ansatz for the SU(2) case, the so called $F_1^A$ bond order \cite{SU2_Dirac_spin_liquid_Hermele_Ran_PRB_2008}.
It corresponds to (any) row of $T_{1g}$, setting $\alpha = 1$ and $\beta = 0$, and also has a single parameter $\delta$. In this ansatz, the hopping amplitudes alternate along parallel lines while the rest of the bonds remain unchanged (different rows of $T_{1g}$ select different directions). Allowing for $\beta \neq 0$ we consider a more general pattern shown in the second row of Fig.~\ref{fig:Local_stability_of_2D_real_ansatze} for $ T_{1g}^3 $, for which the hoppings are given in Tab.~\ref{tab:hoppings_with_colors_for_local_2_parameter_ansatzes}. This includes the $F_1^B$ bond order in Ref.~\onlinecite{SU2_Dirac_spin_liquid_Hermele_Ran_PRB_2008}. 

Both the ungerade David star ansatz $- T_{1u}^1 +  T_{1u}^2 +  T_{1u}^3$ and a row of the same irrep $T_{1u}^3$ are shown in the third and fourth rows of Fig.~\ref{fig:Local_stability_of_2D_real_ansatze}. They have only two free parameters $\alpha \delta$ and $\beta \delta$, despite parametrizing six and four different hoppings, respectively, which are given in Tab.~\ref{tab:hoppings_with_colors_for_local_2_parameter_ansatzes}.

For the gerade $E_g$, to take the normalization into account, we multiply the two rows with different coefficients. We considered the ansatz $v_1 E^1_g  + v_2 E^2_g $, shown in the fifth row of Fig.~\ref{fig:Local_stability_of_2D_real_ansatze}, for which the hoppings are given in Tab.~\ref{tab:hoppings_with_colors_for_local_2_parameter_ansatzes}.
For $v_1 = 0$ and $v_2 = 1$ it describes anisotropic chains with hoppings $t_{i,j} = t^{\text{DSL}}_{i,j} (1 + 2 \delta)$ along one direction and $t_{i,j} = t^{\text{DSL}}_{i,j} (1 -  \delta)$ along the other two. 
The linear combinations with $v_1 = \pm \sqrt{3} v_2$ rotate the chains with different hoppings.
For $v_1 = 1$ and $v_2 = 0$, the lines in the three different directions have three different strengths $t_{i,j} = t^{\text{DSL}}_{i,j} (1 + \xi \delta)$, with $\xi \in \lbrace -1, 0, 1 \rbrace$. Here again, the $v_2 = \pm \sqrt{3} v_1$ rotates the unequivalent chains.

For the ungerade $E_{u}$ we show the ansatz  $v_1 E^1_u  + v_2 E^2_u $ in the sixth row of Fig.~\ref{fig:Local_stability_of_2D_real_ansatze} and Tab.~\ref{tab:hoppings_with_colors_for_local_2_parameter_ansatzes}. 

The $A_{1g}$, $A_{1u}$, $E_{g}$, $E_{u}$ ansätze are gapless in the mean field and have a two-fold degenerate Dirac Fermi point at $\mathbf{k} = 0$, similarly to the DSL. Among the real perturbations, only the $T_{1g}$ and $T_{2g}$ open a gap. However, for the $T_{1g}$ the gap closes along a curve going through the $t_H=t_T=t$ and the $(t_T=0,t_H=-t/\sqrt{2})$ point (shown in Fig.~\ref{fig:Global_stability_of_David_star} as a green dashed line), and a Dirac cone appears. 
The $T_{1u}$ and $T_{2u}$ have two Dirac Fermi points at finite $\mathbf{k}_D$ and $-\mathbf{k}_D$ values, each non-degenerate. 
In the projective sense, the translational symmetry is restored for the $A_1$ and $E$ ansätze, but they break the point group symmetries. The $T_{1g}$ and $T_{2g}$ keep some of the point group projective symmetries but break translation symmetries, while the $T_{1u}$ and $T_{2u}$ break both the translations and point group symmetries. 


Let us also mention that the $E_g$, $T_{2u}$, $T_{1g}$, and $T_{1u}$ ansätze introduce charge imbalance between the sites, meaning the average occupancy of the sites differs. This imbalance can be compensated for by introducing on-site chemical potentials.

As we will show in Figs.~\ref{fig:Local_stability_of_1D_real_ansatze} and \ref{fig:Local_stability_of_2D_real_ansatze}, all of these real perturbations raise the variational energy, so the Dirac spin liquid is the lowest energy state.

\subsection{Complex-valued $\tilde t_{i,j}$ perturbations of the DSL}
\label{sec:complex_instabilities}

In the previous section, we studied the stability of the DSL against real-valued perturbations that respect time reversal invariance. Here, we consider the simplest perturbations that break the time-reversal symmetry. For complex $\tilde t_{i,j}$ in Eq.~(\ref{eq:perturbation_of_the_hoppings})  the $\pi_{\hexagon} \pi_{\triangle} \pi_{\bigtriangledown}$ flux structure of the DSL changes. Unlike the real ansätze, the direction of the hoppings matters, since $t_{i,j} f^{\dagger}_{i,\sigma} f^{\phantom{\dagger}}_{j,\sigma} = (t_{j, i} f^{\dagger}_{j,\sigma} f^{\phantom{\dagger}}_{i,\sigma})^{\dagger}$. Consequently, $t_{i,j} = t_{j,i}^{*}$, and $\tilde t_{i,j} = \tilde t_{j,i}^{*}$.
For the sake of simplicity, we consider purely imaginary  $\tilde t_{i,j}$ perturbations below.  Exponentializing the imaginary perturbations, we get $t_{i,j} =  t^{\text{DSL}}_{i,j}(1 + \delta \tilde t_{i,j}) \approx t^{\text{DSL}}_{i,j} e^{\delta \tilde t_{i,j}}$, and we can introduce the phases $\varphi_{i,j} \equiv \delta \tilde t_{i,j}/i \in \mathbb{R}$ to denote the hoppings as $t_{i,j} = t^{\text{DSL}}_{i,j} e^{i \varphi_{i,j}}$. Therefore, for $\varphi_{i,j} = 0$, we get back the hoppings of the DSL, with the $\pi_{\hexagon} \pi_{\triangle} \pi_{\bigtriangledown}$ flux structure. 
We tabulate the irreducible representations of the complex-valued $\tilde t_{i,j}$ in Tab.~\ref{tab:complex_perturbations}.

\begin{figure}[t]
\centering
\includegraphics[width=0.45\textwidth]{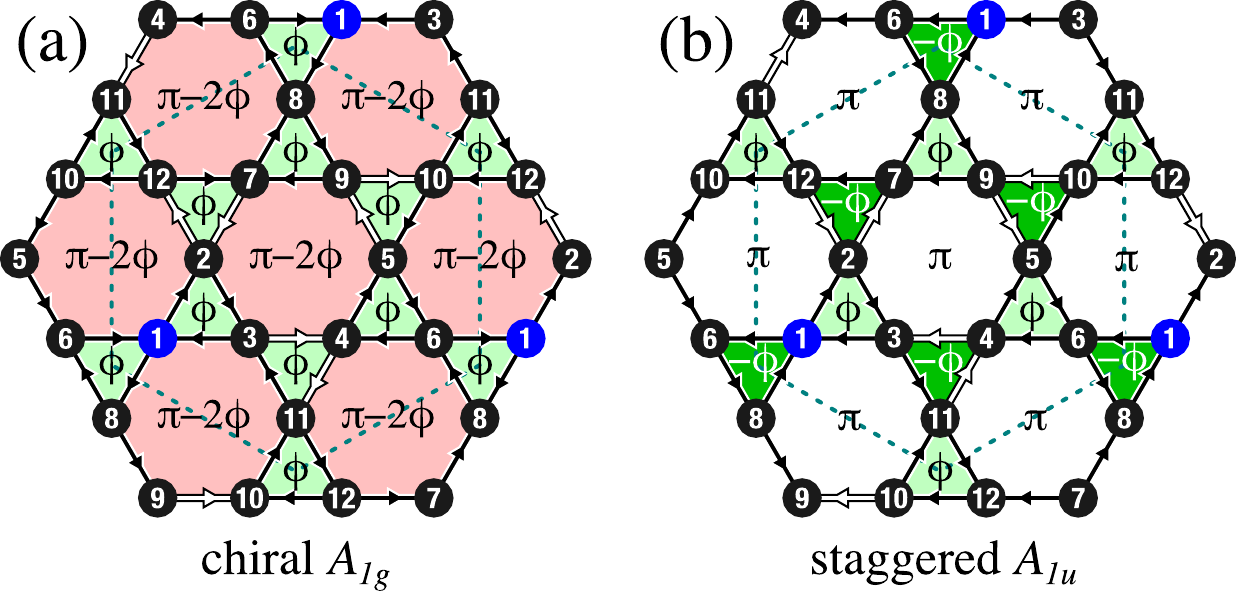}
\caption{(a) the chiral $A_{1g}$ ansatz and (b) the staggered $A_{1u}$ ansatz. Since $t_{i,j} = t_{j,i}^{*}$, the direction  matters for the complex hopping amplitudes. The white arrows pointing from $j$ to $i$ denote $t_{i,j} = e^{i \varphi} $, while the black arrows correspond to $t_{i,j} = -e^{i \varphi}$. The fluxes of upward pointing triangles are $\phi = \pi + 3 \varphi$ in both of them since the product of the hoppings around every triangle in the clockwise direction is $\propto -e^{i 3 \varphi}$. However, the fluxes in the downward-pointing triangles and the hexagons differ in the two cases.
\label{fig:chiral_and_staggered_ansatzes}
}
\end{figure}

The chiral $A_{1g}$ ansatz, shown in Fig.~\ref{fig:chiral_and_staggered_ansatzes}(a), has equal fluxes $\phi = \pi + 3\varphi = \pi + 3 \delta$ in both up- and downward pointing triangles, the product of the hoppings around every triangle in the clockwise direction is $-e^{i 3 \varphi}$.
The fluxes of the hexagons in the clockwise direction is $\pi -6 \varphi = \pi - 6 \delta = 3 \pi - 2 \phi \leftrightarrow \pi - 2\phi $, so we denote the flux structure as $(\pi - 2\phi)_{\hexagon} \phi_{\triangle} \phi_{\bigtriangledown}$.

The staggered $A_{1u}$ ansatz 
has clockwise fluxes $\phi = \pi + 3\varphi = \pi + 3 \delta$ in upward pointing triangles, $\pi - 3\varphi$ (being equivalent to $-\pi - 3\varphi = -\phi$) in downward pointing triangles, and $\pi$ in the hexagons (see Fig.~\ref{fig:chiral_and_staggered_ansatzes}(b)), therefore we will denote its flux structure as $\pi_{\hexagon} \phi_{\triangle} (-\phi)_{\bigtriangledown}$.

The stability of the DSL against the extended chiral $A_{1g}$ and the extended staggered $A_{1u}$ ansätze is shown in Fig.~\ref{fig:global_stability_chiral_staggered} and will be discussed in Sec.~\ref{sec:global_stability_of_chiral_staggered}.

We will not consider the three-dimensional irreducible representation in greater detail. Let us only mention that the $T_{1g}$ breaks the degeneracy and shifts the Dirac cones up and down with a circular Fermi line in the mean-field energy spectrum. The $T_{1u}$ generally opens a gap, but more complex situations are possible.

\begin{table*}[bt]
\caption{The irreducible representation of the complex-valued perturbations of the hopping amplitudes. Every $\tilde t_{i,j}$ is purely imaginary and the hoppings can be written as $t_{i,j} = t^{\text{DSL}}_{i,j}(1 + \delta \tilde t_{i,j}) \approx t^{\text{DSL}}_{i,j} e^{\delta \tilde t_{i,j}} = t^{\text{DSL}}_{i,j} e^{i \varphi}$, where $\varphi_{i,j} \equiv \delta \tilde t_{i,j}/i \in \mathbb{R}$. 
Since the hermiticity requires $t_{i,j} = t_{j,i}^{*}$, consequently $\tilde t_{i,j} = \tilde t_{j,i}^{*}$, the directions of the hoppings matter (as shown in Fig.~\ref{fig:chiral_and_staggered_ansatzes}). Therefore, the $\langle i,j \rangle$ pairs (directed bonds) denote the site ordering in $t_{i,j} f^{\dagger}_{i, \sigma} f^{\phantom{\dagger}}_{j,\sigma}$. \label{tab:complex_perturbations}
}
\begin{ruledtabular}
\begin{tabular}{cccccccccccccc}
	&\multicolumn{12}{c}{$\tilde t_{i,j}$}	&deg.\\
$\langle i,j \rangle$	
&$\langle 2, 1 \rangle$	&$\langle 1, 3 \rangle$	&$\langle 3, 2 \rangle$	&$\langle 5, 4 \rangle$	&$\langle 4, 6 \rangle$	&$\langle 6, 5 \rangle$	&$\langle 8, 7 \rangle$	&$\langle 7, 9 \rangle$	&$\langle 9, 8 \rangle$	&$\langle 11, 10 \rangle$	&$\langle 10, 12 \rangle$	&$\langle 12, 11 \rangle$\\
$C_2\langle i,j \rangle$	
&$\langle 5, 10 \rangle$	&$\langle 10,9 \rangle$	&$\langle 9, 5 \rangle$	&$\langle 2, 7 \rangle$	&$\langle 7, 12 \rangle$	&$\langle 12, 2 \rangle$	&$\langle 11, 4 \rangle$	&$\langle 4, 3 \rangle$	&$\langle 3, 11 \rangle$	&$\langle 8, 1 \rangle$	&$\langle 1, 6 \rangle$	&$\langle 6, 8 \rangle$\\
\hline
chiral $A_{1g}$	&$i$	&$i$	&$i$	&$i$	&$i$	&$i$	&$i$	&$i$	&$i$	&$i$	&$i$	&$i$	&$\times2$\\
\\
staggered $A_{1u}$	
&$\pm i$	&$\pm i$	&$\pm i$	&$\pm i$	&$\pm i$	&$\pm i$	&$\pm i$	&$\pm i$	&$\pm i$	&$\pm i$	&$\pm i$	&$\pm i$	&$\times2$\\
\\
\multirow{3}{*}{chiral $T_{1g}$}	
&$i\beta$	&$i\beta$	&$i\alpha$	&$-i\beta$	&$-i\beta$	&$-i\alpha$	&$-i\beta$	&$-i\beta$	&$-i\alpha$	&$i\beta$	&$i\beta$	&$i\alpha$	&\multirow{3}{*}{$\times1$}\\
&$i\beta$	&$i\alpha$	&$i\beta$	&$i\beta$	&$i\alpha$	&$i\beta$	&$-i\beta$	&$-i\alpha$	&$-i\beta$	&$-i\beta$	&$-i\alpha$	&$-i\beta$	&\\
&$i\alpha$	&$i\beta$	&$i\beta$	&$-i\alpha$	&$-i\beta$	&$-i\beta$	&$i\alpha$	&$i\beta$	&$i\beta$	&$-i\alpha$	&$-i\beta$	&$-i\beta$	&\\
\\
\multirow{3}{*}{staggered $T_{1u}$}	
&$\pm i\beta$	&$\pm i\beta$	&$\pm i\alpha$	&$\mp i\beta$	&$\mp i\beta$	&$\mp i\alpha$	&$\mp i\beta$	&$\mp i\beta$	&$\mp i\alpha$	&$\pm i\beta$	&$\pm i\beta$	&$\pm i\alpha$	&\multirow{3}{*}{$\times1$}\\
&$\pm i\beta$	&$\pm i\alpha$	&$\pm i\beta$	&$\pm i\beta$	&$\pm i\alpha$	&$\pm i\beta$	&$\mp i\beta$	&$\mp i\alpha$	&$\mp i\beta$	&$\mp i\beta$	&$\mp i\alpha$	&$\mp i\beta$	&\\
&$\pm i\alpha$	&$\pm i\beta$	&$\pm i\beta$	&$\mp i\alpha$	&$\mp i\beta$	&$\mp i\beta$	&$\pm i\alpha$	&$\pm i\beta$	&$\pm i\beta$	&$\mp i\alpha$	&$\mp i\beta$	&$\mp i\beta$	&\\
\end{tabular}
\end{ruledtabular}
\end{table*}%

\section{Stability of the DSL}
\label{sec:second_nearest_neighbor_and_ring_exchange}

\begin{figure*}[t]
\centering
\includegraphics[width=1.0\textwidth]{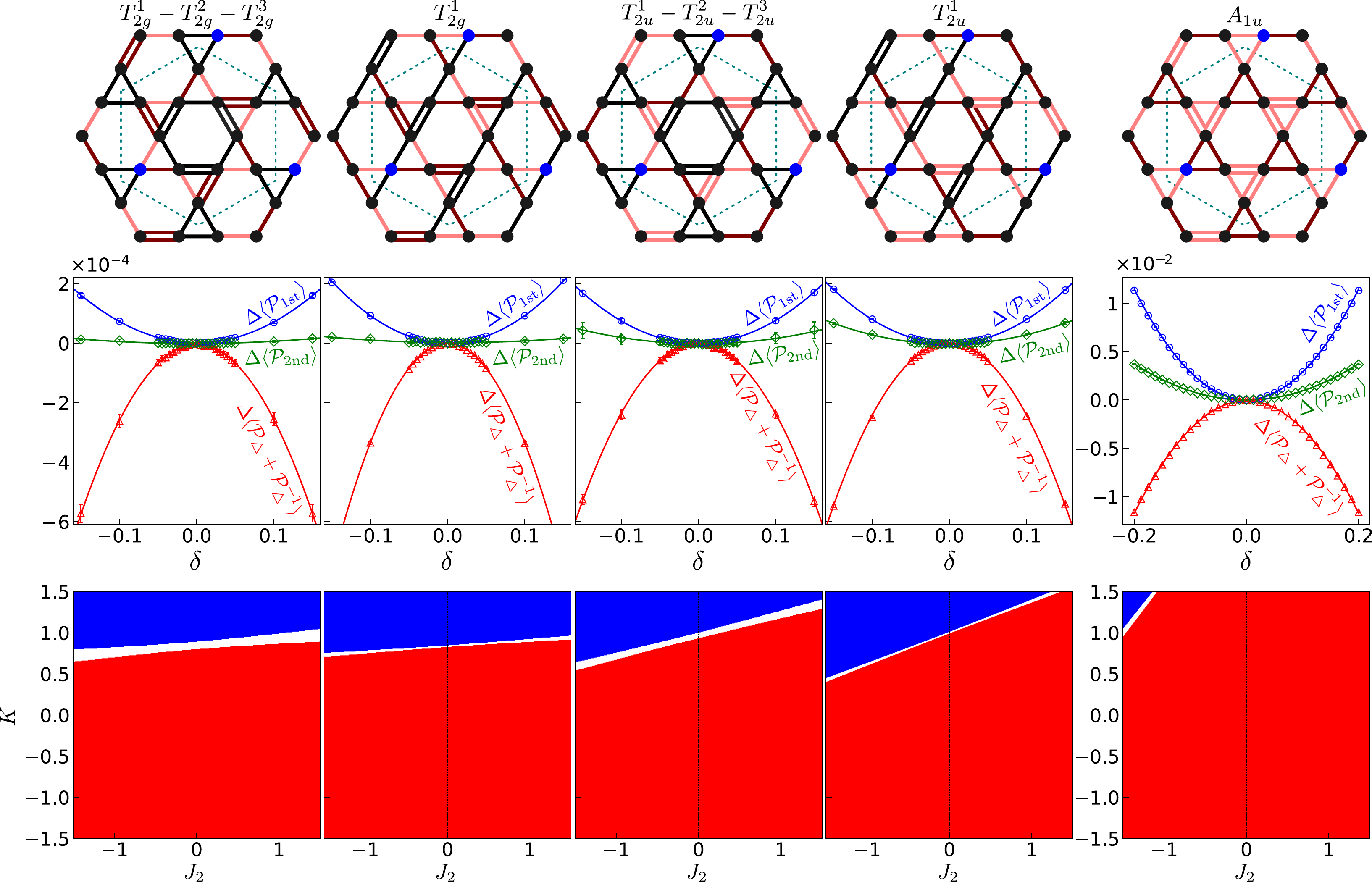}
\caption{In the first row, we show the hopping structure of the real perturbations of the Dirac spin liquid having a single free parameter $\delta$. Different shades represent different absolute values of the hoppings. The white bonds show positive hoppings (each ansatz has a $\pi_{\hexagon} \pi_{\triangle} \pi_{\bigtriangledown}$ flux structure, just as the DSL). The black bonds have amplitude 1, the dark red hoppings have amplitude $1 + \delta$, and the light reds $1 - \delta$. In the midle row, the red points show the $\Delta \langle \mathcal{P}_{\triangle} + \mathcal{P}^{-1}_{\triangle} \rangle$, the blue points $\Delta \langle \mathcal{P}_{\text{1st}} \rangle$, and the green points $\Delta \langle \mathcal{P}_{\text{2nd}} \rangle$ defined in Eq.~(\ref{eq:expectation_values_relative_to_the_DSL}), while the solid lines are the fitted parabolas. The bottom row shows the local stability of these ansätze (as explained in sec.~\ref{sec:local_stability_against_real_perturbations}), as a function of $K$ and $J_2$, fixing $J_1 = 1$. The DSL is the lowest energy state in the red region, and the perturbation wins in the blue region.}
\label{fig:Local_stability_of_1D_real_ansatze}
\end{figure*}

\begin{figure*}[t]
\centering
\includegraphics[width=0.92\textwidth]{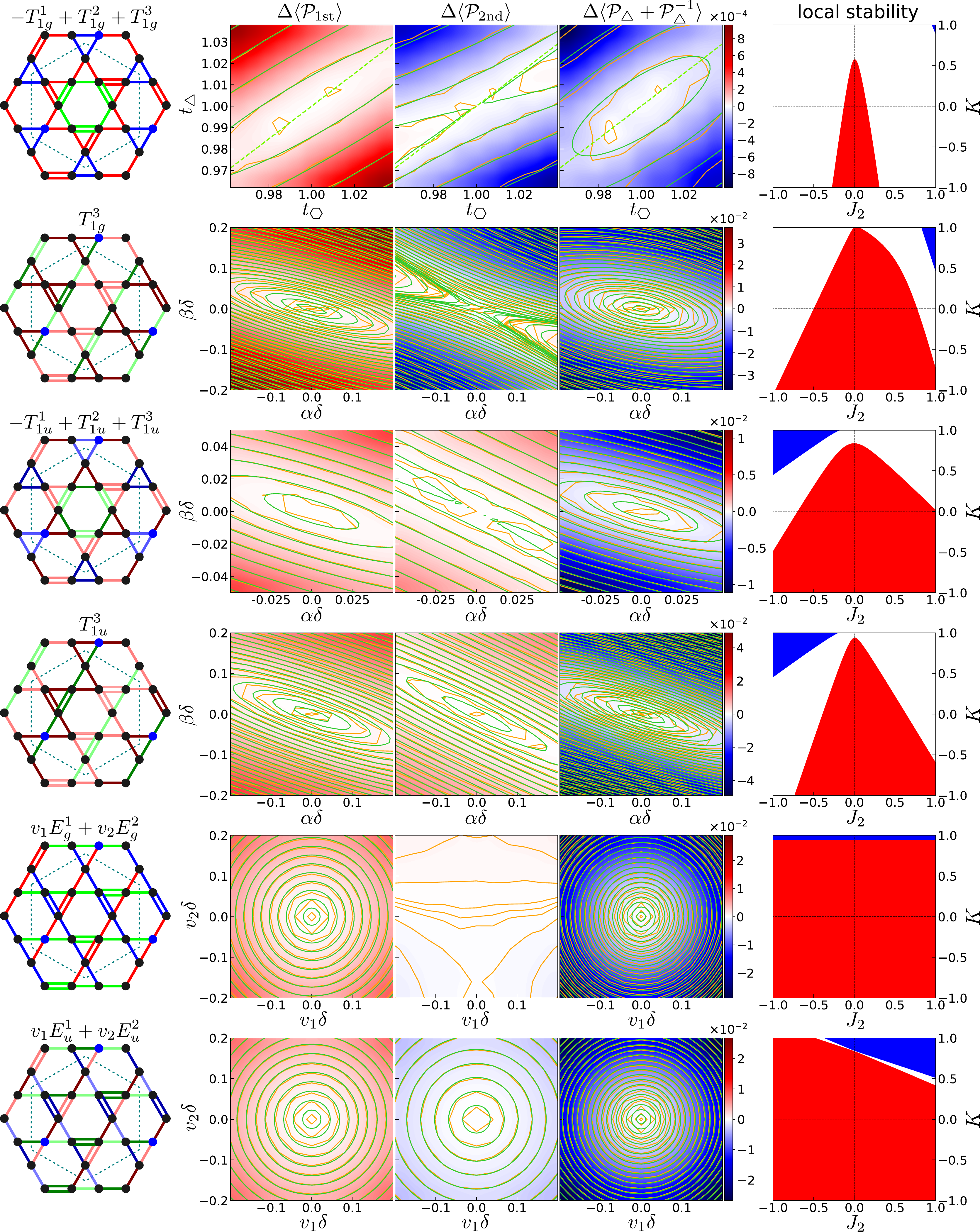}
\caption{Stability analysis of different real ansätze with $\pi_{\hexagon} \pi_{\triangle} \pi_{\bigtriangledown}$ fluxes and nonuniform hopping amplitudes, having two free parameters.
In the first column, different colors represent different hopping amplitudes listed in Tab.~\ref{tab:hoppings_with_colors_for_local_2_parameter_ansatzes} for the denoted irreducible representation (the solid bonds stand for negative, and the inverted bonds for positive $t_{i,j}$). 
The second column shows the 
$\Delta \langle \mathcal{P}_{\text{1st}} \rangle$, the third $\Delta \langle \mathcal{P}_{\text{2nd}} \rangle$, and the fourth $\Delta \langle \mathcal{P}_{\triangle} + \mathcal{P}^{-1}_{\triangle} \rangle$ defined in Eq.~(\ref{eq:expectation_values_relative_to_the_DSL}), calculated by VMC for a cluster of 192 sites with APBC, the $\delta=0$ is the DSL. 
The effect of the APBC is visible only in the $\Delta \langle \mathcal{P}_{\text{2nd}} \rangle$ of $\alpha E^{1}_g + \beta E_{g}^2$ (which becomes flat after averaging over all APBC orientations). 
The contours of the fitted ellipsoids are shown in light green, while the contours of the data are orange. The fifth column shows the local stability of these ansätze (see Sec.~\ref{sec:local_stability_against_real_perturbations}), as a function of $K$ and $J_2$, keeping $J_1 = 1$. The DSL has the lowest energy in the red region and the perturbed ansatz of the given row in the blue region.}
\label{fig:Local_stability_of_2D_real_ansatze}
\end{figure*}

The Heisenberg Hamiltonian of Eq.~(\ref{eq:Heisenberg_Hamiltonian}) is a result of the leading order perturbation theory of the repulsive Hubbard Hamiltonian of Eq.~(\ref{eq:Hubbard_Hamiltonian}) for $U/t \to \infty$. However, we can incorporate further terms so that the effective Hamiltonian reads
\begin{equation}
    \mathcal{H} = J_1 \sum_{\langle i,j \rangle } \mathcal{P}_{i,j} + J_2 \sum_{\langle \langle i,j \rangle \rangle } \mathcal{P}_{i,j} + K \sum_{ \langle i,j,k \rangle} ( \mathcal{P}_{i,j,k} + \mathcal{P}_{i,j,k}^{-1} ),
\label{eq:extended_Heisenberg_Hamiltonian}
\end{equation}
where the $\langle \langle i,j \rangle \rangle $ denotes second nearest neighbor sites, $\mathcal{P}_{i,j,k}$ is the ring exchange operator acting on elementary triangles $\langle i,j,k \rangle$, and the $J_1$ is of the order $\frac{t^2}{U}$ and usually positive (antiferromagnetic), $K  \propto  \frac{t^3}{U^2}$, and $J_2 \propto \frac{t^4}{U^3}$ in case it arises from nearest neighbor hoppings. The $\mathbf{T}_{i} \cdot \mathbf{T}_{j}$ relates to the  exchange operator $\mathcal{P}_{i,j}$ as
\begin{equation}
    \mathbf{T}_{i} \cdot \mathbf{T}_{j} =  \frac{1}{2}\mathcal{P}_{i,j} - \frac{1}{2N} \mathcal{I}, 
    \label{eq:TTP}
\end{equation}
in the fundamental representation, where $\mathcal{I}$ is the identity operator (see also Sec.~\ref{sec:relations_between_permuations_and_the_diagonal_terms_in_the_Fermi_sea} for details).
Written explicitly, $P_{ij}$ and $P_{ijk}$ are defined through their action on the local basis states, $P_{ij}|A_i B_j \rangle = |B_i A_j \rangle$  and $P_{ijk}|A_i B_j C_k \rangle = |C_i A_j B_k \rangle$, for a fixed orientation of the triangle $i$, $j$, $k$. 
The variational energy per lattice site of the effective Hamiltonian (\ref{eq:extended_Heisenberg_Hamiltonian}) can be written as
\begin{align}
  \frac{E}{N_s} &= \frac{1}{N_s} \frac{\langle \psi|\mathcal{H}| \psi \rangle}{\langle \psi| \psi \rangle} \nonumber\\
  &= 2 J_1 \langle \mathcal{P}_{\text{1st}} \rangle + 2 J_2 \langle \mathcal{P}_{\text{2nd}} \rangle + \frac{2}{3} K \langle \mathcal{P}_{\triangle} + \mathcal{P}^{-1}_{\triangle} \rangle,
\label{eq:variational_energy_with_J_2_and_K_per_site}
\end{align}
where the $\langle \mathcal{P}_{\text{1st}} \rangle$ is the averaged expectation value of the permutation operator between the nearest neighbor sites, the  $\langle \mathcal{P}_{\text{2nd}} \rangle$ between the second neighbor sites, and $\langle \mathcal{P}_{\triangle} + \mathcal{P}^{-1}_{\triangle} \rangle$ of the ring exchange on the triangles. The coefficients consider that there are twice as many first- and second-neighbor bonds than sites and two triangles for every three sites.
The expectation values of these exchange operators were evaluated by Monte Carlo sampling the wavefunctions $|\psi \rangle $ of the ansätze discussed in section \ref{sec:possible_instabilities_of_the_Dirac_spin_liquid}, for details see Appendix \ref{sec:relations_between_permuations_and_the_diagonal_terms_in_the_Mott_phase}. 
In Figs.~\ref{fig:Local_stability_of_1D_real_ansatze}, \ref{fig:Local_stability_of_2D_real_ansatze}, and \ref{fig:VMC_chiral_staggered} we show the expectation values relative to the DSL, i.e., 
\begin{subequations}
\label{eq:expectation_values_relative_to_the_DSL}
\begin{align}
\Delta &\langle \mathcal{P}_{\text{1st}} \rangle = \langle \mathcal{P}_{\text{1st}} \rangle - \langle \mathcal{P}_{\text{1st}} \rangle_{\text{DSL}} \\
\Delta &\langle \mathcal{P}_{\text{2nd}} \rangle = \langle \mathcal{P}_{\text{2nd}} \rangle - \langle \mathcal{P}_{\text{2nd}} \rangle_{\text{DSL}} \\
\Delta &\langle \mathcal{P}_{\triangle} + \mathcal{P}^{-1}_{\triangle} \rangle = \langle \mathcal{P}_{\triangle} + \mathcal{P}^{-1}_{\triangle} \rangle - \langle \mathcal{P}_{\triangle} + \mathcal{P}^{-1}_{\triangle} \rangle_{\text{DSL}} 
\end{align}
\end{subequations}
for all ansätze as a function of the perturbation strength. 
On the one hand, we will see that the $\langle \mathcal{P}_{\text{1st}} \rangle$ is minimal for the DSL for small perturbations for all the ansätze we considered. We call this 
the local stability of the DSL. On the other hand, comparing with the expectation values of the $\langle \mathcal{P}_{\text{2nd}} \rangle$ and $\langle \mathcal{P}_{\triangle} + \mathcal{P}^{-1}_{\triangle} \rangle$ we find that finite values of $J_2$ and $K$ are needed to destabilize the DSL.

\subsection{Local stability against all real perturbations of the DSL}
\label{sec:local_stability_against_real_perturbations}

To assess the stability of the DSL, we calculated the minimum values of $J_2$ and $K$ needed to locally destabilize the DSL in favor of the real perturbations discussed in Sec.~\ref{sec:real_instabilities}.
Specifically, we determined the threshold values of $J_2$ and $K$ in the Hamiltonian (\ref{eq:extended_Heisenberg_Hamiltonian}) where the variational energy of the DSL becomes higher than that of a perturbed ansatz. 
The results, presented in the last row of Fig.\ref{fig:Local_stability_of_1D_real_ansatze} and the last column of Fig.\ref{fig:Local_stability_of_2D_real_ansatze}, indicate that the DSL remains locally stable over a relatively large parameter region. 
Global stability ot the DSL against certain ansätze will be discussed in the subsections \ref{sec:global_stability_of_David_star} and \ref{sec:global_stability_of_chiral_staggered} below.

First, we considered the ansätze with two free parameters. We fitted a quadratic surface to the expectation values of the exchanges $X=\langle \mathcal{P_{\text{1st}}} \rangle$, $\langle \mathcal{P}_{\text{2nd}} \rangle$, and $\langle \mathcal{P}_{\triangle} + \mathcal{P}^{-1}_{\triangle}\rangle$ for small $\delta$ values around the DSL point, 
\begin{equation}
    f(\alpha \delta, \beta \delta)_X = (a_X \alpha^2  + b_X \beta^2 + 2 c_X \alpha \beta) \delta^2\;.
\label{eq:ellipsoid_fitting_for_local_stability}
\end{equation}
The data sets consisted of $11\times 11$ points in the $\alpha \delta, \beta \delta$ plane, all having a Monte Carlo error, which we considered in the fitting procedure and transferred to the parameters $a_X$, $b_X$, and $c_X$.
Fig.~\ref{fig:Local_stability_of_2D_real_ansatze} shows the results of this analysis. To facilitate the comparison between the calculated and fitted values, the contours of the Monte Carlo data are shown in orange, and the contours of the fitted surface in green (we use the same quadratic contour level spacing for every figure, both for the contours of the fitted ellipsoids and the contours of the Monte Carlo data). For the nearest neighbor exchange, the expectation value is the smallest for the $\delta=0$, thus providing numerical evidence for the local stability of DSL. However, the expectation values of the second neighbor and ring exchanges show a different picture: the $\langle \mathcal{P}_{\text{2nd}}\rangle$ can lose or gain energy depending on the perturbation, but the $\langle \mathcal{P}_{\triangle} + \mathcal{P}^{-1}_{\triangle}\rangle$ will always gain energy with perturbations (assuming $K>0$). It raises the question of how large the $J_2$ and  $K$ must be to win over the DSL ground state.

To answer this question, we calculated the eigenvalues $\lambda_{1,2}$ of the Hessian of the quadratic surface describing the energy surface for finite values of $J_1$, $J_2$, and $K$,
\begin{equation}
\left(
\begin{array}{cc}
J_1 a_{1} + J_2 a_{2} + K a_{\triangle} & J_1 c_{1} + J_2 c_{2} + K c_{\triangle}  \\
J_1 c_{1} + J_2 c_{2} + K c_{\triangle}  & J_1 b_{1} + J_2 b_{2} + K b_{\triangle} 
\end{array}
\right).
\end{equation}
The eigenvalues characterize the curvatures of the ellipsoid, and the DSL is stable when both eigenvalues are positive. The last column of Fig.~\ref{fig:Local_stability_of_2D_real_ansatze} shows the region of the local stability of the DSL in the parameter space of $J_2$ and $K$, with $J_1$ set to 1. Considering the Monte Carlo error, we regard the DSL as stable (red color) if the eigenvalues, including the error bars, are positive. Similarly, the DSL is unstable (blue region) if any eigenvalues are more negative than the error. Otherwise, in the parameter region where the eigenvalues are zero within the error bars, the fate of the DSL is uncertain (white region). 
The DSL is stable for relatively large values of the $ J_2 $ and $ K $.

For the ansätze with a single free parameter $\delta$, shown in Fig.~\ref{fig:Local_stability_of_1D_real_ansatze}, we fit a simple parabola to  $\Delta E/ N_s$ instead of a quadratic surface, and the coefficient of the $\delta^2$ takes the role of the eigenvalues discussed above. We repeat the same procedure to get the regions of stability.

For the David star shown in the first row of Fig.~\ref{fig:Local_stability_of_2D_real_ansatze}, we used $t_{\hexagon}$, and $t_{\triangle}$ (given in Tab.~\ref{tab:hoppings_with_colors_for_local_2_parameter_ansatzes}) instead of $\alpha \delta$ and $\beta \delta$. For this ansatz, there are sizeable contributions in $\delta^3$ by Eq.~(\ref{eq:c2c3c4}), so we considered only a tiny neighborhood around the $t_{\hexagon}=t_{\triangle}=1$ DSL point to fit the quadratic surface. The energy differences were also tiny, with large relative errors, explaining the large white region. However, we expect the same local stability figure as we obtained for the $T^{3}_{1g}$, where the $\delta^3$ are absent ($u_1=u_2=0$ in  Eq.~(\ref{eq:c2c3c4})). Indeed, the calculations give a smaller relative error in that case and a smaller white region. 

As shown in the last column of Figs.~\ref{fig:Local_stability_of_2D_real_ansatze}
and bottom row in Fig.~\ref{fig:Local_stability_of_1D_real_ansatze}, only a positive $K$ can destabilize the DSL locally, which originates from DSL being a local maximum for all ansätze.
However, the plot of the David star ansatz with extended parameters $t_{\triangle},t_{\hexagon} \in \{ -1, 1\}$ in Fig.~\ref{fig:Global_stability_of_David_star} shows that the DSL is in a local but not global maximum. Therefore, a negative $K$ can destabilize the DSL globally.

As stated in sec.~\ref{sec:possible_instabilities_of_the_Dirac_spin_liquid}, the David star ansatz opens a gap for every $t_{\triangle}$ and $t_{\hexagon}$, except for the green dashed curve shown in the first row of Fig.~\ref{fig:Local_stability_of_2D_real_ansatze}, where the lowest two bands touch at the Dirac Fermi point at the $\Gamma$ point in reciprocal space ($\mathbf{k} = \mathbf{0}$), just like for the DSL.

Let us note that in the mean-field level, all the real ansätze except the David star show the stability of the DSL ($\Delta \langle \mathcal{P}_{\text{1st}} \rangle > 0$). 
For the David star ansatz, we get $\Delta \langle \mathcal{P}_{\text{1st}} \rangle < 0$, suggesting that already for $K = 0$ and $J_2 = 0$, the David star ansatz has lower nearest neighbor energy than the DSL. However, introducing site-dependent chemical potentials to enforce $\langle n_i \rangle = 1$, $\forall$ $i$ the $\Delta \langle \mathcal{P}_{\text{1st}} \rangle$ becomes positive, making the DSL locally stable, similarly as was found with the SU(2) projected mean-field calculation in \cite{hastings2000_gap_opening_with_anisotropy}.

\subsection{Global stability of the DSL against the David star}
\label{sec:global_stability_of_David_star}

We introduced the David star ansatz as a perturbation of the Dirac spin liquid in Eq.~(\ref{eq:DavidStar}), where we assumed that the changes in the hopping amplitudes $t_{\DavidStarOut}$, $t_{\hexagon}$, and $t_{\triangle}$ relative to the uniform hopping amplitudes of the DSL are small. However, we can extend the domain of these hoppings to arbitrary values. The DSL is globally stable against the David star formation if its variational energy is lower than that of the possible David star ansatz.

\begin{figure*}[t]
\centering
\includegraphics[width=1.0\textwidth]{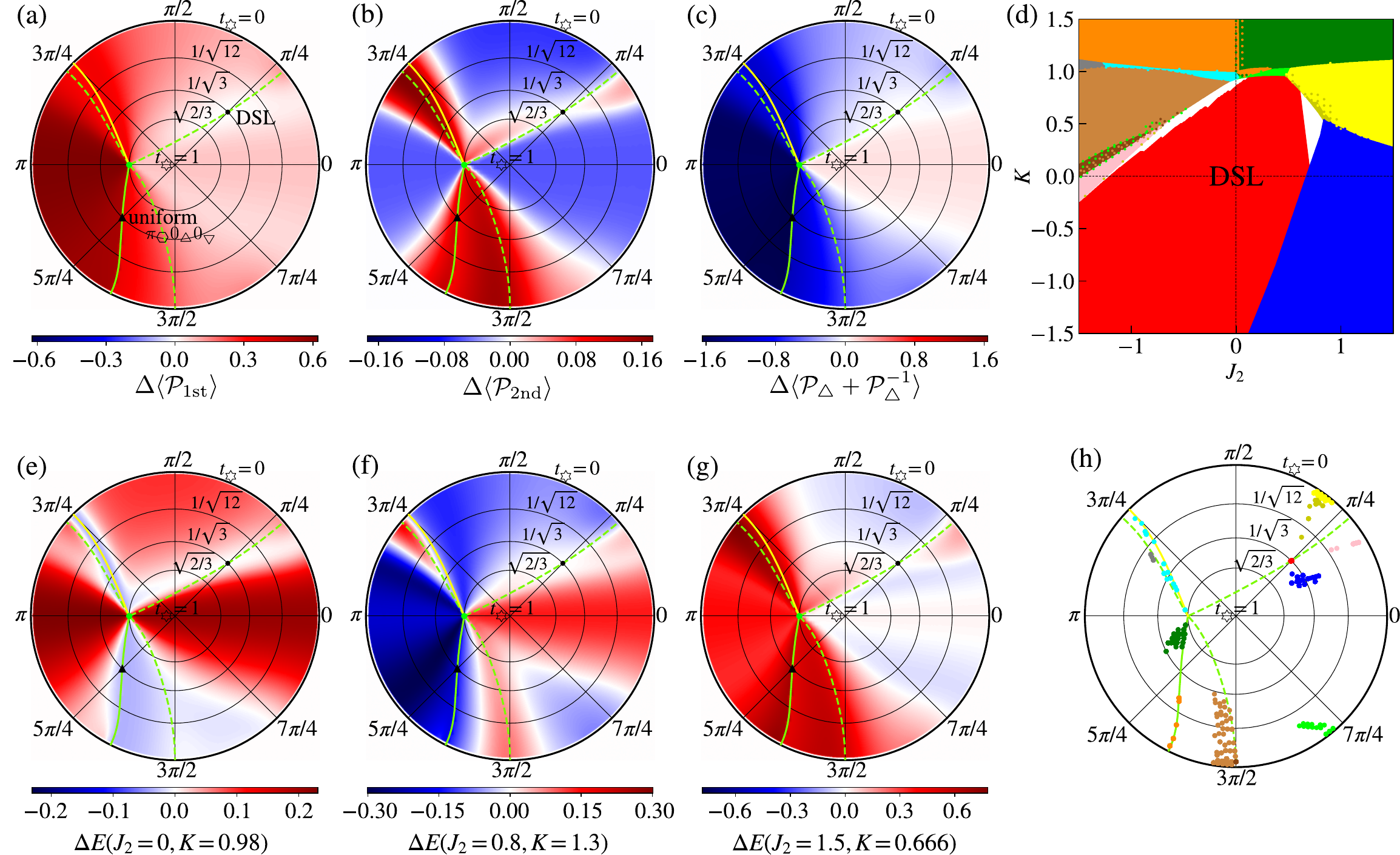}
\caption{\label{fig:Global_stability_of_David_star}
The relative expectation values of the 
(a) first neighbor $\Delta \langle \mathcal{P}_{\text{1st}} \rangle$, 
(b) second neighbor $\Delta \langle \mathcal{P}_{\text{2nd}} \rangle$, and 
(c) ring exchange $\Delta \langle \mathcal{P}_{\triangle} + \mathcal{P}^{-1}_{\triangle} \rangle$ defined in Eq.~(\ref{eq:expectation_values_relative_to_the_DSL}), for the David star ansatz with hopping amplitudes $t_{\DavidStarOut}$, $t_{\triangle}$, and $t_{\hexagon}$ defined in Eq.~(\ref{eq:t_global}).
%
In this stereographic map, the radius determines $t_{\DavidStarOut}$ and $t_{\hexagon} \propto \cos \Theta$, $t_{\triangle} \propto \sin \Theta$, subject to $t^2_{\DavidStarOut} + t^2_{\hexagon} + t^2_{\triangle} = 1$. 
The black dot indicates the Dirac spin liquid ($t_{\DavidStarOut}= t_{\hexagon} = t_{\triangle}= 1/\sqrt{3}$). 
The green dot denotes the singular point at $(t_{\DavidStarOut},t_{\hexagon}, t_{\triangle}) = (\sqrt{2/3}, -1/\sqrt{3},0)$, where the mean-field ground state is in the manifold of highly degenerate localized states. 
(e), (f) and (g) show the variational energy per site,  Eq.~(\ref{eq:variational_energy_with_J_2_and_K_per_site}), relative to the DSL state, $(E-E_{\text{DSL}})/N_s$,  for selected $K$ and $J_{2}$ values (keeping $J_{1} = 1$). 
(d) The global stability of the DSL against the David star ansatz in the parameter space of $J_2$  and $K$. The Dirac spin liquid wins in the red region. Monte Carlo errors do not allow determining the lowest energy state in white areas.
The different colors correspond to the optimal $t_{\DavidStarOut}$, $t_{\hexagon}$ and $t_{\triangle}$ values shown in (h). We consider the clustered points in (h) as the same phase and denote them with the same color. 
If the energies for different hopping amplitudes overlap within the Monte Carlo error, we also plot the higher-energy states as appropriately colored dots in panel (d). 
}
\end{figure*}

We can choose the hopping amplitudes such that we keep $t_{\DavidStarOut} = 1$, or we can normalize the hoppings as 
\begin{equation}
    t_{\DavidStarOut}^2 + t_{\hexagon}^2 + t_{\triangle}^2 = 1,
\label{eq:normalization_of_the_hoppings_for_the_global_David_star}
\end{equation}
putting them on a sphere, in which case $t_{\DavidStarOut}, t_{\hexagon}, t_{\triangle} \in [ -1, 1 ]$. 
 The $t_{\DavidStarOut} > 0$, $t_{\hexagon} > 0$ and $t_{\triangle} > 0$ gives the $\pi_{\hexagon} \pi_{\triangle} \pi_{\bigtriangledown}$ fluxes. The $t_{\triangle} < 0$ changes the fluxes of the interstar triangles from $\pi$ to $0$, and the $t_{\hexagon} < 0$ changes the fluxes of the instar triangles from $\pi$ to $0$, leaving the fluxes of the hexagons unaltered. The sign of $t_{\DavidStarOut}$ does not affect the fluxes, so we choose $t_{\DavidStarOut}$ to be positive. 
The locus of the hopping amplitudes with the normalization condition (\ref{eq:normalization_of_the_hoppings_for_the_global_David_star}) is then a half sphere of unit radius (say, the north hemisphere). 
We use a stereographic projection to map a point in the northern hemisphere onto the plane that intersects the sphere through the equator by connecting the point to the south pole and plotting the intersection of the line and the plane.  
The coordinate transformation is then 
\begin{subequations}
\label{eq:t_global} 
\begin{align}
    t_{\DavidStarOut} &= \frac{1-r^2}{1+r^2} ,
\label{eq:t_edge_global} \\
    t_{\hexagon} &= \sqrt{1 - t_{\DavidStarOut}^2} \cos \Theta
\label{eq:t_hexagon_global} \\
    t_{\triangle} &= \sqrt{1 - t_{\DavidStarOut}^2} \sin \Theta.
\label{eq:t_triangle_global}
\end{align}
\end{subequations}
where $\Theta$ is a polar angle and $r$ is the distance from the origin. The DSL is located at $t_{\DavidStarOut} = t_{\hexagon} = t_{\triangle} = \frac{1}{\sqrt{3}}$, shown with a black dot in Fig.~\ref{fig:Global_stability_of_David_star}.

In Fig.~\ref{fig:Global_stability_of_David_star}(a), (b) and (c) we show the $\Delta \langle \mathcal{P}_{\text{1st}} \rangle$, $\Delta \langle \mathcal{P}_{\text{2nd}} \rangle$, and $\Delta \langle \mathcal{P}_{\triangle} + \mathcal{P}^{-1}_{\triangle} \rangle$ for the David star ansatz on the stereographic plane. Fig.~\ref{fig:Global_stability_of_David_star}(a)  shows that the nearest neighbor exchange $\langle \mathcal{P}_{\text{1st}} \rangle$ is minimal for the Dirac spin liquid. This is not true for the second neighbor and ring exchanges, Figs.~\ref{fig:Global_stability_of_David_star}(b) and (c), and it indicates the instability of the DSL phase for larger values of $J_2$ and $K$. We will address this question in Sec.~\ref{sec:DavidStarPhaseDiagram}. But before doing that, below, we describe the points and curves along which the mean-field band structure is particular.

\subsubsection{Special points and special curves}
\label{sec:specialPoints}

Except for particular points or curves, the David star ansatz has six doubly degenerate dispersive energy bands, where the lowest band is separated from the rest by an energy gap. Thus, the Fermi sea fills the lowest band and is non-degenerate. 

As mentioned in Sec.~\ref{sec:local_stability_against_real_perturbations}, there is a special curve going through the DSL (shown with dashed green in Fig.~\ref{fig:Global_stability_of_David_star}, that is the same curve as in the top row of Fig.~\ref{fig:Local_stability_of_2D_real_ansatze}) along which the gap closes at the $\Gamma$ point, defined by 
\begin{multline}
0 = t_{\DavidStarOut}^8-4 t_{\DavidStarOut}^6 \left(t_{\hexagon}^2-2 t_{\hexagon} t_{\triangle}+3 t_{\triangle}^2\right)\\
+ 2 t_{\DavidStarOut}^4 \left(2 t_{\hexagon}^4-8 t_{\hexagon}^3 t_{\triangle}-t_{\hexagon}^2 t_{\triangle}^2+12 t_{\triangle}^4\right)  \\
-12 t_{\DavidStarOut}^2 t_{\hexagon} t_{\triangle}^2 (t_{\hexagon}-2 t_{\triangle}) (t_{\hexagon}-t_{\triangle}) (t_{\hexagon}+t_{\triangle})  \\
+3 t_{\hexagon}^2 t_{\triangle}^4 \left(3 t_{\hexagon}^2-4 t_{\triangle}^2\right) .
\end{multline}
 This curve ends at a singular point $t_{\DavidStarOut} = \sqrt{2/3}$ and $t_{\hexagon}/t_{\DavidStarOut} = -\frac{1}{\sqrt{2}}$, and from this point another solution of the polynomial equation emerges for positive $t_{\triangle}$ and negative $t_{\hexagon}$.
 
 For negative values of $t_{\triangle}$ and $t_{\hexagon}$, a gapless Dirac point appears, with the Dirac cones being at the $\text{K}$ points. For this to happen, the hoppings need to satisfy the
\begin{equation}
t_{\DavidStarOut}^4 - 2 t_{\DavidStarOut}^2 t_{\hexagon}^2 - 5 t_{\DavidStarOut}^2 t_{\hexagon} t_{\triangle} + 6 t_{\hexagon}^2 t_{\triangle}^2 =0 
\end{equation}
equation, shown by a dashed green line in the stereographic map.  More precisely, all the dashed green curves in the map specify gapless ansatz with Dirac points. Those in the upper half plane have a Dirac cone at the $\Gamma$ point, while those at the lower half at the $\text{K}$ points.

 At the singular point, the energy spectrum of the mean-field Hamiltonian (\ref{eq:mean_field_Hamiltonian}) 
consists of a 6-fold degenerate flat band and three two-fold degenerate flat bands above it, which makes the Fermi sea highly degenerate. Actually, along the horizontal equator with $t_{\triangle} = 0$, the system corresponds to disconnected 12-site David stars, resulting in six two-fold degenerate flat bands (see Appendix \ref{sec:flatBands} for a detailed analysis of flat bands). The flatness of the bands will make the special lines of gapless Dirac points meet at the singular point.
The ground state of this 12-site mean-field Hamiltonian is an SU(6) singlet Fermi sea, with a filled lowest energy flat band, the $|\text{FS} \rangle$ in real space being a product of 12-site singlets. 
The separation of the flat bands depends on $t_{\hexagon}/t_{\DavidStarOut}$. Changing the  $t_{\hexagon}/t_{\DavidStarOut}$ ratio, the lowest three two-fold degenerate bands cross and become six-fold degenerate at the $t_{\hexagon}/t_{\DavidStarOut} = -\frac{1}{\sqrt{2}}$.
As a consequence of the crossing flat bands, the value of the $\langle \text{FS}| \mathcal{P}_{\text{1st}}|\text{FS} \rangle$, $\langle \text{FS}| \mathcal{P}_{\text{2nd}} |\text{FS}\rangle$, and $\langle \text{FS} | \mathcal{P}_{\triangle} + \mathcal{P}^{-1}_{\triangle} |\text{FS} \rangle$ is one constant on the equator left to the singular point and a different constant on the right. 
(Let us mention that there is another similar point at $t_{\hexagon}/t_{\DavidStarOut} = +\frac{1}{\sqrt{2}}$, where the three highest energy bands collapse. This point does not influence the lowest energy band in any way).

The lowest bands remain flat also along the solid yellow curve, defined by
\begin{equation}
    t_{\DavidStarOut}^2 - 2t_{\hexagon}^2 - 2 t_{\hexagon} t_{\triangle} = 0 \,,
\end{equation}
and the solid green curve, defined by
\begin{equation}
    t_{\DavidStarOut}^2 - 2t_{\hexagon}^2  + t_{\hexagon}t_{\triangle} = 0.
    \label{eq:greenline}
\end{equation}
The difference is that the yellow curve gives a gapped spectrum, providing a non-degenerate Fermi sea at one-sixth filling, while along the green curve, the flat band is four-fold degenerate and gives a gapless ansatz. In the latter case, the gap closes for any momentum in the Brillouin zone,  just like for the singular point, so we call it a singular curve. The name is also motivated by the sudden change in $\Delta \langle \mathcal{P}_{\triangle} + \mathcal{P}^{-1}_{\triangle} \rangle$, as shown in Fig.~\ref{fig:Global_stability_of_David_star}(c). 
This sudden change is the consequence of the flat bands. As we cross the singular line, the gap closes in the entire Brillouin zone, and the Fermi sea reorganizes. On the singular line, the Fermi sea is highly degenerate, making the evaluation of the expectation values unfeasible. 
Note that the uniform $\pi_{\hexagon}0_{\triangle}0_{\bigtriangledown}$ ansatz (shown with a black triangle in Fig.~\ref{fig:Global_stability_of_David_star}) is also located on this curve.

\subsubsection{Phase diagram of VMC results}
\label{sec:DavidStarPhaseDiagram}

Having calculated the expectation values of the $\Delta \langle \mathcal{P}_{\text{1st}} \rangle$, $\Delta \langle \mathcal{P}_{\text{2nd}} \rangle$, and $\Delta \langle \mathcal{P}_{\triangle} + \mathcal{P}^{-1}_{\triangle} \rangle$ of Eq.~(\ref{eq:expectation_values_relative_to_the_DSL}) for the different hopping amplitudes, we are ready to construct the phase diagram as a function of the exchange couplings in the Hamiltonian (\ref{eq:extended_Heisenberg_Hamiltonian}).
Namely, for a given $J_2$, $K$ pair (assuming $J_1 = 1$), we calculate the energy provided by Eq.~(\ref{eq:variational_energy_with_J_2_and_K_per_site}) and determine for which values of hopping parameters it is minimal. 
We find that in the area colored by red in Fig.~\ref{fig:Global_stability_of_David_star})(d), the energy of the DSL is the lowest, including the Monte Carlo error. In other words, the DSL is stable against introducing small values of $J_2$ and $K$. We leave the area white if the Monte Carlo error is too big to conclude.
Otherwise, the color of the region in the phase diagram corresponds to the hoppings $t_{\DavidStarOut}$, $t_{\triangle}$, $t_{\hexagon}$ shown by the same color in Fig.~\ref{fig:Global_stability_of_David_star}(h) that minimize the energy, and we treat the points close to each other in Fig.~\ref{fig:Global_stability_of_David_star}(h) as the same phase.

\subsubsection{The interpretation of different phases}
\label{sec:phase_diagram_of_David_star}

From the perspective of the ultracold atomic experiments described by the SU(6) Hubbard model (\ref{eq:Hubbard_Hamiltonian}), we expect both $J_2$ and $K$ in Eq.~(\ref{eq:extended_Heisenberg_Hamiltonian}) to be small compared to $J_1$, and while $J_2\propto t^4/U^3$ is generally positive, $K\propto t^3/U^2$ can be both positive and negative. 
Thus, the experimentally relevant phases are likely the four phases in the right half (i.e., $J_2>0$) of the phase diagram shown in Fig.~\ref{fig:Global_stability_of_David_star}(d). 
The blue and the dark green regions represent weakly coupled David stars ($|t_{\triangle}| < t_{\DavidStarOut}$). In the yellow and light green regions, the instar hexagons and interstar triangles become nearly isolated from each other (the points are very close to the outer edge of the stereographic map defined by $t_{\DavidStarOut}=0$). They differ in the fluxes of the triangles 
(as can be read off from Fig.~\ref{fig:Global_stability_of_David_star}(h), the green and yellow points are related by $t_{\triangle} \rightarrow -t_{\triangle}$, and the blue and dark green by $t_{\hexagon} \rightarrow -t_{\hexagon}$ and $t_{\triangle} \rightarrow -t_{\triangle}$, in both cases the fluxes through the triangles change by $\pi$).
 As both $| \text{FS} \rangle $ and $P_{\text{G}}| \text{FS} \rangle $ are SU(6) singlets, the weakly coupled David stars support local SU(6) singlets. In the case of disconnected hexagons and triangles, the spins in the hexagon form a singlet, and the decoupled triangles form higher dimensional self-conjugate irreducible representations that eventually combine into a singlet. 
 However, we note that one cannot construct a nondegenerate Fermi sea from completely decoupled hexagons and triangles with single occupancy: the spectrum consists of flat bands made from either hexagons or triangles, 
 and even after applying site-dependent chemical potential, the Fermi sea remains gapless in the entire Brillouin zone.

As shown in Fig.~\ref{fig:Global_stability_of_David_star}(d), weakly coupled David stars are stabilized for positive (dark green regions) and negative (dark blue) values of $K$. 
However, the $\Delta\langle P_{\triangle} + P_{\triangle}^{-1} \rangle$ in Fig.~\ref{fig:Global_stability_of_David_star}(c) seems not to be sensitive to the $t_{\hexagon} \rightarrow -t_{\hexagon}$ and $t_{\triangle} \rightarrow -t_{\triangle}$ (connecting the blue and dark green phases), but instead to their position relative to the singular point/line, implying that their Fermi sea consists of wave functions coming from different bands, just like for the completely disconnected David stars at the equator.

Considering the left half (i.e., $J_2<0$) of the phase diagram, the instability of the DSL phase is toward the cyan phase very close to the yellow continuous curve, followed by the orange phase close to the green singular curve.
As shown in Fig.~\ref{fig:Global_stability_of_David_star}(e), setting $K > 0.96$ and $J_2 = 0$ creates a valley of minima in energy along these curves emerging from the singular point and having flat bands.

\subsection{Global stability of the DSL against time-reversal symmetry breaking ansätze}
\label{sec:global_stability_of_chiral_staggered}

\begin{figure}[t]
\centering
\includegraphics[width=0.8\columnwidth]{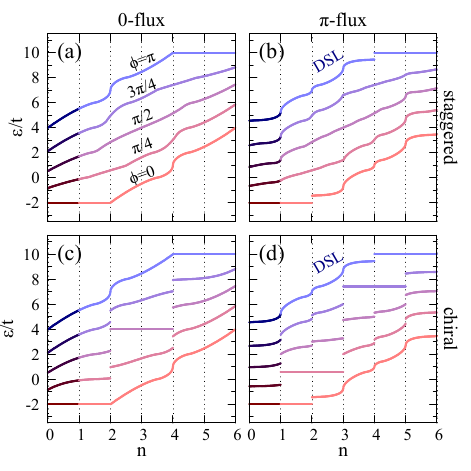}
\caption{
The integrated density of states for the (a) staggered 0-flux $0_{\hexagon} \phi_{\triangle} (-\phi)_{\bigtriangledown}$, (b) staggered $\pi$-flux 
$\pi_{\hexagon} \phi_{\triangle} (-\phi)_{\bigtriangledown}$, (c) chiral $0$-flux $(-2\phi)_{\hexagon} \phi_{\triangle} \phi_{\bigtriangledown}$, and (d) chiral $\pi$-flux $(\pi-2\phi)_{\hexagon} \phi_{\triangle} \phi_{\bigtriangledown}$ ansätze for a few $\phi$ values, shifted vertically by $8\phi t/\pi$ for clarity. $n$ denotes the average number of fermions on a site; the states up to $n=1$ are filled for SU(6) (darker colors). The $\phi=0$ and $\pi$ are the time-reversal invariant ansätze. The discontinuities denote gaps in the spectrum, which happen mostly for the chiral ansätze. The infinite slopes on the continuous curves are the Dirac points [e.g., for $n=1$, $2$, and $3$ for $\phi=\pi$ in (b)]. The horizontal lines of width $\Delta n=2$ show the contribution of flat bands, as they can accommodate two fermions per site without an energy cost.
\label{fig:IDOS}
}
\end{figure}

\begin{figure}
\includegraphics[width=0.85\columnwidth]{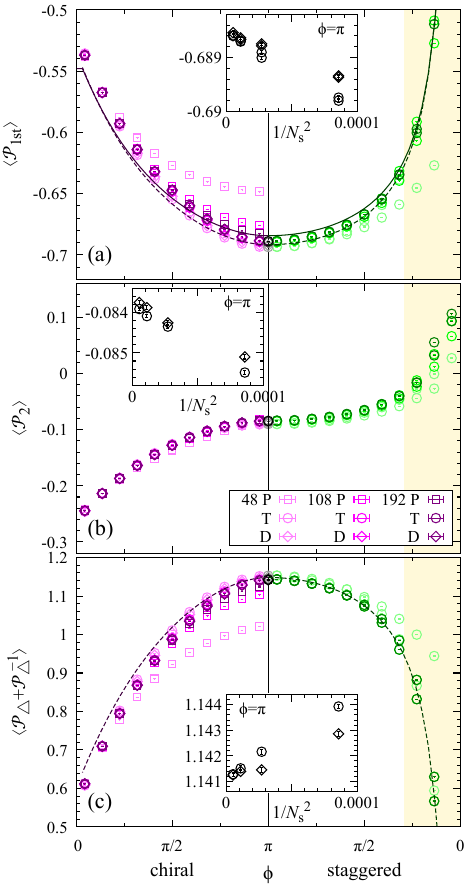}
\caption{The expectation values of the nearest neighbor exchange $\langle \mathcal{P}_{\text{1st}} \rangle
$, the second neighbor exchange $\langle \mathcal{P}_{\text{2nd}} \rangle$, and the ring exchange $\langle \mathcal{P}_{\triangle} + \mathcal{P}^{-1}_{\triangle} \rangle$, for the chiral $A_{1g}$ ansatz with $(\pi - 2 \phi)_{\hexagon} \phi_{\triangle} \phi_{\bigtriangledown}$ flux structure on the left and the staggered $A_{1u}$ ansatz with $\pi_{\hexagon} \phi_{\triangle} (-\phi)_{\bigtriangledown}$ fluxes on the right as a function of the flux $\phi$, calculated by VMC. 
The $\phi = \pi$ corresponds to the DSL,
and the $\phi = 0$ to the uniform $\pi_{\hexagon} 0_{\triangle} 0_{\bigtriangledown}$ ansatz with a highly degenerate mean-field ground state.
For $\phi < 0.29387 \pi$ (yellow region), a band crosses the Fermi energy for the staggered case, and a Fermi surface appears. 
The insets show the finite size scaling of the DSL for each exchange interaction up to $N_s=432$.
The squares, circles, and diamonds represent periodic (P), twisted (T, with twisting angles that ensure threefold symmetry of the $\mathbf{k}$ points around the $\Gamma$ point), and antiperiodic-periodic (D, with twofold symmetry) boundary conditions for clusters of 48, 108, and 192 sites.
\label{fig:VMC_chiral_staggered}
}
\end{figure}

In Sec.~\ref{sec:complex_instabilities}, we considered complex perturbations to the hopping amplitudes. They lead to time-reversal symmetry breaking ansätze and chiral spin liquids \cite{Marston-Zeng_chiral_kagome_1991,hastings2000_gap_opening_with_anisotropy,SUN_square_Heisenberg_large_N_limit_chiral_spin_liquid_with_semiclassical_argument_against_magnetic_order_PRL_2009,SU3_Kagom_chiral_2016}. 
In particular, we introduced the chiral $A_{1g}$ (No. 15 in Ref.~\onlinecite{Bieri_Kagome_chiral_PRB_2015}) and staggered $A_{1u}$ (No. 11 in Ref.~\onlinecite{Bieri_Kagome_chiral_PRB_2015}) ansätze with uniform $(\pi - 2\phi)_{\hexagon} \phi_{\triangle} \phi_{\bigtriangledown}$ and $\pi_{\hexagon} \phi_{\triangle} (-\phi)_{\bigtriangledown}$ flux structures, see Fig.~\ref{fig:chiral_and_staggered_ansatzes}.
Both of these ansätze can be extended to an arbitrary value of $\phi$ such that they interpolate between the DSL ($\phi = \pi$) and the uniform $\pi_{\hexagon}0_{\triangle}0_{\bigtriangledown}$ ansatz ($\phi = 0$). 
The expectation values of the average nearest neighbor exchange $\langle \mathcal{P}_{\text{1st}} \rangle
$, the second neighbor exchange $\langle \mathcal{P}_{\text{2nd}} \rangle$, and the ring exchange $\langle \mathcal{P}_{\triangle} + \mathcal{P}^{-1}_{\triangle} \rangle$ are shown in Fig.~\ref{fig:VMC_chiral_staggered}.

These two time-reversal breaking ansätze have different band structures.
The chiral $A_{1g}$ ansatz opens a gap for all $\phi \notin \{0, \pi \}$ in the mean-field spectrum [Fig.~\ref{fig:IDOS}(d)].
The staggered $A_{1u}$ ansatz preserves the Dirac-Fermi point for $0.294 \pi < |\phi| \leq \pi$, while for $0  < \phi < 0.294 \pi$, a band crosses the Fermi energy, and a Fermi surface appears instead of a Dirac point [Fig.~\ref{fig:IDOS}(b)].

As we already mentioned in Sec.~\ref{sec:specialPoints}, for the $\pi_{\hexagon}0_{\triangle}0_{\bigtriangledown}$ ansatz, the lowest band is flat, without a gap at the Fermi level in the entire Brillouin zone. The expectation values, therefore, depend non-analytically as $\phi \rightarrow 0$ approaches this singularity. 

Fig.~\ref{fig:VMC_chiral_staggered}(a) shows the expectation value of the $\langle \mathcal{P}_{\text{1st}} \rangle$ nearest neighbor exchange, which is minimal for $\phi=\pi$, the DSL ansatz. Thus, the DSL is locally stable against these flux phases. However, the  $\langle \mathcal{P}_{\text{2nd}} \rangle$ second neighbor exchange  and the  $\langle \mathcal{P}_{\triangle} + \mathcal{P}^{-1}_{\triangle} \rangle$ ring exchange terms  are not minimal for $\phi=\pi$, suggesting the instability of the DSL for finite values of $J_2$ and $K$. 

\begin{figure}
\includegraphics[width=0.85\columnwidth]{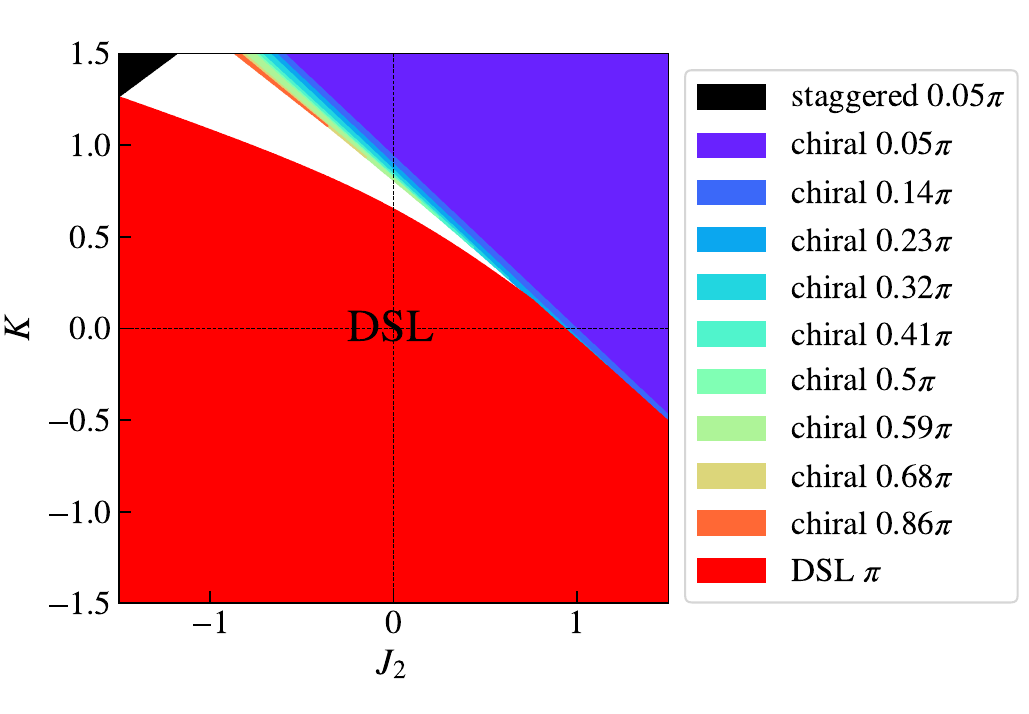
}
\caption{The red region shows the stability of the DSL ($\phi = \pi$) against the chiral $A_{1g}$ ansatz with $(\pi - 2 \phi)_{\hexagon} \phi_{\triangle} \phi_{\bigtriangledown}$  and the staggered $A_{1u}$ ansatz with $\pi_{\hexagon} \phi_{\triangle} (-\phi)_{\bigtriangledown}$ fluxes. The colors shown in the legend denote the optimal flux $\phi$ minimizing the variational energy (\ref{eq:variational_energy_with_J_2_and_K_per_site}).
As $\phi \rightarrow 0$ (blue and black regions), we approach the uniform $\pi_{\hexagon} 0_{\triangle} 0_{\bigtriangledown}$ ansatz with half-filled flat bands, where the large degeneracy of the mean-field ground state prevents VMC. 
The size of the Monte Carlo error does not unambiguously allow us to determine the lowest energy state in the white region.
\label{fig:global_stability_chiral_staggered}
}
\end{figure}

We considered the global stability of the DSL against the chiral $A_{1g}$ and the staggered $A_{1u}$ ansätze in Fig.~\ref{fig:global_stability_chiral_staggered}. Again, the DSL is stable in a finite region of $J_2$ and $K$. 
The chiral $A_{1g}$ destabilizes the DSL for $K > 0$, where the $\phi$ specifying the lowest energy ansatz decreases rapidly from $\pi$ (DSL) to a small value, eventually ending with the uniform $\pi_{\hexagon} 0_{\triangle} 0_{\bigtriangledown}$ ansatz for $\phi=0$. 
Comparing with the global stability of the David star in Fig.~\ref{fig:Global_stability_of_David_star}(d), the DSL becomes unstable against the chiral $A_{1g}$ at a smaller $K >0$ and $J_2 > 0$, implying that increasing $K$ and $J_2$ the first ansatz to stabilize is probably the chiral $A_{1g}$. Therefore, experimental realization of the chiral $A_{1g}$ may be possible if one can tune $t$ and $U$ such that the $K$ and $J_2$ happen to be in the narrow region where a $\phi \notin \{0, \pi\}$ gives the lowest energy ansatz. Further increasing the $K > J_2 > 0$ the lowest energy state is probably an almost disconnected David star with  $\pi_{\hexagon} 0_{\triangle} 0_{\bigtriangledown}$ flux structure (dark green phase in Fig.~\ref{fig:Global_stability_of_David_star}(d)), which is not equivalent to the uniform $\pi_{\hexagon} 0_{\triangle} 0_{\bigtriangledown}$ ansatz.

Fig.~\ref{fig:VMC_chiral_staggered}(b) and (c) suggest that the staggered $A_{1u}$ state becomes favorable for $J_2 < 0$ and $K > 0$. However, the $\langle \mathcal{P}_{\text{2nd}} \rangle$ and $\langle \mathcal{P}_{\triangle} + \mathcal{P}^{-1}_{\triangle} \rangle$ vary much slower in $\phi$ as the $\langle \mathcal{P}_{\text{1st}} \rangle$, thus the lowest energy state is pushed toward $\phi=0$, the uniform $\pi_{\hexagon} 0_{\triangle} 0_{\bigtriangledown}$ ansatz. This is consistent with the black phase in Fig.~\ref{fig:global_stability_chiral_staggered} and the orange phase in Fig.~\ref{fig:Global_stability_of_David_star}(d). In short, there seems to be no region where the staggered $A_{1u}$ ansatz with an intermediate $\phi \notin \{0, \pi \}$ would be the lowest in energy, making experimental realization impossible. 

We also compared the expectation values against the projected mean-field approximation \cite{Hsu_1990,hastings2000_gap_opening_with_anisotropy}. This approximation only considers single occupancy at two and three sites, as described in the Appendix~\ref{sec:proj_mean_field}.  The solid line in Fig.~\ref{fig:VMC_chiral_staggered}(a) is the projected mean field result for two sites, Eq.~(\ref{eq:TdT_projectedMF_2sites}), and the dashed line for the three sites of a triangle, Eq.~(\ref{eq:P12_projectedMF_3sites}). Likewise, we compared the ring exchange expectation values to Eq.~(\ref{eq:P123_projectedMF_3sites}). They are surprisingly close to the VMC values, likely due to the significant value of $N=6$.

\subsection{Chiral states with $\Phi=2\pi/3$ and $\pi/2$ flux}
\label{sec:chirals}

So far we have considered ansätze with $0$ and $\pi$ fluxes through the elementary unit cell consisting of the hexagon and the two adjacent triangles. When described by the mean-field Hamiltonian, the $\Phi=\pi$ flux state requires a doubled unit cell. However, the large value of $N$ may necessitate considering chiral states with fluxes that require larger unit cells \cite{SUN_square_Heisenberg_large_N_limit_chiral_spin_liquid_with_semiclassical_argument_against_magnetic_order_PRL_2009,Szirmai_SU6_honeycomb_chiral_spin_liquid_with_Gutzwiler_projection_PhysRevA_2011, SUN_triangular_lattice_chiral_spin_liquid_Karlo_ED_VMC_PRL_2016, SU3_Kagom_chiral_2016}. Here, we will examine the tripled unit cell, which allows $(2\pi/3)_{\hexagon} \pi_{\triangle} \pi_{\bigtriangledown}$  
flux states, and the quadrupled unit cell for the $\Phi=\pi/2$ flux states  of the form $(\pi/2)_{\hexagon} \pi_{\triangle} \pi_{\bigtriangledown}$. 
For 1/6 filling, the mean-field spectrum of the $(2\pi/3)_{\hexagon} \pi_{\triangle} \pi_{\bigtriangledown}$ is gapless, and is gapped for $(\pi/2)_{\hexagon} \pi_{\triangle} \pi_{\bigtriangledown}$. 
Table.~\ref{tab:chirals} provides their VMC energies, and they are all conveniently above the VMC energy of the DSL, $\langle \mathcal{P_\text{1st}} \rangle=-0.6885$. 

\begin{table}[bt]
\caption{The energy of the Gutzwiller projected Fermi sea for the $(\pi/2)_{\hexagon} \pi_{\triangle} \pi_{\bigtriangledown}$ and $(2\pi/3)_{\hexagon} \pi_{\triangle} \pi_{\bigtriangledown}$ ansatze. We used twisted boundary conditions (BC) to remove possible degeneracy of the ground state and adjust the symmetry of the $\mathbf{k}$-points in the Brillouin zone: the D denotes a two-fold symmetry, T a threefold, and H a sixfold.
\label{tab:chirals}
}
\begin{ruledtabular}
\begin{tabular}{ccccc}
$\Phi$ & $N_s$ & BC & $\langle \mathcal{P_\text{1st}} \rangle$ & $\langle \mathcal{P}_{\triangle} + \mathcal{P}^{-1}_{\triangle} \rangle$ \\
\hline
$2\pi /3$ & 108 & D & $-0.64279 \pm 0.00003$ & $1.0253 \pm 0.0001$ \\
$2\pi /3$ & 432 & D & $-0.64185 \pm 0.00003$ & $1.02488 \pm 0.00007$ \\
$\pi /2$ & 192 & H & $-0.64192 \pm 0.00003$ & $1.02513 \pm 0.00008$\\
$\pi /2$ & 192 & D & $-0.64287 \pm 0.00003$ & $1.0274 \pm 0.0001$\\
$\pi /2$ & 192 & T & $-0.64281 \pm 0.00002$ & $1.02726 \pm 0.00009$
\end{tabular}
\end{ruledtabular}
\end{table}%

\section{Structure factor and correlation function}
\label{sec:structure_factor}

\begin{figure}[t]
	\centering
	\includegraphics[width=0.9\columnwidth]{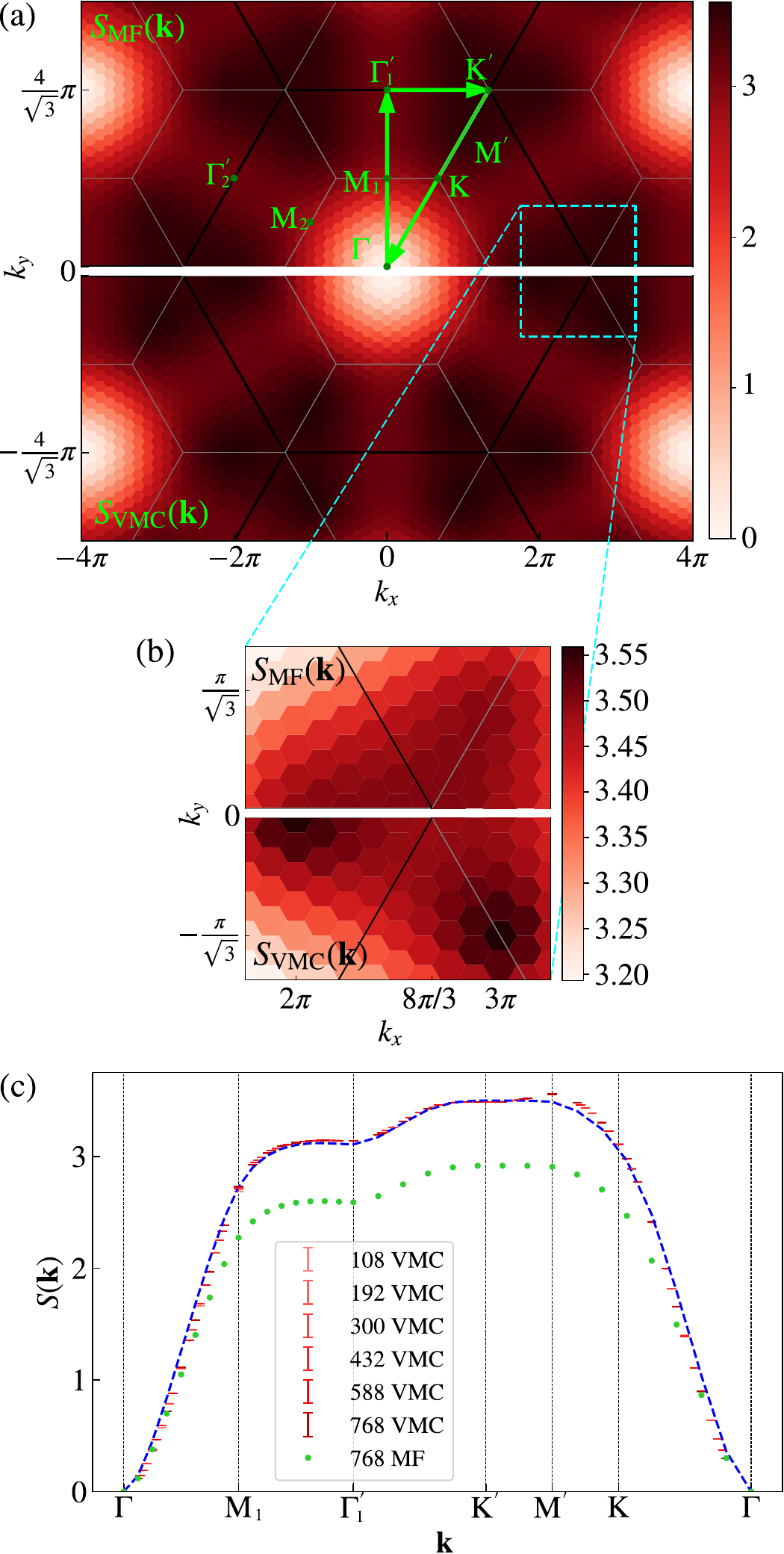}
	\caption{(a) The structure factor $S(\mathbf{k})$, Eq.~(\ref{eq:Structure_factor}), for a cluster of 768 sites in the mean field approach (upper half) and calculated by VMC (lower half). Both show triangular-shaped plateaus around the $\text{K}'$ points.
The mean-field structure factor is multiplied by $N/(N-1)=6/5$ to satisfy the sum rule, Eq.~(\ref{eq:TdTMFcorr}).
(b) While the $S_{\text{MF}}(\mathbf{k})$ is smooth, the $S_{\text{VMC}}(\mathbf{k})$ has weak maxima at the $\mathrm{M}'$ points.  (c) The $S_{\text{MF}}(\mathbf{k})$ and the $S_{\text{VMC}}(\mathbf{k})$ for different cluster sizes along the high-symmetry lines in the extended Brillouin zone [the green path in (a)]. The mean-field result multiplied by $N/(N-1)=6/5$ (blue dashed line) follows the VMC result except around the $\mathrm{M}'$ point.}
 \label{fig:structure_factors}
\end{figure}

In this section, we compute the structure factor 
\begin{equation}
    S(\mathbf{k}) \equiv \frac{1}{N_s}  \sum_{i,j} e^{i \mathbf{k} \cdot (\mathbf{r}_j - \mathbf{r}_i)} \langle \psi| \mathbf{T}_i \cdot \mathbf{T}_j |\psi \rangle \;. 
\label{eq:Structure_factor}
\end{equation}
For comparison, we calculated both $S_{\text{VMC}}$ and $S_{\text{MF}}$, using $ | \psi \rangle = P_{\text{G}} | \text{FS} \rangle$ and $| \psi \rangle = | \text{FS} \rangle$, respectively. 
In the VMC calculation we use Eq.~(\ref{eq:T_i_T_j_related_to_n_i_n_j})
to relate the structure factor to diagonal matrix elements of fermion densities.
 %
 In the mean field case, we evaluated $\langle \text{FS}|  \mathbf{T}_i \cdot \mathbf{T}_j  |\text{FS} \rangle$ using Eq.~(\ref{eq:T_i_T_j_related_to_n_i_n_j_mean_field}) derived from Wick's theorem.
 We show our results in Fig.~\ref{fig:structure_factors}.
 Due to the sublattice structure of the kagome lattice, the structure factor is periodic
 in the extended Brillouin zone (eBZ,  black hexagons in Fig.~\ref{fig:structure_factors}), equivalent to the Brillouin zone of the triangular lattice one gets by adding lattice sites in the centers of the hexagons. 
Both the $S_{\text{MF}}(\mathbf{k})$ and the $S_{\text{VMC}}(\mathbf{k})$ have triangular-shaped plateaus around the $\text{K}'$ points,
the only difference being the minor peaks at the $\mathrm{M}'$ points in the VMC calculation. 
Since in the VMC, we do not know the norm of the $ | \psi \rangle = P_{\text{G}} | \text{FS} \rangle$,  we normalize the $S_{\text{VMC}}(\mathbf{k})$ to fulfill the sum rule
\begin{equation}
    \sum_{\mathbf{k}} S(\mathbf{k}) = N_{\mathbf{k}} \langle \mathbf{T} \cdot \mathbf{T} \rangle = \frac{4}{3} N_s C_2 = \frac{35}{9}N_s\;,
\end{equation}
where the summation is over the $N_{\mathbf{k}}=\frac{4}{3} N_s$ $\mathbf{k}$-points in the extended Brillouin zone and  $C_2= \langle \mathbf{T} \cdot \mathbf{T} \rangle = \frac{N^2-1}{2N}$ is the eigenvalue of the quadratic Casimir operator in the fundamental representation on any site.
We also note that the ratio of the sum rules $\sum_{\mathbf{k}} S_{\text{MF}}(\mathbf{k})  / \sum_{\mathbf{k}} S_{\text{VMC}}(\mathbf{k})  = 1 - \frac{1}{N}$ is due to the charge fluctuations in the mean field case, which affect the value of the quadratic Casimir operator $\langle \mathbf{T} \cdot \mathbf{T} \rangle' = \left( 1 - \frac{1}{N} \right) \langle \mathbf{T} \cdot \mathbf{T} \rangle$ \cite{Mi_SU4_cikkunk}, see also Eq.~(\ref{eq:TdTMFcorr}).

\begin{figure}[t]
	\centering
	\includegraphics[width=0.45\textwidth]{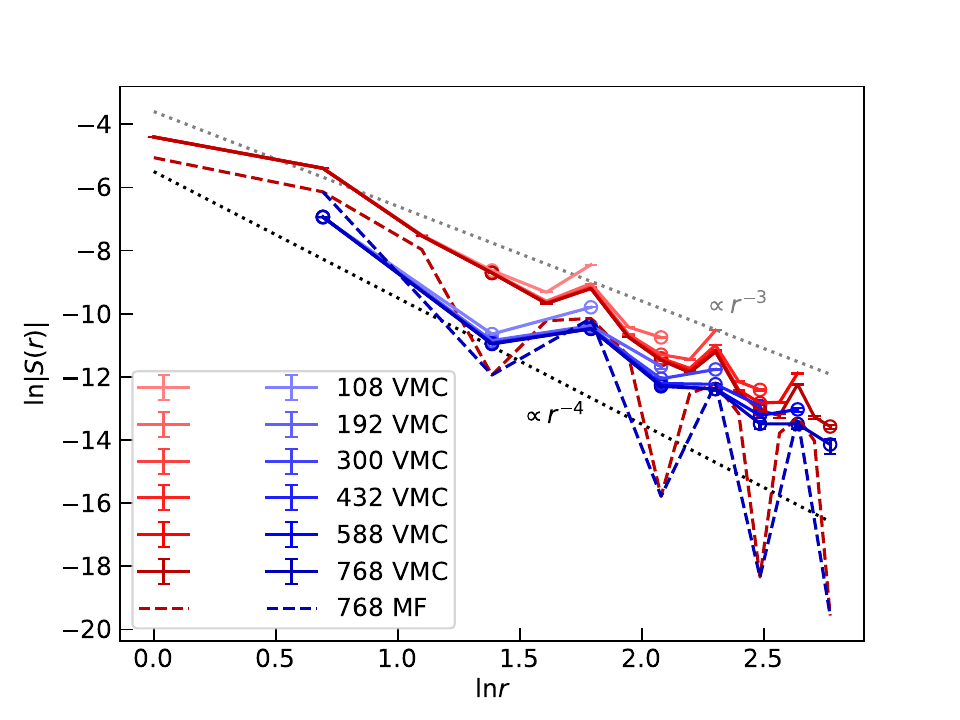}
	\caption{
	The decay of the real space flavor-flavor correlation function $|S(r_{i,j})|$, defined in Eq.~(\ref{eq:Srij}), calculated by VMC for different system sizes (solid lines with error bars) and the mean-field result (dashed lines).
	The colors encode different directions: the red along the edges of the triangles and the blue in the directions crossing the centers of the hexagons.
	We average over equivalent directions to eliminate the effect of the antiperiodic boundary condition. 
 The absolute values of $S(r)$ take care of alternating sign structure along the edges of the triangles (we put empty circles for $S(r) > 0$). At the same time, it is always positive along the directions through the centers of the hexagons. 
 The mean-field results are all negative, following Eq.~(\ref{eq:T_i_T_j_related_to_n_i_n_j_mean_field}).
 The dotted $\propto r^{-\alpha}$ lines with $\alpha=3$ and $4$ are guides for the eye. The VMC results decay algebraically with a power between 3 and 4, while the mean field $S_{\text{MF}}(r) \propto r^{-4}$ for large enough $r$, typical for Dirac fermions.
	}
 \label{fig:power_law_decay_of_correlations}
\end{figure}

In real space, the correlation function 
\begin{equation}
  S_{\text{VMC}}(r_{i,j}) = \langle \text{FS}|P_{\text{G}} \mathbf{T}_i \cdot \mathbf{T}_j P_{\text{G}}|\text{FS} \rangle
 \label{eq:Srij}
 \end{equation}
decays algebraically with the distance $r_{i,j} = | \mathbf{r}_i - \mathbf{r}_j|$ between the two sites, with a power between 3 and 4 (see Fig.~\ref{fig:power_law_decay_of_correlations}). The $S_{\text{MF}}(r)$ for a system with a Dirac-Fermi point is expected to approach the $S_{\text{MF}}(r) \propto r^{-4}$ behavior for large enough $r$ \cite{Affleck_pi_flux,rantner2002, Mi_SU4_cikkunk}. Interestingly, the alternating sign structure of $S_{\text{VMC}}(r)$ in the directions $\mathbf{r}_i - \mathbf{r}_j$ along the edges of the triangles (not entering the hexagons) can be reproduced by summing the phases coming from the six peaks at the $\mathrm{M}'$ points, $\sum_{l = 1}^6 e^{i \mathbf{k}_{\mathrm{M}_{l}'} \cdot (\mathbf{r}_i - \mathbf{r}_j)}$.

\section{Dynamical structure factor}
\label{sec:dynamical_structure_factor}

The momentum-resolved dynamical spin structure factor at zero temperature is defined by 
\begin{equation}
\label{eq:S^33_k_omega}
     S^{33}(\mathbf{k},\omega) =  \sum_{f} \left| \langle f| T^{3}_{\mathbf{k}}| \text{GS} \rangle  \right|^2
     \delta(\omega + E_{\text{GS}} - E_{f})\;,
\end{equation}
where the sum is over the final states $|f\rangle$ of $\mathcal{H}$ with energies $E_{f}$, $|\text{GS}\rangle$ denotes the ground state with energy $E_{\text{GS}}$, and 
\begin{equation}
 T^{3}_{\mathbf{k}} = \frac{1}{\sqrt{N_s}} \sum_{\mathbf{R},d} e^{i \mathbf{k} \cdot (\mathbf{R}+\boldsymbol{\delta}_d)} T^{3}_{\mathbf{R},d}
\end{equation}
 is the Fourier transform of the 
\begin{equation}
 T^{3}_{\mathbf{R},d} = \frac{1}{2} \left( f^{\dagger}_{\mathbf{R},d,1} f^{\phantom{\dagger}}_{\mathbf{R},d,1} - f^{\dagger}_{\mathbf{R},d,2} f^{\phantom{\dagger}}_{\mathbf{R},d,2} \right)
 \label{eq:T3def}
\end{equation}
diagonal spin operator on the sublattice $d$, as defined in Eq.~(\ref{eq:spin_operator_with_partons_SU_4}).
Here $\mathbf{R} = R_1 \mathbf{a_1} + R_2 \mathbf{a_2}$ is a Bravais lattice vector, $d \in \{1,2,3\}$ are sublattice indices with basis vectors 
$\boldsymbol{\delta}_1 = (0,0)$, 
$\boldsymbol{\delta}_2 =(1/2, 0)$, and  
$\boldsymbol{\delta}_3 =(1/4, \sqrt{3}/4)$, so that the position of site $j$ is $\mathbf{r}_j = \mathbf{R}_j + \boldsymbol{\delta}_{d_j}$.
 Due to the SU(6) symmetry of both the Heisenberg (\ref{eq:Heisenberg_Hamiltonian}) and the mean-field Hamiltonian (\ref{eq:mean_field_Hamiltonian}), and since the ground state $| \text{GS} \rangle$ is a singlet, replacing $T^{3}_{\mathbf{k}}$  with any of the SU(6) spin operators $T^{a}_{\mathbf{k}}$ ($a \in \{ 1,\dots, N^2 - 1 = 35 \}$) in the matrix elements of Eq.~(\ref{eq:S^33_k_omega}) results in $S^{aa}(\mathbf{k},\omega) = S^{33}(\mathbf{k},\omega)$. 
So the full dynamical structure factor is $S(\mathbf{k},\omega) = \sum_{a} S^{aa}(\mathbf{k},\omega) =  (N^2 - 1) S^{33}(\mathbf{k},\omega)$, and the static structure factor in Eq.~(\ref{eq:Structure_factor}) is $\int S(\mathbf{k},\omega) d\omega = S(\mathbf{k})$.
 
In the following, we calculate $S^{33}_{P_{\text{G}}}(\mathbf{k},\omega)$ using Gutzwiller projected particle-hole states built upon the approximating ground state $P_{\text{G}} | \text{FS} \rangle$ \cite{First_S_q_w_latter,*Yang_Li_2011PhRvB..83f4524Y, Excited_states_above_Fermi_sea, Becca, Mi_SU3_cikkunk, Mi_SU4_cikkunk}. We compare it with the mean field case $S^{33}_{\text{MF}}(\mathbf{k},\omega)$, for which $|\text{GS}\rangle = |\text{FS} \rangle$ and $|f \rangle \in \{ f^{\dagger}_{\mathbf{k} + \mathbf{q}, n, \sigma} f^{\phantom{\dagger}}_{\mathbf{q}, n', \sigma'} |\text{FS} \rangle \} $ are the exact ground- and excited states, explained in detail in \cite{Mi_SU4_cikkunk}.

\subsection{Calculation of $S^{33}_{P_{\text{G}}}(\mathbf{k},\omega)$ using VMC}

For the $S^{33}_{P_{\text{G}}}(\mathbf{k},\omega)$, following Refs.~\onlinecite{First_S_q_w_latter,*Yang_Li_2011PhRvB..83f4524Y, Excited_states_above_Fermi_sea, Becca, Mi_SU3_cikkunk, Mi_SU4_cikkunk} we construct an approximation to the eigenstates $|f \rangle$ by diagonalizing the Hamiltonian in a Hilbert subspace spanned by Gutzwiller-projected particle-hole states of flavor $\sigma$ with momenta $\mathbf{k}$, 
\begin{align}
|\mathbf{k};\sigma; \mathbf{R}, d;\bar{d}\rangle &= P_{\text{G}} \sum_{\mathbf{R'}} e^{i \mathbf{k} \cdot \mathbf{R'}}
(-1)^{(R_1 + R_2)R'_1} \nonumber \\  
&\times (-1)^{\xi (\mathbf{R},\mathbf{R'})} f^{{\dagger}}_{\mathbf{R+R'},d,\sigma} f^{\phantom{\dagger}}_{\mathbf{R'},\bar{d},\sigma} |\pi\text{FS} \rangle ,
\label{eq:particle-hole_excited_states}
\end{align}
where the $(-1)^{(R_1 + R_2)R'_1}$ comes from the gauge transformations
$G_{\mathsf{T}_1}:f^{{\dagger}}_{j,\sigma} \rightarrow (-1)^{R_{j1} + R_{j2}} f^{{\dagger}}_{j,\sigma}$ and $G_{\mathsf{T}_2}:f^{{\dagger}}_{j,\sigma} \rightarrow f^{{\dagger}}_{j,\sigma}$ for the doubled unit cell ansatz shown in Fig.~\ref{fig:gauge_transformations_connecting_the_old_and_new_ansatze}(a), as explained in \cite{Mi_SU4_cikkunk}.
The $(-1)^{\xi (\mathbf{R},\mathbf{R'})}$ accounts for the boundary conditions, it is always $+1$ for periodic boundaries, while for antiperiodic boundaries we get $-1$ when $\mathbf{R}$ is inside the cluster, but $\mathbf{R+R'}$ crosses antiperiodic boundaries an odd number of times \cite{Mi_SU4_cikkunk}. 
The particle belongs to the fundamental $\mathbf{6}$ and the hole to the antifundamental $\mathbf{\bar 6}$ irreducible representation, and since $\mathbf{6} \otimes \mathbf{\bar 6} = \mathbf{1} + \mathbf{35}$, the state itself is not an irreducible representation \cite{Mi_SU3_cikkunk}. 
We then consider 
\begin{equation}
|\mathbf{k};\mathbf{35}_3; \mathbf{R}, d;\bar{d}\rangle \equiv
 \frac{1}{2} \left( |\mathbf{k};A; \mathbf{R}, d;\bar{d}\rangle - |\mathbf{k};B; \mathbf{R}, d;\bar{d}\rangle \right)
\end{equation}
as the final eigenstates $|f\rangle$, which transform as the adjoint irreducible representation $\mathbf{35}$ of SU(6) \cite{Mi_SU3_cikkunk}, just like the $T^{3}_{\mathbf{k}}| \text{GS} \rangle$.
We calculate the overlap 
$\langle \mathbf{k}; \mathbf{35}_3; \mathbf{R}, d;\bar{d} | \mathbf{k}; \mathbf{35}_3; \mathbf{R'}, d';\bar{d}'  \rangle$ 
and the Hamiltonian 
$\langle \mathbf{k}; \mathbf{35}_3; \mathbf{R}, d;\bar{d} | \mathcal{H}|\mathbf{k}; \mathbf{35}_3; \mathbf{R'}, d';\bar{d}'  \rangle$ matrices by Monte Carlo sampling. 
Solving the generalized eigenvalue problem provides the excitation energies $E_f$ and the states $|f\rangle$, from which we can calculate $\langle f | T^3_{\mathbf{k}} | \text{GS} \rangle =  \sum_{d} e^{i \mathbf{k} \cdot \delta_d}\langle f | \mathbf{k}; \mathbf{35}_3; \mathbf{0}, d;d  \rangle $, as Eq.~(\ref{eq:T3def}) implies. We refer the reader to Refs.~\onlinecite{Mi_SU3_cikkunk,Mi_SU4_cikkunk} for details.

\subsection{Comparison of $S^{33}_{P_{\text{G}}}(\mathbf{k},\omega)$ and $S^{33}_{\text{MF}}(\mathbf{k},\omega)$}
 
\begin{figure}[t]
	\centering
	\includegraphics[width=0.95\columnwidth]{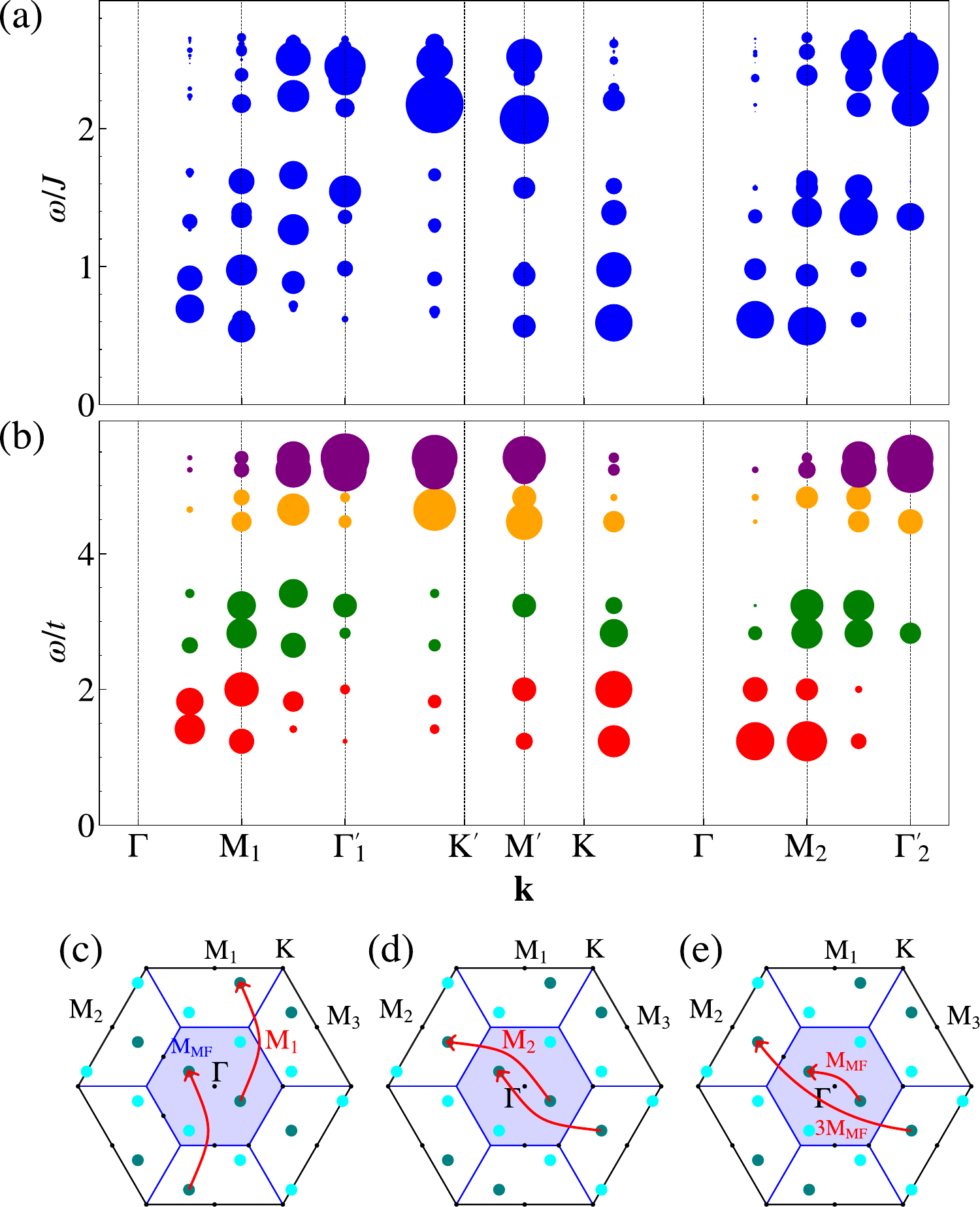}
	\caption{
	The dynamical spin structure factor: (a) calculated by VMC, $S_{P_{\text{G}}}^{33}(\mathbf{k},\omega)$, and (b) in the mean-field approximation, $S_{\text{MF}}^{33}(\mathbf{k},\omega)$, for a 48-site cluster. 
	The area of the circles is proportional to the spectral weights in $S^{33}(\mathbf{k},\omega)$.
	To avoid the degeneracy of the mean-field ground state, we imposed an antiperiodic boundary condition that shifts the momenta away from the Dirac point, opening a finite-size gap. 
		In (c)-(e), we show the shifted momenta for the 48-site cluster; four of them are in the reduced Brillouin zone of the quadrupled unit cell. The lowest energy excitations originate from particle-hole excitations (red arrows) between the momenta closest to the Dirac point (colored teal). 
		All the relative momenta connecting these points give equal excitation energies. 
		We extended the path shown in Fig.~\ref{fig:structure_factors}(a) by the relative momenta $\mathbf{k} =  \mathrm{M}_2$ and $\Gamma'_2$ to see this effect. 
	The Gutzwiller projection lifts the degeneracy, similarly to the SU(4) case \cite{Mi_SU4_cikkunk}.}
 \label{fig:dynamical_structure_factor_48}
\end{figure}

\begin{figure}[t]
	\centering
	\includegraphics[width=0.95\columnwidth]{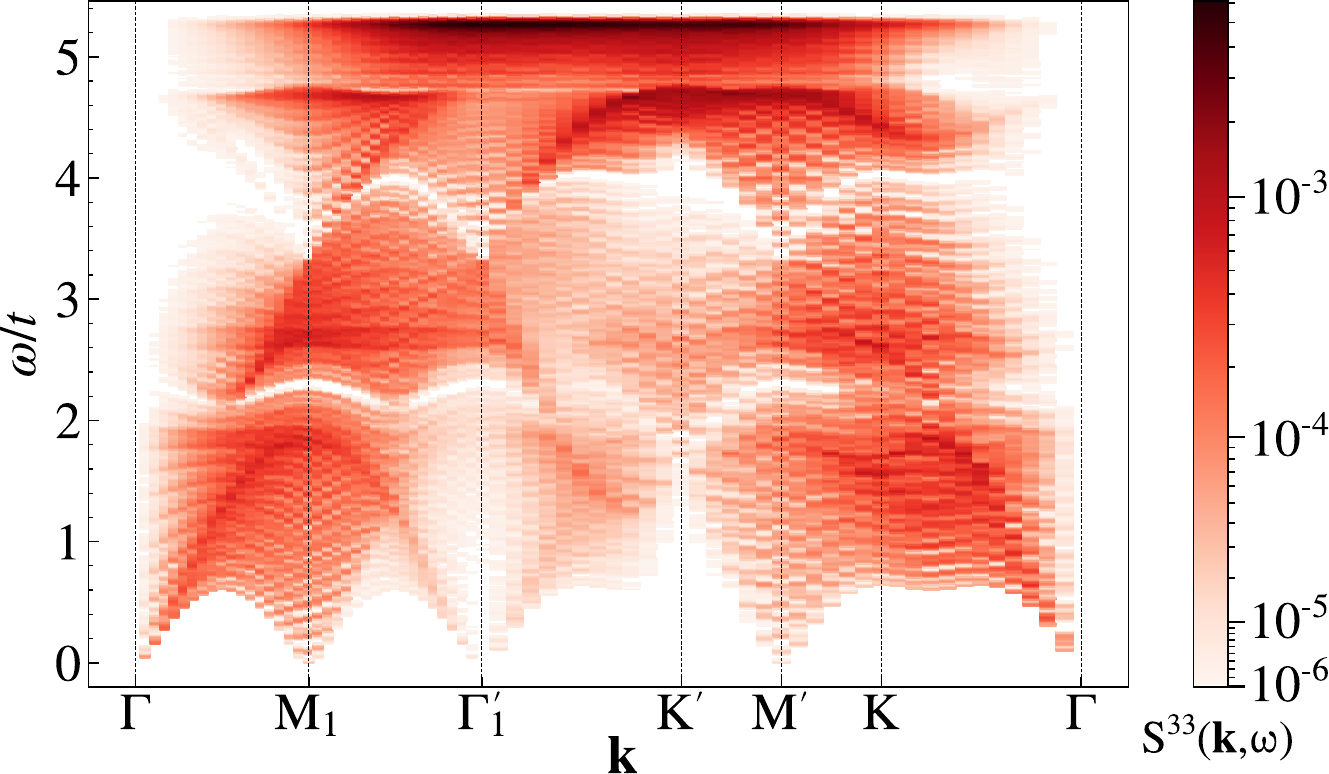}
	\caption{The dynamical spin structure factor of the DSL ansatz, $S_{\text{MF}}^{33}(\mathbf{k},\omega)$, calculated in the mean-field approach for a $3\times 36^2 = 3888$ site cluster, along the path shown in Fig.~\ref{fig:structure_factors}(a). 
 The gapless towers of fractionalized excitations (flavorons), located at the $\Gamma$, $\mathrm{M}_1$, $\Gamma'_1$, and  $\mathrm{M'}$ points in the (extended) Brillouin zone originate from low-energy particle-hole excitations across the Dirac cones and closely follow the momenta of the mean-field Fermi points [see  Fig.~\ref{fig:DSL_bands}(a)]. 
 The APBC has the same effect on the states in the tower as in Fig.~\ref{fig:dynamical_structure_factor_48}(b), namely additional degenerate peaks appear next to the towers shifted to the nearest finite size momentum, with vanishing spectral weight in the thermodynamic limit.
 \label{fig:dynamical_structure_factor_continuum}}
\end{figure}

Since the dimension of the local Hilbert space is $N$, the dimension of the Hilbert space, $N_s!/[(N_s/N)!]^N=N_s!/[(N_s/6)!]^6$, quickly increases with the system size $N_s$, the largest system size for which we could calculate $S^{33}_{P_{\text{G}}}(\mathbf{k},\omega)$ with small Monte Carlo errors had 48 sites. 
Although it would be difficult to conclude the thermodynamic limit from these results, Figs.~\ref{fig:dynamical_structure_factor_48}(a) and (b) show that the distribution of the spectral weights (i.e., the matrix elements $\left| \langle f| T^{3}_{\mathbf{k}}| \text{GS} \rangle  \right|^2$) in $S^{33}_{P_{\text{G}}}(\mathbf{k},\omega)$ and $S^{33}_{\text{MF}}(\mathbf{k},\omega)$ is quite similar at low energies. 
 This is in contrast to what happens in the SU(2) case \cite{ 2019PhRvX...9c1026F} (where $P_{\text{G}}$ can even create new gapless excitations), although the tendency was already there for the SU(4) case \cite{Mi_SU4_cikkunk}.
 Thus, assuming that the mean-field method gives qualitatively, perhaps even quantitatively, good results for the low energy behavior in the thermodynamic limit, we calculate the $S^{33}_{\text{MF}}(\mathbf{k},\omega)$ for an $N_s=3888$ site cluster in Fig.~\ref{fig:dynamical_structure_factor_continuum}.
 The gapless excitation towers at the $\mathrm{M}$, $\mathrm{M'}$, $\Gamma$ and $\Gamma'$ points originate from the Dirac Fermi point shown in Fig.~\ref{fig:DSL_bands}(b), and can be understood as particle-hole excitations, where a fermion having momentum $\mathbf{q}$ is transferred from the top of the occupied band to an unoccupied state
having momentum $\mathbf{k} + \mathbf{q}$ at the bottom of the Dirac cone in the second band. 
 Thus, a gapless tower is expected at each relative momenta $\mathbf{k}$ connecting two $\Gamma_{\text{MF}}$ points, as shown in Fig.~\ref{fig:DSL_bands}(a). Furthermore, as shown in Ref.~\cite{Mi_SU4_cikkunk} for the SU(4) case, the position of the excitation towers in the extended Brillouin zone is not modified by the Gutzwiller projector, and the projected spectrum also seems to be gapless. 
In principle, the dynamical spin structure factor of cold atoms in optical
lattices is accessible by Bragg spectroscopy using coherent momentum transfer mapping so that
a direct comparison may be feasible \cite{Hoinka_spin_Bragg_PRL_2012,Ruwan_Science_2022}.

\section{Conclusion}
\label{sec:conclusion}

In this study, we proposed the Dirac spin liquid (DSL) as the ground state for the Mott insulating phase of fermions with six flavors on the Kagome lattice, which can be realized using optically trapped $^{173}$Yb atoms. 
To reach this conclusion, we have investigated the energetical stability of the DSL as the ground state for the SU(6) Heisenberg model in the fundamental representation using mean-field theory and variational Monte Carlo (VMC) to evaluate Gutzwiller-projected wavefunctions. 
After classifying the possible perturbations of the hopping amplitudes in the mean-field Hamiltonian within a unit cell of 12 sites, including chiral ones, we confirmed that the DSL state remained the lowest energy state. 
Furthermore, we found that finite values of the second-neighbor ($J_2$) and ring ($K$) exchange are necessary to destabilize the DSL, highlighting its resilience to further interactions. However, our study cannot exclude an SU(6) symmetry-breaking ground state of some form.

To characterize the DSL, we studied the correlation function and dynamical spin structure factor in the mean-field and VMC approaches. 
The spin-spin correlations decays algebraically with the distance, with an exponent between 3 and 4, only slightly different from the mean-field exponent four coming from the Dirac points (Fig.~\ref{fig:power_law_decay_of_correlations}). 
The Fourier transformed correlation function $S(\mathbf{k})$ (i.e., the spin structure factor) shows increased spectral weights in the form of triangular plateaus around the $\text{K}'$ points in the extended Brillouin zone. 
There are only slight differences between the mean-field and the VMC results in the form of barely noticeable humps appearing for the latter at the $\mathrm{M}'$ points in Fig.~\ref{fig:structure_factors}. 

We have also calculated the dynamical spin structure factor $S(\mathbf{k},\omega)$ in a 48-site cluster by VMC in a reduced Hilbert space spanned by Gutzwiller projected particle-hole excitations. The system is too small to get the details of $S(\mathbf{k},\omega)$, but we were able to compare it to the mean-field approximation and, in contrast to the SU(2) case, found a substantial similarity (see Fig.~\ref{fig:dynamical_structure_factor_48}). Based on this similarity, we have studied the $S(\mathbf{k},\omega)$ in the mean-field approach for an extensive system with 3888 sites and found a continuum of fractionalized excitations [flavorons, the equivalent of spinons in SU(2)] with gapless towers centered at the $\Gamma$ and $\mathrm{M}$ points in the Brillouin zone, Fig.~\ref{fig:dynamical_structure_factor_continuum}. We attribute the applicability of the mean-field approximations to the large, $N=6$, number of fermionic flavors.

In addition to the main results described above, we also obtained several side results:
(i) We provided a lower and an upper bound on the ground state energy (Appendix~\ref{sec:lower_bound});
(ii) We derived an expression to efficiently calculate the expectation value of the three-site ring exchange in the mean-field approximation in Appendix ~\ref{sec:relations_between_permuations_and_the_diagonal_terms_in_the_Fermi_sea}, and in the VMC using diagonal operators in Appendix~\ref{sec:relations_between_permuations_and_the_diagonal_terms_in_the_Mott_phase}; 
(iii) We determined the boundary of the ferromagnetic states from the gap closing of one- and two-magnon excitations in Appendix~\ref{sec:ferro};
(iv) We have extended the projected mean-field expressions to the calculation of off-diagonal operators and three site terms in SU(N) models in Appendix~\ref{sec:proj_mean_field}. The method gives expectation values that are very close to the VMC values;
(v) We have identified the non-topological nature of the flat bands in the DSL band structure (Appendix~\ref{sec:flatBands}).

Our study provides a comprehensive understanding of the DSL in the SU(6) Heisenberg model on the Kagome lattice. 

%
%
%


\begin{acknowledgments}

We thank Yasir Iqbal for the helpful discussions.
We acknowledge the financial support of the Hungarian NKFIH Grant No. K 142652 and, in part, the NSF PHY-1748958 grant to the Kavli Institute for Theoretical Physics (KITP). 

\end{acknowledgments}

\appendix

\section{Gauge transformations}
\label{sec:gauge_trans}

\subsection{Gauge transformation restoring the symmetries of the kagome lattice}
\label{App:projective_symmetries}

\begin{figure}[t]
\centering
\includegraphics[width=0.8\columnwidth]{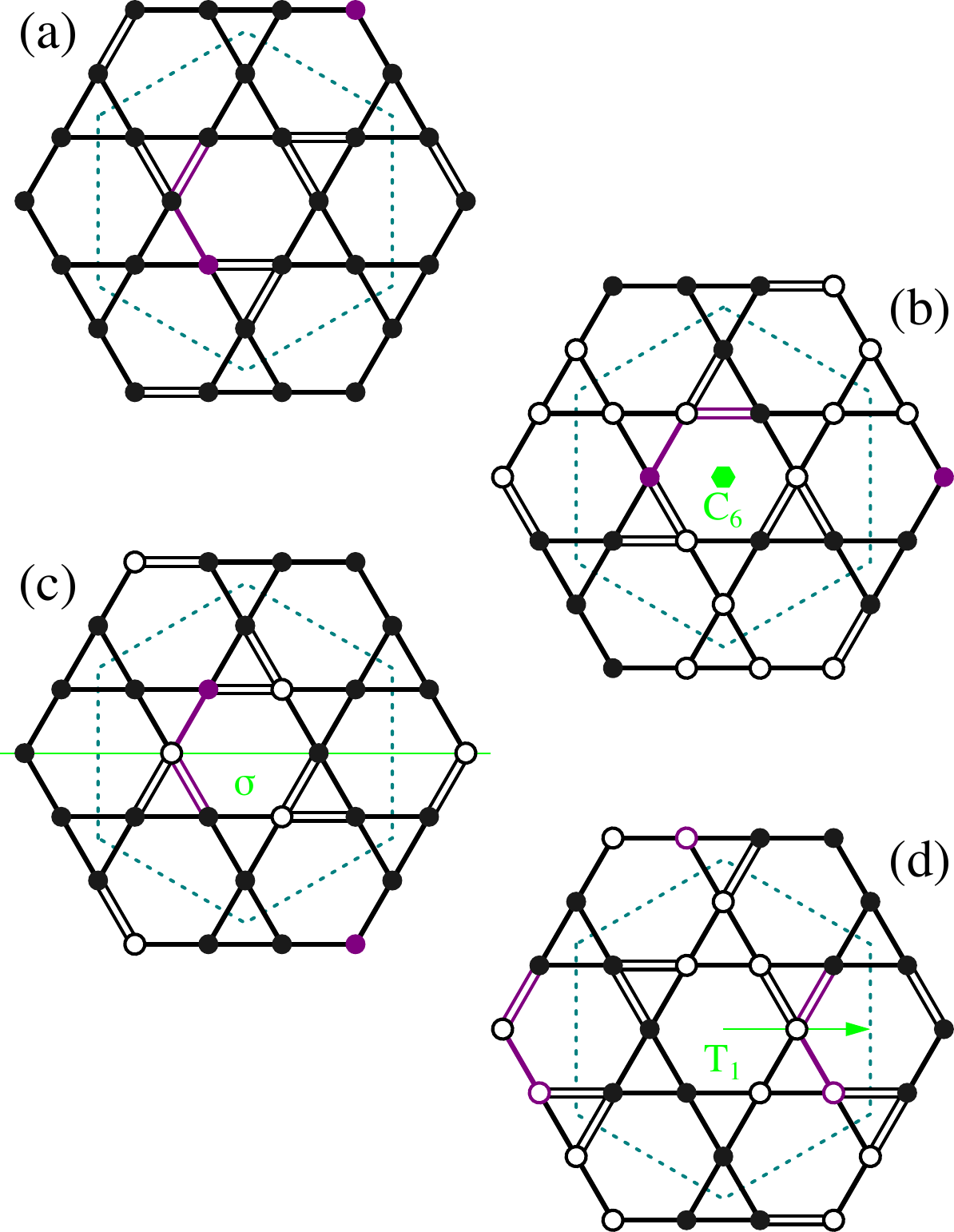}
\caption{
The site-dependent but flavor-independent gauge transformations 
$G:f^{{\dagger}}_{j,\sigma} \rightarrow e^{i \phi(j)}f^{{\dagger}}_{j,\sigma}$ 
needed to restore the symmetries of $\mathcal{H}_{\text{MF}}$, under the symmetry operations $\mathsf{g}$ of the kagome lattice, as $\mathcal{H}_{\text{MF}} = G_{\mathsf{g}} \mathsf{g} \mathcal{H}_{\text{MF}} \mathsf{g}^{-1} G^{-1}_{\mathsf{g}}$. 
The phases $e^{i \phi(j)}$ turn out to be $\pm 1$ signs, the $f^{{\dagger}}_{j,\sigma}$ on the sites marked with white circles acquire a minus sign. Consequently, the minus sign can be transferred to the hopping $t_{i,j}$, provided that only one of the $f^{{\dagger}}_{i,\sigma}$ and $f^{{\dagger}}_{j,\sigma}$ acquires a minus sign. The gauge transformations $G_{\mathsf{C}_6}$, $G_{\mathsf{\sigma}}$, $G_{\mathsf{T}_1}$ are shown in (b), (c), (d), respectively. 
(a) shows the hopping structure of the DSL of Fig.~\ref{fig:DSL_with_APBC}, for easier comparision. The action of the symmetry operations $\mathsf{g}$ is visualized by transforming the purple bonds and disks.
\label{fig:gauge_transformations_for_projective_symmetry}
}
\end{figure}

The gauge transformations $G:f^{{\dagger}}_{j,\sigma} \rightarrow e^{i \phi(j)}f^{{\dagger}}_{j,\sigma}$ needed to restore the symmetries of the kagome lattice are shown in Fig.~\ref{fig:gauge_transformations_for_projective_symmetry}, for the case of periodic boundary conditions. As it turns out, the $e^{i \phi(j)}$ are simple $\pm 1$ signs, which changes the hoppings $t_{i,j}$ connected to site $j$. In the case of the antiperiodic boundary condition, the projective symmetries of $C_6$ and $\sigma$ are broken (though they are restored in the thermodynamic limit), and the gauge transformation of the translations should be modified at the boundary, as explained in \cite{Mi_SU4_cikkunk}.

\subsection{Connection between the DSL ansätze}
\label{App:connection_between_the_DSL_ansatze}

\begin{figure}[t]
\centering
\includegraphics[width=0.9\columnwidth]{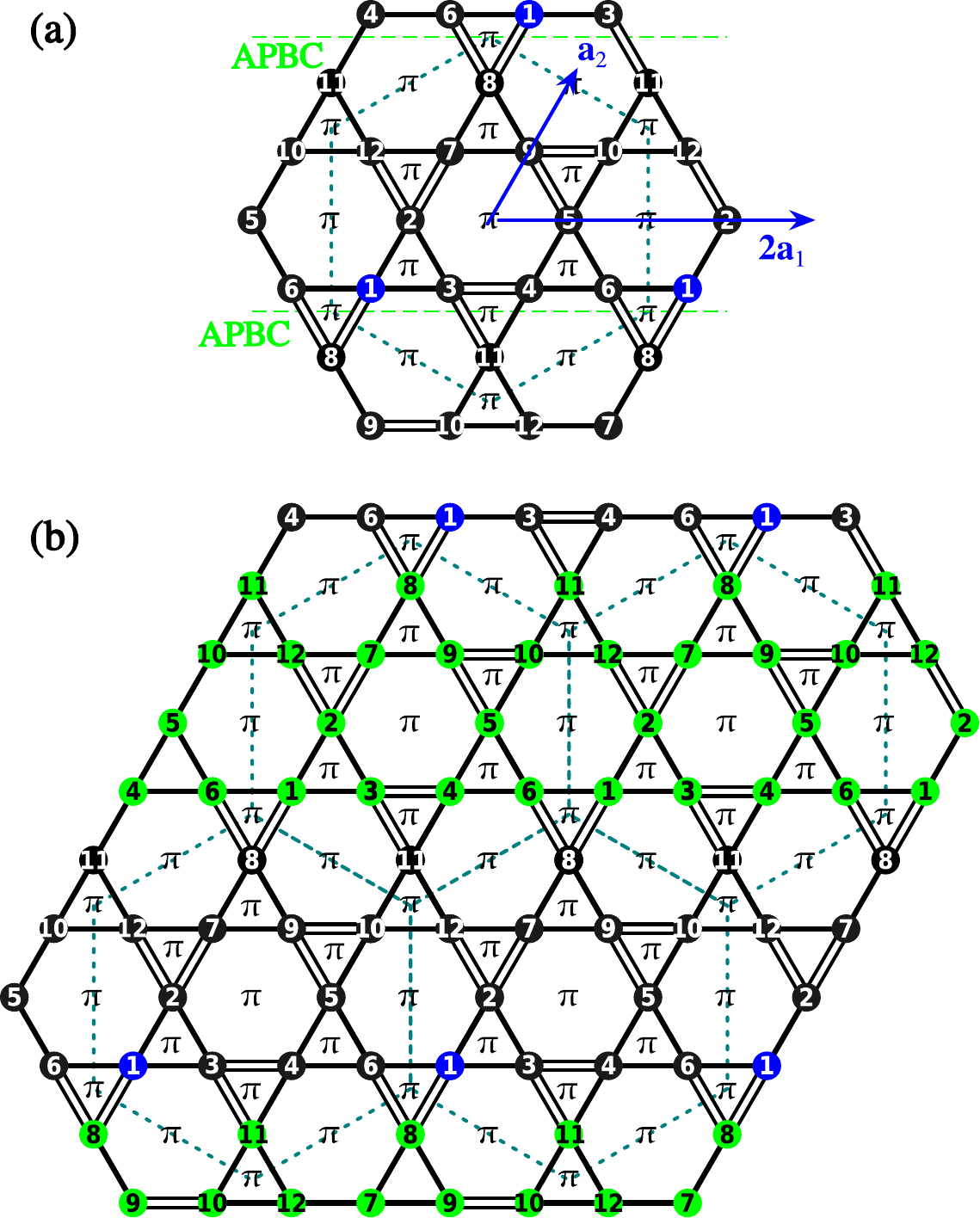}
\caption{(a) The DSL ansatz in Ref.~\onlinecite{SU2_Dirac_spin_liquid_Ran_PRL_2007} with doubled unit cell. Using antiperiodic boundary conditions, we can transform this ansatz to the one shown in Fig.~\ref{fig:DSL_with_APBC} with periodic boundaries. (b) The gauge transformation in a 48-site cluster changes this ansatz to the one in Fig.~\ref{fig:DSL_with_APBC}.  APBC is required for systems with an odd number of stripes.
\label{fig:gauge_transformations_connecting_the_old_and_new_ansatze}
}
\end{figure}

The DSL ansatz used in \cite{SU2_Dirac_spin_liquid_Ran_PRL_2007} with a doubled unit cell is shown in Fig.~\ref{fig:gauge_transformations_connecting_the_old_and_new_ansatze}(a). The gauge transformation $G:f^{{\dagger}}_{j,\sigma} \rightarrow e^{i \phi(j)}f^{{\dagger}}_{j,\sigma}$ connecting this ansatz with the one shown in Fig.~\ref{fig:DSL_with_APBC} is complicated by its dependence on the cluster size. As shown in Fig.~\ref{fig:gauge_transformations_connecting_the_old_and_new_ansatze}(b), the gauge transformation corresponds to sign changes of the fermionic operators along stripes containing whole 12-site unit cells. If the number of such stripes is even, as in Fig.~\ref{fig:gauge_transformations_connecting_the_old_and_new_ansatze}(b), then this gauge transformation is enough to connect the two ansätze. However, when the number of stripes is odd, one has to impose an extra antiperiodic boundary condition, as shown in Fig.~\ref{fig:gauge_transformations_connecting_the_old_and_new_ansatze}(a). The effect of the APBC is to change the sign of the hoppings it goes through.
All sites are located within one such stripe for the smallest 12-site unit cell in Fig.~\ref{fig:gauge_transformations_connecting_the_old_and_new_ansatze}(a). Therefore, no fermionic operator acquires a minus sign, and the APBC alone connects this ansatz to the one shown in Fig.~\ref{fig:DSL_with_APBC}.

\section{Isomorphy of the 12-site unite cell and the $O_h$ group}
\label{App:isomorphy_of_the_12_site_unit_cell_and_the_O_h_group}

\begin{figure}[t]
\centering
\includegraphics[width=0.4\columnwidth]{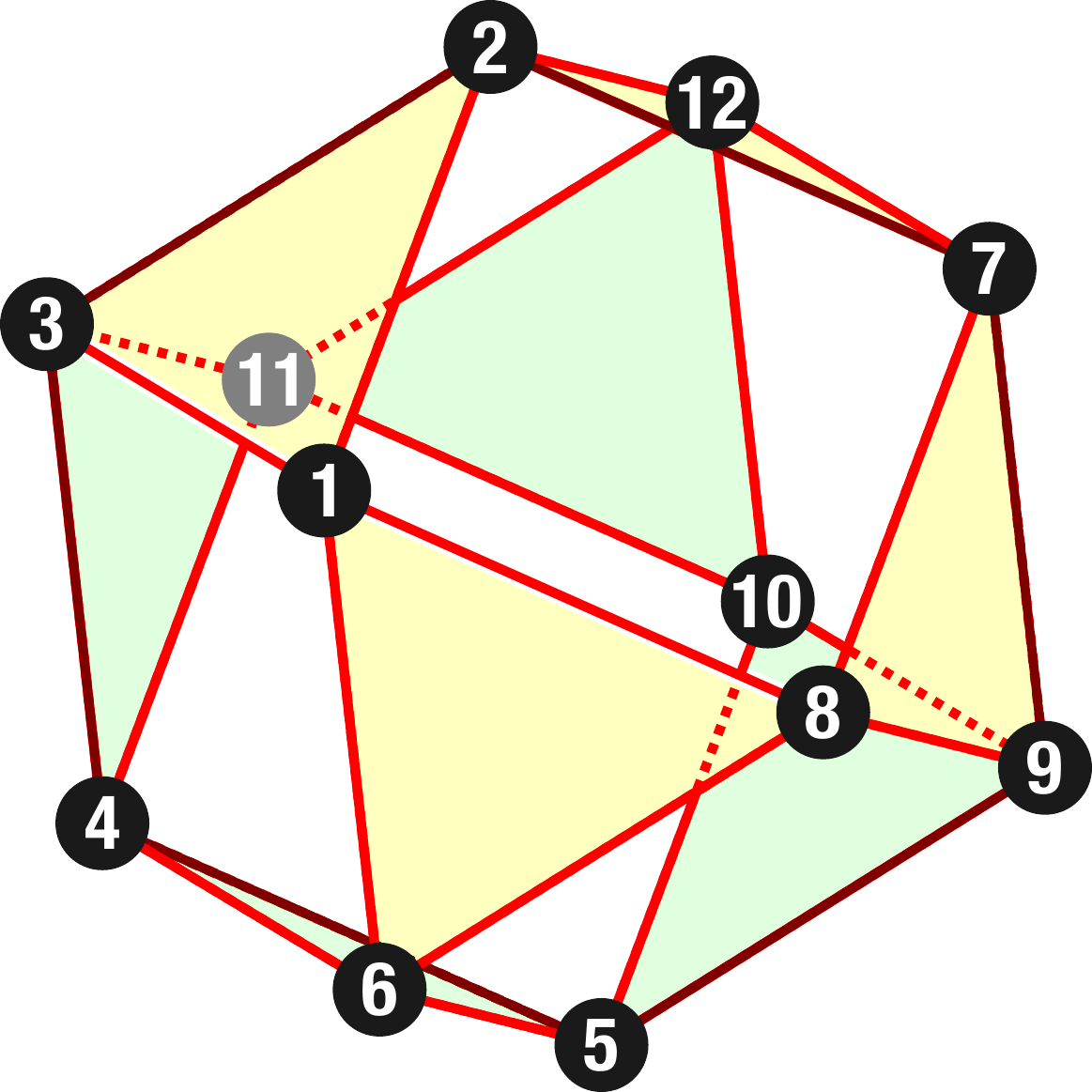}
\caption{The truncated octahedron showing the topology of the 12-site unit cell. The site numbering follows Fig.~\ref{fig:DSL_with_APBC}.
\label{fig:truncated_octahedron}
}
\end{figure}


The 12-site unit cell is equivalent to a truncated octahedron (see Fig.~\ref{fig:truncated_octahedron}), and its symmetry group is isomorphic to the $O_h$ point group. The $C_2$ in the wallpaper group becomes the inversion $i$ in $O_h$ with cyclic permutation $(1\>  10)(2\>5)(3\>9)(4\>7)(6\>12)(8\>11)$. The glide reflection $(1\>2\>7\>8)(3\>12\>9\>6)(4\>11\>10\>5)$ becomes the $C_4$, and the $(1\>2\>3)(5\>9\>10)(6\>7\>11)(4\>8\>12)$ is a threefold rotation $C_3$ in both cases. The $i$, $C_4$ and $C_3$ are the generators of the $O_h$.

\section{Relations between the permutations and the diagonal terms}
\label{sec:perms_diag}


\subsection{Expectation values in the Mott phase}
\label{sec:relations_between_permuations_and_the_diagonal_terms_in_the_Mott_phase}

The rotational invariance of an SU($N$) singlet state allows us to express expectation values of an off-diagonal operator using diagonal operators only. For example, the two-site permutation operator $\mathcal{P}_{i,j}$ reads
\begin{equation}
  \langle  \mathcal{P}_{i,j} \rangle = N(N + 1) \langle n_{i,\sigma}  n_{j,\sigma} \rangle-1 
  \label{eq:P_ij_related_to_n_i_n_j}
\end{equation}
and
\begin{equation}
  \langle \mathbf{T}_{i} \cdot \mathbf{T}_{j} \rangle = \frac{(N + 1)N}{2} \left( \langle n_{i,\sigma}  n_{j,\sigma} \rangle- \frac{1}{N^2} \right) ,
\label{eq:T_i_T_j_related_to_n_i_n_j}
\end{equation}
where $\langle n_{i,\sigma}  n_{j,\sigma} \rangle$ is the expectation value of the density-density operator of a single flavor $\sigma$. These expressions are instrumental as calculating diagonal matrix elements is much faster in the variational Monte Carlo method than off-diagonal ones.

We also evaluated the expectation values of the three-site terms and density correlations for a few examples of singlet wave functions for different values of $N$ and sites. By inspection, we arrived at the 
\begin{equation}
  \langle  \mathcal{P}_{i,j,k}  + \mathcal{P}_{i,k,j}   \rangle = N^2 p_{3}  
- N p_{21} + p_{111} + 1
   \label{eq:P123_w_densities} 
\end{equation}
expression, where $p_{3}=\sum_\sigma \langle n_{i,\sigma}  n_{j,\sigma} n_{\sigma,k} \rangle$ is the probability that the three sites are occupied with the same flavors, $p_{111}=\sum_{\sigma\neq\sigma'\neq\sigma''} \langle n_{i,\sigma}  n_{\sigma',j} n_{\sigma'',k} \rangle$  if they are all different, and $ p_{21}$ if there are precisely two identical flavors, satisfying $p_{3} + p_{21} + p_{111} = 1$.

We can check Eq.~(\ref{eq:P123_w_densities}) in the following limiting cases:
(i) If all three sites are in a fully antisymmetrical $N$-site SU($N$) singlet, then $\langle  \mathcal{P}_{i,j,k}  + \mathcal{P}_{i,k,j}   \rangle = 2$. At the same time, there are no two sites with the same color within the singlet, so $p_{3}  = p_{2,1} =0$ and $p_{111} = 1$, fulfilling  Eq.~(\ref{eq:P123_w_densities}). (ii) If the three sites belong to three disjunct singlets, we get contribution only if they are occupied with the same colors, having probability $1/N^3$, but since any of the $N$ colors will do it, $\langle  \mathcal{P}_{i,j,k}  + \mathcal{P}_{i,k,j}   \rangle = 2N\times 1/N^3 = 2/N^2$. The density correlations are also easy to calculate, 
\begin{align}
 p_{3} &= N \frac{1}{N^3} = \frac{1}{N^2} \;, \\
 p_{21}  &= 3 N (N-1)\frac{1}{N^3}  = \frac{3}{N}- \frac{3}{N^2} \;,   \\
 p_{111}  & = N(N-1)(N-2) \frac{1}{N^3} = 1- \frac{3}{N} + \frac{2}{N^2} \;.
\end{align}
In this case, we also satisfy Eqs.~(\ref{eq:P123_w_densities}).

\subsubsection{Monte Carlo sampling}
\label{sec:VMC}

We use a Monte Carlo method to evaluate the expectation values of different operators in a Gutzwiller projected Fermi sea \cite{1987JPSJ...56.1490Y}. 
First, we express the $\langle \mathcal{P}_{i,j} \rangle$ and $ \langle  \mathcal{P}_{i,j,k}  + \mathcal{P}_{i,k,j} \rangle$ with the diagonal operators $n_{i,\sigma}$, $p_{3}$, $p_{21}$ and $ p_{111}$ using Eqs.~(\ref{eq:P_ij_related_to_n_i_n_j}) and (\ref{eq:P123_w_densities}). Then, the expectation values of diagonal operators can be easily calculated with Monte Carlo sampling of fermionic configurations. For example,
\begin{multline}
\langle \text{FS}|P_{\text{G}} n_i n_j P_{\text{G}}|\text{FS} \rangle =\\
=\sum_x \langle x| n_i | x \rangle \langle x | n_j| x \rangle | \langle x | P_{\text{G}}|\text{FS} \rangle|^2,
\end{multline}
where we inserted the identity operator $\sum_{x} |x \rangle \langle x|$ and the configurations $|x\rangle$ runs over all the possible product states of the fermions in real space. We can use $P(x) \propto | \langle x |P_{\text{G}}|\text{FS} \rangle|^2$  as a probability distribution (since the Monte Carlo steps follow each other in a Markov chain, there is no need to impose $\sum_x P(x) = 1$).  The effect of the Gutzwiller projector is to select configurations $|x\rangle$ with singly occupied sites exclusively. Furthermore, as the $P_{\text{G}}|\text{FS} \rangle$ is an SU($N$) singlet, the configurations giving non-zero contribution have an equal number of flavors. For example, in a 6-site system, there are 6! = 720 such configurations, $| x \rangle \in \{ | A,B,C,D,E,F \rangle, | B,A,C,D,E,F \rangle, \dots \}$. For a system with $N_s = 6 M_s$ sites, the number of such configurations is $N_s!/(M_s!)^6$. In the evaluation of the $| \langle x | P_{\text{G}}|\text{FS} \rangle|^2$ we use the standard determinant updates using the Sherman-Morrison formula, see Ref.~\onlinecite{Becca_Sorella_2017}.

Every Monte Carlo step involves many elementary steps corresponding to the correlation time of the given system size. Following the thermalization, we made $1,5 \times 10^{8}$ Monte Carlo steps for the $S_{\text{VMC}}(\mathbf{k})$ of the 768-site cluster in Fig.~\ref{fig:structure_factors}(a), $2.5 \times 10^{7}$ for the 192-site David star ansatz in the stereographic map in Fig.~\ref{fig:Global_stability_of_David_star} (with a total of 5703 points), and $4 \times 10^{9}$ for the real-valued perturbations of the DSL in Figs.~\ref{fig:Local_stability_of_1D_real_ansatze} and \ref{fig:Local_stability_of_2D_real_ansatze}, with 192-site clusters.

The Monte Carlo sampling of the Hamiltonian and overlap matrices for the calculation of the dynamical structure factor $S^{33}_{\text{VMC}}(\mathbf{k},\omega)$ of a 48-site cluster was explained in detail in Ref.~\cite{Mi_SU3_cikkunk}, and involved $1.5 \times 10^{9}$ Monte Carlo steps, with which the error became negligible.

In the case of the chiral and staggered ansätze in Fig.~\ref{fig:VMC_chiral_staggered} and in Tab.~\ref{tab:chirals}, we performed $10^{6}$ Monte Carlo steps separated with $4N_s$ elementary steps. For the DSL in the insets of Fig.~\ref{fig:VMC_chiral_staggered}, we increased the number of steps to $8 \times 10^{6}$.

\subsection{Expectation values in the mean-field approximation}
\label{sec:relations_between_permuations_and_the_diagonal_terms_in_the_Fermi_sea}

For free fermions, the relationship between the exchange of fermions 
\begin{equation}
 \mathcal{P}_{i,j}= \sum_{\sigma,\sigma'} 
f^{\dagger}_{i, \sigma'}  
f^{\dagger}_{j, \sigma}  
f^{\phantom{\dagger}}_{j, \sigma'}  
f^{\phantom{\dagger}}_{i, \sigma}  
\end{equation}
and $\mathbf{T}_{i} \cdot \mathbf{T}_{j}$, where Eq.~(\ref{eq:spin_operator_with_partons_SU_4}) defines $\mathbf{T}_{i}$,  modifies to 
\begin{equation}
  \mathcal{P}_{i,j} = 2 \mathbf{T}_{i} \cdot \mathbf{T}_{j} + \frac{1}{N} \hat n_i \hat n_j \,.
\end{equation}
This follows from Eq.~(\ref{eq:spin_operator_with_partons_SU_4}) when we take into account that  
\begin{equation}
 \sum_{a=1}^{N^2-1} \lambda^a_{\alpha\beta} \lambda^a_{\mu\nu} = 2 \delta_{\alpha,\nu} \delta_{\mu,\beta} - \frac{2}{N}  \delta_{\alpha\beta} \delta_{\mu\nu}
 \end{equation}
for the SU($N$) generalizations of the Gell-Mann matrices \cite{SUrels_SciPostPhysLectNotes.21}. In the fundamental representation $\hat n_i=1$, we recover Eq.~(\ref{eq:TTP}).

We shall be careful when we compare the noninteracting expectation values to the projected ones, as they include charge fluctuations. We can use Wick's theorem to calculate the expectation values:
\begin{align}
  \langle \mathcal{P}_{i,j} \rangle' 
  & = \sum_{\sigma,\sigma'} \langle f^{\dagger}_{i,\sigma} f^{\dagger}_{j,\sigma'} f^{\phantom{\dagger}}_{j,\sigma} f^{\phantom{\dagger}}_{i,\sigma'} \rangle' \nonumber\\
  & = \sum_{\sigma,\sigma'} \left(
   \langle f^{\dagger}_{i,\sigma} f^{\phantom{\dagger}}_{i,\sigma'} \rangle' 
   \langle f^{\dagger}_{j,\sigma'} f^{\phantom{\dagger}}_{j,\sigma} \rangle' 
  -
   \langle f^{\dagger}_{i,\sigma} f^{\phantom{\dagger}}_{j,\sigma} \rangle'   
   \langle f^{\dagger}_{j,\sigma'} f^{\phantom{\dagger}}_{i,\sigma'} \rangle'
       \right)
  \nonumber\\
  & = N 
   \langle f^{\dagger}_{i} f^{\phantom{\dagger}}_{i} \rangle' 
   \langle f^{\dagger}_{j} f^{\phantom{\dagger}}_{j} \rangle' 
  -
   N^2 \left|\langle f^{\dagger}_{i} f^{\phantom{\dagger}}_{j} \rangle' \right|^2 
  \nonumber\\
  & =  \frac{1}{N}- N^2  |\langle f^{\dagger}_{i} f^{\phantom{\dagger}}_{j} \rangle'|^2 \;,
  \label{eq:PijMFWick}
\end{align}
where we introduced the $ \langle \text{FS} |\hat A | \text{FS} \rangle =  \langle  \hat A \rangle'$ shorthand notation for the expectation values in the Fermi sea. Furthermore, $\langle f^{\dagger}_{i,\sigma} f^{\phantom{\dagger}}_{i,\sigma'} \rangle' = \delta_{\sigma,\sigma'}/N$ in the fundamental representation and $\langle f^{\dagger}_{i} f^{\phantom{\dagger}}_{j} \rangle'$ denotes $\langle f^{\dagger}_{i,\sigma} f^{\phantom{\dagger}}_{j,\sigma} \rangle'$ that is independent of $\sigma$ in a singlet (closed shell) Fermi sea. Similarly,
\begin{align}
 \langle\mathbf{T}_{i} \cdot \mathbf{T}_{j} \rangle' 
  & =  \frac{1}{2} \langle \mathcal{P}_{i,j}\rangle' - \frac{1}2{N} \langle\hat n_i \hat n_j \rangle'
  \nonumber\\
    & =  - \frac{N^2-1}{2}  |\langle f^{\dagger}_{i} f^{\phantom{\dagger}}_{j} \rangle'|^2   \,.
\label{eq:T_i_T_j_related_to_n_i_n_j_mean_field}
\end{align}
Let us also calculate the expectation value of having two fermions of the same flavor on neighboring sites:
\begin{equation}
  \langle \hat n_{i,\sigma} \hat n_{j,\sigma} \rangle' = \frac{1}{N^2} - |\langle f^{\dagger}_{i} f^{\phantom{\dagger}}_{j} \rangle'|^2 \;.
\end{equation}
We can associate $|\langle f^{\dagger}_{i} f^{\phantom{\dagger}}_{j} \rangle'|^2$ with the exchange hole.  
Since $\hat n_{i,\sigma} \hat n_{j,\sigma} = (f^{\phantom{\dagger}}_{i,\sigma} f^{\phantom{\dagger}}_{j,\sigma})^\dagger f^{\phantom{\dagger}}_{i,\sigma} f^{\phantom{\dagger}}_{j,\sigma} $ is a positive semidefinite operator, we find that the $\langle f^{\dagger}_{i,\sigma} f^{\phantom{\dagger}}_{j,\sigma} \rangle'$ is bounded as
\begin{equation}
   |\langle f^{\dagger}_{i,\sigma} f^{\phantom{\dagger}}_{j,\sigma} \rangle'| \leq \frac{1}{N} \;.
\end{equation}
On one hand, the flavor-flavor correlations are absent in the absence of the exchange hole, so that 
$\langle\mathbf{T}_{i} \cdot \mathbf{T}_{j} \rangle' =0$ for $|\langle f^{\dagger}_{i,\sigma} f^{\phantom{\dagger}}_{j,\sigma} \rangle'|=0$.
On the other hand, for maximally entangled fermions with $|\langle f^{\dagger}_{i,\sigma} f^{\phantom{\dagger}}_{j,\sigma} \rangle'| = \frac{1}{N}$ we get $\langle\mathbf{T}_{i} \cdot \mathbf{T}_{j} \rangle' 
=- \frac{N^2-1}{2N^2} $, instead of the $\langle\mathbf{T}_{i} \cdot \mathbf{T}_{j} \rangle 
= -\frac{N+1}{2N}$, the minimal value which can be achieved within an SU($N$) singlet. So we can approximate
\begin{equation}
\langle\mathbf{T}_{i} \cdot \mathbf{T}_{j} \rangle
\approx \frac{N}{(N-1)}
\langle\mathbf{T}_{i} \cdot \mathbf{T}_{j} \rangle' \;,
\label{eq:TdTMFcorr}
\end{equation}
which we will refer to as the corrected mean field. This result is analogous to the relation of the Casimir operators $\langle\mathbf{T}_{i} \cdot \mathbf{T}_{i} \rangle
= \frac{N}{(N-1)}
\langle\mathbf{T}_{i} \cdot \mathbf{T}_{i} \rangle'$ derived in Ref.~\onlinecite{Mi_SU4_cikkunk}.

For the 3-site ring exchange on a triangle, we get
\begin{multline}
\langle 
\mathcal{P}_{123}
\rangle'
=
\langle 
f^{\dagger}_{1,\alpha}
f^{\dagger}_{2,\gamma}
f^{\dagger}_{3,\beta}
f^{\phantom{\dagger}}_{3,\alpha}
f^{\phantom{\dagger}}_{2,\beta}
f^{\phantom{\dagger}}_{1,\gamma}
\rangle' = 
\\
N^3 |\xi|^3 e^{i \phi} +N (|\xi|^3 e^{-i \phi} - 3|\xi|^2) + \frac{1}{N^2}
\end{multline}
where
$\xi_{1,3} \xi_{3,2} \xi_{2,1}  = |\xi|^3 e^{i \phi_\triangle}$. The U(1) gauge invariant ring exchange is thus sensitive to the phase of the $\xi$'s via their product around the triangle.
For  $\xi=0$, we get the correct
$\langle \mathcal{P}_{123} \rangle' = 1/N^2$, while for $\xi=1/N$ it takes the 
$\langle \mathcal{P}_{123} \rangle' = (N-1)(N-2)/N^2$ value instead of 1. 
%

\section{Projected mean field}
\label{sec:proj_mean_field}

 Here, we will closely follow the projected mean-field calculation of Hsu in Ref.~\onlinecite{Hsu_1990}, where the SU(2) mean-field calculation was extended to enforce single occupancy on nearest neighbor sites only. Below, we summarize the method and extend it to off-diagonal quantities, the SU($N$) case, and three sites when calculating ring exchanges.

  \subsection{Projected mean field for 2-sites}

 Let us denote by $P(XY)$ the probability that site $1$ has a flavor $X$ and site $2$ has a flavor $Y$ fermion after Gutzwiller projection. For the SU($6$), the flavors are $X,Y \in \{ A,B,\dots,F \}$. We can express the probabilities as
 \begin{subequations}
\begin{align}
  P(AA) &= P_0(AA) P_0(\bar B \bar B) \times \cdots \times P_0(\bar F \bar F) \label{eq:PAA}\\ 
  P(AB) &= P_0(A \bar A) P_0(\bar B B)  P_0(\bar C \bar C)  
  \phantom{=} \times \cdots \times P_0(\bar F \bar F) \label{eq:PAB}
 \end{align}
  \end{subequations}
where $P_0(XX)$ denotes the probability of finding two fermions of flavor $X$ at sites $1$ and $2$ in the Fermi sea before the projection, 
$P_0(X\bar X)$ denotes the probability of finding a fermion with flavor $X$ at site $1$ and not finding it at site $2$, and so on (i.e., we denote by $\bar X$ the flavor-$X$ hole, i.e. the absence of that flavor). 
The product form is the consequence of the fermions of different colors being independent in the Fermi-sea.
We can use Wick's theorem to calculate the probabilities in the Fermi-sea:
 \begin{subequations}
  \label{eq:expX}
 \begin{align} 
  P_0(XX) &=  \langle n_{1,X} n_{2,X} \rangle'  
  = \langle f^\dagger_{1,X} f^{\phantom{\dagger}}_{1,X} f^\dagger_{2,X} f^{\phantom{\dagger}}_{2,X} \rangle'
  \nonumber\\ 
   & = \langle f^\dagger_{1,X} f^{\phantom{\dagger}}_{1,X}\rangle'  
      \langle f^\dagger_{2,X} f^{\phantom{\dagger}}_{2,X} \rangle' 
    - \langle f^\dagger_{1,X} f^{\phantom{\dagger}}_{2,X} \rangle' 
       \langle f^\dagger_{2,X} f^{\phantom{\dagger}}_{1,X} \rangle'
   \nonumber\\ 
   &= \begin{vmatrix}
  \nu^X_1 & \xi^X_{1,2}  \\
 \xi^X_{2,1} & \nu^X_2  \\
\end{vmatrix}
   = \frac{1}{N^2}- \xi^X_{1,2}\xi^X_{2,1} \\
 P_0(X\bar X) &=  \langle n_{1,X} \bar n_{2,X} \rangle'  
   = \begin{vmatrix}
  \nu^X_1 & -\xi^X_{1,2}  \\
\xi^X_{2,1} & 1\!-\!\nu^X_2  \\
\end{vmatrix}
   = \frac{N\!-\!1}{N^2} +  \xi^X_{1,2}\xi^X_{2,1} 
   \\
    P_0(\bar X\bar X) &=  \langle \bar n_{1,X} \bar n_{2,X} \rangle'  
   = \begin{vmatrix}
1 \!-\! \nu^X_1 & -\xi^X_{1,2}  \\
-\xi^X_{2,1} & 1 \!-\! \nu^X_2  \\
\end{vmatrix}
\\
 &= \frac{(N\!-\!1)^2}{N^2} -  \xi^X_{1,2}\xi^X_{2,1} \nonumber \; ,
\end{align}
\end{subequations}
where
\begin{align}
  \bar n_{i,X} &= 1-n_{i,X} \,,
\end{align}
and we introduced the $\nu^X_i \equiv \langle n_{i,X} \rangle'$, $\bar \nu^X_i \equiv \langle \bar n_{i,X} \rangle'$ (which take on $\nu^X_i = 1/N$ and $\bar \nu^X_i = 1-1/N$ in a density-uniform state) and the
\begin{equation}
  \xi^X_{i,j} = \langle f^\dagger_{i,X} f^{\phantom{\dagger}}_{j,X}  \rangle'
\end{equation}
off-diagonal expectation values, for which $\xi^X_{i,j}=\bar\xi^X_{j,i}$ (the bar denotes complex conjugation).
These probabilities satisfy the 
\begin{equation}
 P_0(XX) + P_0(\bar X X) + P_0(X\bar X) + P_0(\bar X\bar X) =1
\end{equation}
condition.
We can then approximate the flavor-flavor correlation function by 
\begin{align}
 \langle T_1^3 T_2^3 \rangle 
 &= \frac{1}{4} \frac{P(A A) + P(B B)- P(A B)- P(B A) }{\sum_{X=A}^F \sum_{Y=A}^F P(X Y)}
 \nonumber \\
 &= \frac{1}{4} \frac{2 P(A A)- 2 P(A B)}{N P(A A) + N(N-1) P(A B)}
 \nonumber \\
 &= \frac{1}{2} \frac{P_0(AA) P_0(\bar B \bar B)- P_0(A \bar A) P_0(\bar B B)}{N P_0(AA) P_0(\bar B \bar B) + N(N-1) P_0(A \bar A) P_0(\bar B B)}
 \nonumber \\
 &=
-\frac{N^2 |\xi|^2}{2 \left[N^4 |\xi|^4 + N^3 |\xi|^2-2 N^2 |\xi|^2 + (N-1)^2\right]}.
\end{align}
Assuming the full SU($N$) symmetry of the ground state,
\begin{equation}
 \langle \mathbf{T}_1 \cdot \mathbf{T}_2  \rangle =
 -\frac{(N^2-1)N^2 |\xi|^2}{2 \left[N^4 |\xi|^4 + N^3 |\xi|^2-2 N^2 |\xi|^2 + (N-1)^2\right]}.
 \label{eq:TdT_projectedMF_2sites}
\end{equation}
For $|\xi|=1/N$ we recover the expected $\langle \mathbf{T}_1 \cdot \mathbf{T}_2  \rangle = -\frac{N+1}{2N}$ in a singlet. 

Applying the expression above to the SU(2) case, we recover the formula in Ref.~\onlinecite{Hsu_1990} by associating the Fermi exchange hole $|\xi|^2$ with $x$: 
\begin{equation}
 \langle \mathbf{S}_1 \cdot \mathbf{S}_2 \rangle =- \frac{6|\xi|^2}{1 + 16|\xi|^4} \,.
\end{equation}
The spin-spin correlation function is a monotonically decreasing function of the Fermi exchange hole. It is 0 for the decoupled sites (where $|\xi|^2 = 0$) and takes $\langle \mathbf{S}_1 \cdot \mathbf{S}_2 \rangle = -3/4$ for maximally entangled pair of sites, where $|\xi|^2=1/2$ is maximal, representing a dimer singlet.

We can generalize the approximation of Hsu to the calculation of expectation values of off-diagonal operators as well, the example being the $\mathcal{P}_{1,2}$ transposition (permutation) operator. We calculated the expectation value of $\mathcal{P}_{1,2}$ in Eq.~(\ref{eq:PijMFWick}) in the mean-field approximation. Now we need to distinguish two possible cases: the two sites are either occupied by the same flavor fermions, in which case the $\mathcal{P}_{1,2}$ acts trivially as the identity operator, or with different flavors, in which case we cannot use the expressions involving the probabilities $P_0$. To overcome this problem, let us introduce the notation $\langle YX | \mathcal{P}_{1,2} | X Y \rangle$ to denote the amplitude of a process where $\mathcal{P}_{1,2}$ exchanges the flavors on sites $1$ and $2$. For identical flavors ($X=Y$) the 
$\langle XX | \mathcal{P}_{1,2} | X X \rangle \equiv   P(X X)$ and following Eq.~(\ref{eq:PAA}), we can write 
\begin{equation}
  \langle XX | \mathcal{P}_{1,2} | X X \rangle = P_0(X X) \prod_{Z\neq X} P_0(\bar Z \bar Z).
\end{equation}
For unlike flavors $X\neq Y$
\begin{align}
  \langle YX | \mathcal{P}_{1,2} | X Y \rangle &=   \langle f^{\dagger}_{1,Y}f^{\dagger}_{2,X}f^{\phantom{\dagger}}_{2,Y}f^{\phantom{\dagger}}_{1,X}\rangle'  \prod_{Z\neq X,Y} P_0(\bar Z \bar Z)
  \nonumber\\
   &=-\langle f^{\dagger}_{2,X}f^{\phantom{\dagger}}_{1,X}\rangle'
  \langle f^{\dagger}_{1,Y}f^{\phantom{\dagger}}_{2,Y}\rangle'
  \prod_{Z\neq X,Y} P_0(\bar Z \bar Z)   \nonumber\\
   &= - |\xi|^2
  \prod_{Z\neq X,Y} P_0(\bar Z \bar Z).
\end{align}
and eventually, we arrive at the
\begin{align} 
\langle \mathcal{P}_{12} \rangle 
&= \frac{\sum_X \langle X X | \mathcal{P}_{12} | X X  \rangle 
  + \sum_{X\neq Y} \langle  Y X |\mathcal{P}_{12}| X Y \rangle}{\sum_{X=A}^F \sum_{Y=A}^F P(X Y)} \nonumber\\
&= \frac{N P(AA) + N(N-1) \langle  B A |\mathcal{P}_{12}| A B \rangle}{N P(AA) + N(N-1) P(AB)} \nonumber\\
&= \frac{N P_0(AA) P_0(\bar B\bar B)- N(N-1) |\xi|^2 }{N  P_0(AA)  P_0(\bar B\bar B) + N(N-1)   P_0(A\bar A)  P_0(\bar B B)} 
\nonumber\\
&=\frac{(N-1)^2 - N^2 (2 - 2 N + N^3) |\xi|^2 + N^4 |\xi|^4}{(N-1)^2 N + (N-2) N^3 |\xi|^2 + N^5 |\xi|^4}
\label{eq:P1_projectedMF_2sites}
\end{align}
This is the same result as Eq.~(\ref{eq:TdT_projectedMF_2sites}) after we take Eq.~(\ref{eq:TTP}) into account.

 \subsection{Projected mean field for 3-sites}

Here, we extend the method to a triangle to consistently treat the three-site ring exchange. Using the same notation as above, the conditional expectation values are
\begin{subequations}
\begin{align}
  \langle \mathcal{P}_{12} \rangle 
  &= 
  \frac{\sum_{X,Y,Z} \langle Y X Z | \mathcal{P}_{12} | X Y Z  \rangle}{\sum_{X,Y,Z} P( X Y Z)} 
 \\
  \langle \mathcal{P}_{123} \rangle &= \frac{\sum_{X,Y,Z} \langle Z X Y  | \mathcal{P}_{123} | X Y Z  \rangle}{\sum_{X,Y,Z} P( X Y Z)}  \;.
\end{align}
\end{subequations}
Assuming a homogenous state, $|\xi_{1,2}|= |\xi_{2,3}| =|\xi_{3,1}|=|\xi|$, we can use the SU($N$) and geometrical symmetries of the system to write the denominator as
\begin{align}
\sum_{X,Y,Z} P(X Y Z) 
  & = N P(AAA) + 3N(N-1) P(AAB) \nonumber\\
  &\phantom{=} + (N-2) (N-1) N P(ABC).
\end{align}
Regarding the numerator, the expectation values of the exchange operators are diagonal for certain configurations, 
\begin{subequations}
\begin{align}
 \langle AAA | \mathcal{P}_{12} | AAA  \rangle &= P(AAA), \\
 \langle AAB | \mathcal{P}_{12} | AAB  \rangle & = P(AAB), \\
 \langle AAA | \mathcal{P}_{123} | AAA  \rangle &=  P(AAA),
\end{align}
and for the off-diagonal, the 
\begin{align}
  \langle BAA | \mathcal{P}_{123} | AAB  \rangle 
&=  
  \langle BAA | \mathcal{P}_{13} | AAB  \rangle,
\\
  \langle ABA | \mathcal{P}_{12} | BAA  \rangle
&=  
  \overline{\langle BAA | \mathcal{P}_{12} | ABA  \rangle} ]  
\end{align}
\end{subequations}
holds. Again, the symmetries allow us to write
\begin{subequations}
\begin{align}
\sum_{X,Y,Z} & \langle YXZ | \mathcal{P}_{12} | XYZ \rangle 
 = N P(AAA ) 
\nonumber\\
& + N(N\!-\!1) P(AAB)
\nonumber\\
&+ N(N\!-\!1) \left( \langle BAA | \mathcal{P}_{12} | ABA  \rangle
 + \langle ABA | \mathcal{P}_{12} | BAA  \rangle \right)
\nonumber\\
 & + (N\!-\!2) (N\!-\!1) \langle BAC | \mathcal{P}_{12} | ABC  \rangle\,,
  \\
\sum_{X,Y,Z} & \langle Z X Y  | \mathcal{P}_{123} | X Y Z  \rangle
   = N  P(AAA ) 
\nonumber\\
    & + 3 N(N\!-\!1) \langle ABA | \mathcal{P}_{12} | BAA  \rangle
\nonumber\\
    &     + (N\!-\!2) (N\!-\!1) N \langle CAB | \mathcal{P}_{123} | ABC  \rangle
  \,.
\end{align}
\end{subequations}
where the probabilities are
\begin{subequations}
\begin{equation}
P(AAA)  = P_0(AAA) 
\prod_{X \neq A} P_0(\bar X \bar X  \bar X)
\end{equation}
\begin{equation}
P(AAB) = 
P_0(AA\bar A) P_0(\bar B \bar B B) 
\prod_{X \neq A,B} P_0(\bar X \bar X  \bar X)
\end{equation}
\begin{multline}
  P(ABC) = 
 P_0(A\bar A\bar A) P_0(\bar B B \bar B) P_0(\bar C \bar C C)
 \\
 \times \prod_{X \neq A,B,C} P_0(\bar X \bar X  \bar X)
\end{multline}
%
\begin{multline}
\langle BAA | \mathcal{P}_{12} | ABA  \rangle =  
  -\langle  f^{\dagger}_{2,A}f^{\dagger}_{3,A}f^{\phantom{\dagger}}_{3,A} f^{\phantom{\dagger}}_{1,A}
 \rangle'
 \langle f^{\dagger}_{1,B} f^{\phantom{\dagger}}_{2,B}
 \rangle' 
  \\
 \times\prod_{X \neq A,B} P_0(\bar X \bar X  \bar X)\,,
\end{multline}
\begin{multline}
\langle BAC | \mathcal{P}_{12} | ABC  \rangle 
=  
 - \langle f^{\dagger}_{2,A} f^{\phantom{\dagger}}_{1,A} \rangle'
 \langle f^{\dagger}_{1,B} f^{\phantom{\dagger}}_{2,B} \rangle'
 P_0(\bar C \bar C  C)
 \\
 \times \prod_{X \neq A,B,C} P_0(\bar X \bar X  \bar X)\,,
\end{multline}
\begin{multline}
\langle CAB | \mathcal{P}_{123} | ABC  \rangle 
=  
 \langle f^{\dagger}_{2,A} f^{\phantom{\dagger}}_{1,A} \rangle'
 \langle f^{\dagger}_{3,B} f^{\phantom{\dagger}}_{2,B} \rangle'
 \\
\times \langle f^{\dagger}_{1,C} f^{\phantom{\dagger}}_{3,C} \rangle'
 \prod_{X \neq A,B,C} P_0(\bar X \bar X  \bar X)\,.
\end{multline}
\end{subequations}

Here again, we can use Wick's theorem to calculate the probabilities in the Fermi-sea of free fermions as
 \begin{subequations}
  \label{eq:expX_2}
 \begin{align} 
  P_0(XXX) &=  \langle n_{1,X} n_{2,X} n_{3,X} \rangle'  
   = \begin{vmatrix}
  \nu^X_1 & \xi^X_{1,2} & \xi^X_{1,3}  \\
  \xi^X_{2,1} & \nu^X_2 & \xi^X_{2,3}  \\
  \xi^X_{3,1} & \xi^X_{3,2} & \nu^X_3   \\
\end{vmatrix} \,,
 \\
  P_0(XX\bar X) &=  \langle n_{1,X} n_{2,X} \bar n_{3,X} \rangle'  
   = \begin{vmatrix}
  \nu^X_1 & \xi^X_{1,2} & -\xi^X_{1,3}  \\
  \xi^X_{2,1} & \nu^X_2 & -\xi^X_{2,3}  \\
  \xi^X_{3,1} & \xi^X_{3,2} & 1-\nu^X_3 \\
\end{vmatrix} 
\\
  P_0(X\bar X\bar X) &=  \langle n_{1,X} \bar n_{2,X} \bar n_{3,X} \rangle'  
     = \begin{vmatrix}
  \nu^X_1 & -\xi^X_{1,2} & -\xi^X_{1,3}  \\
  \xi^X_{2,1} & 1- \nu^X_2 & - \xi^X_{2,3}  \\
  \xi^X_{3,1} & - \xi^X_{3,2} & 1- \nu^X_3 \\
\end{vmatrix}  \,,
\\
  P_0(\bar X \bar X\bar X) &=  \langle \bar n_{1,X} \bar n_{2,X} \bar n_{3,X} \rangle'  
   = \begin{vmatrix}
  1- \nu^X_1 & - \xi^X_{1,2} & - \xi^X_{1,3}  \\
  - \xi^X_{2,1} & 1- \nu^X_2 & - \xi^X_{2,3}  \\
  - \xi^X_{3,1} & - \xi^X_{3,2} & 1- \nu^X_3 \\
\end{vmatrix} \,,
\end{align}
and the off-diagonal expectation values
\begin{align}
%
 \langle f^{\dagger}_{2,X} f^{\dagger}_{3,X} f^{\phantom{\dagger}}_{3,X}f^{\phantom{\dagger}}_{1,X} \rangle'
   &= \begin{vmatrix}
 \xi^X_{2,1} & \xi^X_{2,3} \\
 \xi^X_{3,1} & \nu^X_{3} \\
\end{vmatrix}
\end{align}
\end{subequations}
Collecting all the terms together and using computer algebra, we arrive at the following expressions:
\begin{align}
  \langle \mathcal{P}_{12} \rangle &= \frac{1}{N}- (N^2 - 1)  N^2  \frac{\Theta_{12}}{\Sigma} \,,
  \label{eq:P12_projectedMF_3sites}\\
  \langle \mathcal{P}_{123}+\mathcal{P}_{132} \rangle &= \frac{2}{N^2}- (N^2 - 1)  N  \frac{\Theta_{123+132}}{\Sigma} \label{eq:P123_projectedMF_3sites} \,,
\end{align}
where
\begin{widetext}
\begin{multline}
\Sigma = 
1
-6 N
-3 \left[3 \xi^2-5\right] N^2
+ \left[6 \xi^3 \cos \phi_\triangle +39 \xi^2-20\right] N^3
+ \left[27 \xi^4-20 \xi^3 \cos \phi_\triangle -66 \xi^2+15\right] N^4
\\
-6 \left[6 \xi^5 \cos \phi_\triangle +13 \xi^4-4 \xi^3 \cos \phi_\triangle -9 \xi^2+1\right] N^5
\\
+ \left[6 \xi^6 \cos 2 \phi_\triangle -21 \xi^6+66 \xi^5 \cos \phi_\triangle +81 \xi^4-12 \xi^3 \cos \phi_\triangle -21 \xi^2+1\right] N^6 
\\
+\xi^2 \left[54 \xi^5 \cos \phi_\triangle -4 \xi^4 \cos 2 \phi_\triangle +57 \xi^4-36 \xi^3 \cos \phi_\triangle -36 \xi^2+2 \xi  \cos \phi_\triangle +3\right] N^7
\\
-6 \xi^4 \left[3 \xi^4 \cos 2 \phi_\triangle +3 \xi^4+13 \xi^3 \cos \phi_\triangle +7 \xi^2-\xi  \cos \phi_\triangle -1\right] N^8
\\
2 \xi^6 \left[3 \xi^3 \cos \phi_\triangle +\xi^3 \cos  3 \phi_\triangle +6 \xi^2 \cos 2 \phi_\triangle  +6 \xi^2+12 \xi  \cos \phi_\triangle +4\right] N^9 \,,
\end{multline}
%
\begin{multline}
\Theta_{12} = 
\xi^2
 -4 \xi^2 N
-2 \xi^2 \left[4 \xi^2-3\right] N^2
 + 2 \xi^2 \left[7 \xi^3 \cos \phi_\triangle  + 9 \xi^2-2\right] N^3
\\
-\xi^2 \left[4 \xi^4 \cos 2 \phi_\triangle  + 5 \xi^4 + 20 \xi^3 \cos \phi_\triangle  + 12 \xi^2-1\right] N^4
\\
 + 2 \xi^4 \left[\xi^3 \cos \phi_\triangle  + \xi^2 \cos 2 \phi_\triangle  + 2 \xi^2 + 3 \xi  \cos \phi_\triangle  + 1\right] N^5 \,,
\end{multline}
and
\begin{multline}
\Theta_{123+132} = 
6 \xi^2 
-8 \xi^2  \left[ \xi  \cos \phi_\triangle  + 3\right] N
-12 \xi^2 \left[  2 \xi^2-2 \xi  \cos \phi_\triangle -3\right] N^2 
\\
+ 2 \xi^2 \left[30 \xi^3 \cos \phi_\triangle  + 30 \xi^2-11 \xi  \cos \phi_\triangle -12\right] N^3 
\\
 -2 \xi^2 \left[8 \xi^4 \cos 2 \phi_\triangle  + 15 \xi^4 + 48 \xi^3 \cos \phi_\triangle  + 27 \xi^2-\xi  \cos \phi_\triangle -3\right] N^4 
\\
 + 6 \xi^3 \left[2 \xi^4 \cos \phi_\triangle  + 2 \xi^3 \cos 2 \phi_\triangle  + 4 \xi^3 + 7 \xi^2 \cos \phi_\triangle  + 4 \xi  + \cos \phi_\triangle \right] N^5
\\
-2 \xi^3 \left[\xi^3 \cos 2 \phi_\triangle  + 3 \xi^2 \cos \phi_\triangle  + 3 \xi  + \cos \phi_\triangle \right] N^6 \,.
\end{multline}
\end{widetext}

\begin{figure}[t]
	\centering
	\includegraphics[width=0.8\columnwidth]{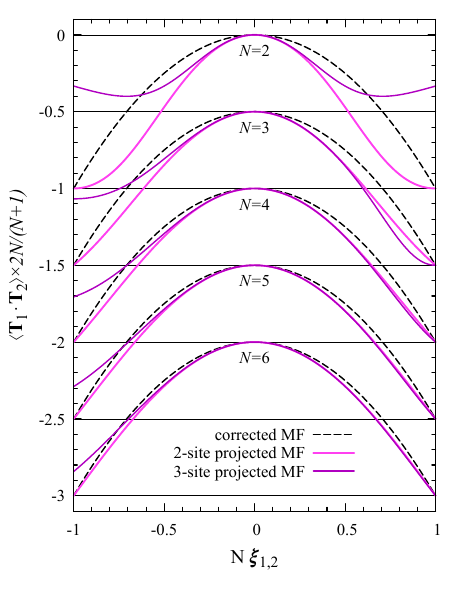}
	\caption{\label{fig:Hsu_projected_MF}
	The nearest neighbor flavor correlation function $\langle \mathbf{T}_1 \cdot \mathbf{T}_2 \rangle$, normalized  by $(N+1)/2N$, as a function of the $\xi_{1,2} = \langle f^{\dagger}_{2,X} f^{\phantom{\dagger}}_{1,X} \rangle'$ calculated using the corrected mean field, Eq.~(\ref{eq:TdTMFcorr}), the two-site projected mean field, Eq.~(\ref{eq:TdT_projectedMF_2sites}), and three-site projected mean field, combining Eqs.~(\ref{eq:TTP}) and (\ref{eq:P12_projectedMF_3sites}).
The expressions for different values of $N$ are shifted vertically for better visibility.	}
\end{figure}

In possession of these expressions, in Fig.~\ref{fig:Hsu_projected_MF} we compare the expectation values of the spin-spin correlation $\langle \mathbf{T}_1 \cdot \mathbf{T}_2 \rangle$ in the corrected mean field, Eq.~(\ref{eq:TdTMFcorr}), and two- and three-site projected mean-field approach, where we divided the $\langle \mathbf{T}_1 \cdot \mathbf{T}_2 \rangle$ by $(N+1)/2N$  to facilitate more straightforward comparison for different SU($N$) cases. The corrected mean-field and the two-site projected mean-field expressions are even functions of the $\xi$. The information about the sign of the $\xi$ first appears for the three-site projected mean field on the triangle.
The $\langle \mathbf{T}_1 \cdot \mathbf{T}_2 \rangle 2N/(N+1)$ is 0 for $\xi=0$ and $-1$ for $\xi=1/N$  in all cases for $N\geq 3$. The 3-site projected mean-field expression is pathological for $N=2$, as it is non-analytical at $\xi=1$. The curves for different expressions collapse as $N\to \infty$. 

In the DSL,  $|\xi| = 0.138051$ in the thermodynamic limit [Eq.~(\ref{eq:fdf})], which is about 83\% of the maximal value $1/6$. 
Than the mean field Eq.~(\ref{eq:T_i_T_j_related_to_n_i_n_j_mean_field}) gives $\langle\mathbf{T}_{i} \cdot \mathbf{T}_{j} \rangle'  = -0.3335$, the corrected mean field is -0.4002 [Eq.~(\ref{eq:TdTMFcorr})], the 2-site projected mean field gives $-0.4255$, the 3-site projected mean field gives $-0.4292$, and the Gutzwiller VMC result is $-0.4276 \pm 0.0001$. The projected mean-field and the VMC results are within 1\%. Similarly close are the three site exchanges, here  $\langle P_{123}+P_{132}  \rangle = 1.1482$ from the 3-site projected mean field compared to $1.141\pm 0.001$ from the VMC [inset in Fig.~\ref{fig:VMC_chiral_staggered}(c)]


\section{Lower bound on the ground state energy}
\label{sec:lower_bound}

\begin{figure}[t]
	\centering
	\includegraphics[width=0.7\columnwidth]{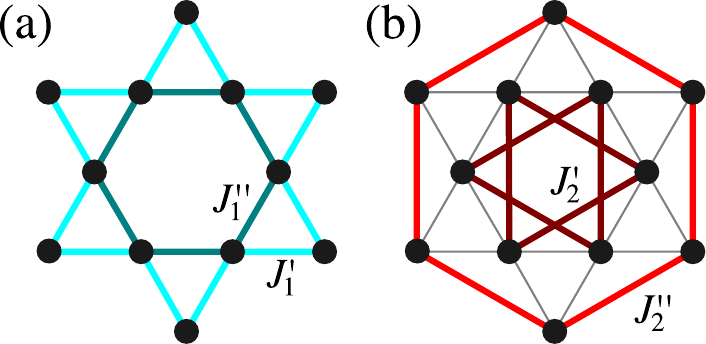}
	\caption{
		The Hamiltonian on the Kagome lattice is the sum over the lattice of the Hamiltonian $\mathcal{H'}$ of the 12-site cluster shown in the figure, such that (a) $J_1 = J''_{1} + 2 J'_{1}$ for the nearest neighbor and (b) $J_2 = J'_{2} + J''_{2}$ for the next-nearest neighbor exchanges. 
		\label{fig:lower_bound_cluster}
	}
\end{figure}

In this section we present a modification of Anderson's method \cite{Anderson, Majumdar, *NishimoriOzekiJPSJ1989, *Tarrach, WittmannStolzePRB1993} to give a lower bound on the ground state energy $E_\text{GS}$ of the Hamiltonian (\ref{eq:extended_Heisenberg_Hamiltonian}) with nearest neighbor exchanges only, 
 \begin{equation}
	\mathcal{H}=J_1 \sum_{\langle i,j \rangle}\mathcal{P}_{i,j}.
	\label{eq:HJ1}
\end{equation}
To this end, we write the $\mathcal{H}$ as a sum of sub-Hamiltonians $\mathcal{H}^\prime$,
\begin{equation}
	\mathcal{H}=\sum_{i}  \mathcal{H}^\prime_i ,
	\label{eq:sum_sub_ham}
\end{equation}
where $\mathcal{H}^\prime$ is defined on a 12-site cluster drawn in Fig.~\ref{fig:lower_bound_cluster}, and the index $i$ runs over every hexagon of the kagome lattice. We assume $\mathcal{H}^\prime$ to manifest the $D_6$ point group of the Kagome lattice, so that  
\begin{equation}
	\mathcal{H}^\prime_i=g \mathcal{H}^\prime_0 g^{-1},
	\label{eq:group_cons}
\end{equation}
where $g$ is an element of the $D_6 \otimes T$ wallpaper group \cite{Uskov}. 
In other words, the $g$ translates  $\mathcal{H}^\prime_0$ into $\mathcal{H}^\prime_i$ if it contains a translation operator.
The nearest neighbor bonds shown in Fig.~\ref{fig:lower_bound_cluster}(a) form a David star, in which Eq.~(\ref{eq:group_cons}) allows for two different exchanges, $J_{1}^{\prime}$ and $J_{1}^{\prime\prime}$. Since the index $i$ in Eq.~(\ref{eq:sum_sub_ham}) runs over every hexagon, these David stars overlap, so that every nearest neighbor bond is a part of three overlapping $\mathcal{H}^\prime_i$, once with exchange $J_{1}^{\prime\prime}$ and twice with $J_{1}^{\prime}$. Therefore, Eqs.~(\ref{eq:HJ1}) and (\ref{eq:sum_sub_ham}) gives the constraint 
\begin{equation}
  J_1=J''_{1} + 2 J'_{1}\;.
  \label{eq:constraintJ1}
\end{equation}
%
We can also include higher-distance exchanges in $\mathcal{H}^\prime$ even though they do not appear in the  Hamiltonian $\mathcal{H}$ [i.e., $J_2 = 0$ in Eq.~(\ref{eq:extended_Heisenberg_Hamiltonian})]. Eq.~(\ref{eq:group_cons}) allows for two different next-nearest neighbor exchanges $J_{2}^\prime$ and $J_{2}^{\prime\prime}$ in $\mathcal{H}^\prime$, shown in Fig.~\ref{fig:lower_bound_cluster}(b). Since every next-nearest neighbor bond is a part of two different $\mathcal{H}^\prime_i$, once with exchange $J'_{2}$ and once with $J''_{2}$, from Eq.~(\ref{eq:sum_sub_ham}) we get the condition
\begin{equation}
 0 = J_2=J'_{2} +  J''_{2}
  \label{eq:constraintJ2}
\end{equation}
constraint.
For the ground state energy per site, the
\begin{equation}
  \frac{E_{\text{GS}}}{N_s} \geq  \frac{E'_\text{GS}}{3}
   \label{eq:EGSE12GS}
\end{equation}
holds, where $E'_\text{GS}$ is the ground state energy of the 12-site Hamiltonian $\mathcal{H}^\prime$ with parameters subject to constraints Eqs.~(\ref{eq:constraintJ1}) and (\ref{eq:constraintJ2}). The factor 1/3 comes from having $N_s/3$ such clusters ($N_s/3$ is the number of hexagons) in an $N_s$ site lattice. Thus, the maximal $E'_\text{GS}$, as we vary  $J'_1$ and $J'_2$, gives a lower bound energy $E_{\text{LB}}$ for the $\mathcal{H}$ Heisenberg model in the thermodynamic limit:
\begin{equation}
   E_{\text{LB}} = \frac{1}{3} \max_{J'_1,J'_2} E'_\text{GS}(J'_1,J'_2).
\end{equation}

\begin{figure}[t]
	\centering
	\includegraphics[width=0.4\textwidth]{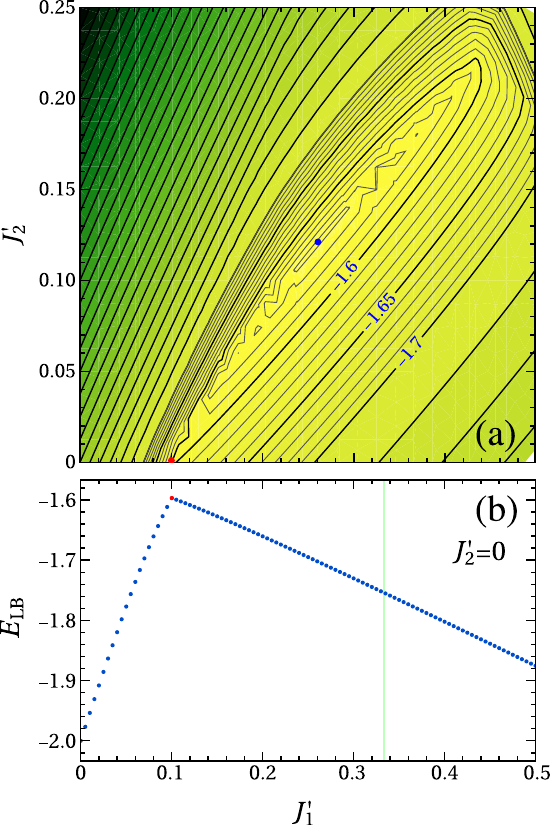}
	\caption{
		\label{fig:lower_bound_plots}
 (a) The lower bound to the ground state energy per site [i.e. the ground state energy of the 12-site cluster in Fig.~\ref{fig:lower_bound_cluster} divided by 3, see Eq.~(\ref{eq:EGSE12GS})], as a function of the free parameters $J'_1$ and $J'_2$ for the nearest neighbor Heisenberg model ($J_1=1$ and $J_2=0$). The blue dot shows the optimal lower bound $E_{\text{LB}} =-1.557157(6)$ for $J'_1 =0.2606476(0)$, $J'_2 = 0.1211317$, and the constraints (\ref{eq:constraintJ1}) and (\ref{eq:constraintJ2}) determine $J''_1$ and $J''_2$.
 (b) The same as (a) with nearest neighbor exchanges only ($J'_2=J''_2=0$). The optimal lower bound $E_\text{LB}=-1.596933(7)$ (red dot) is reached where ground state energies of different symmetries (even and odd under reflections) cross at $J'_1=0.100148(3)$. The thin green vertical line denotes the cluster with equal exchanges, $J'_1 = J''_1 =1/3.$}
 \end{figure}

Using the Lánczos method, we diagonalized the 12-site Hamiltonian for $J_1=1$ and  $J_2=0$ in the Hilbert subspace where each of the six flavors occupies two sites and plot $E'_\text{GS}(J'_1, J'_2)/3$ in Fig~\ref{fig:lower_bound_plots}(a). Applying the Nelder-Mead method, the maximum  ground state energy is at 
$J_{1}^{\prime}=0.2606476(0)$, $J''_{1} = 0.4787048(0)$
and 
$J_{2}^\prime=-J_{2}^{\prime\prime}=0.1211317(1)$, and the 
 the spectrum of the Hamiltonian (\ref{eq:HJ1}) is bounded from below by the per-site energy
\begin{equation}
 E_{\text{LB}} =-1.557157(6) J_1.
\end{equation}
The lower bound for the expectation value of the permutation operator $\mathcal{P}_{\text{1st}}$ between nearest neighbors is half of the energy (there are two bonds for each site in the kagome lattice), $\langle \mathcal{P}_{\text{1st}} \rangle \geq-0.778579$. It appears that the optimal energy is reached where the exchanges at the boundary are weaker than in the center, thus reducing the boundary effects.

Excluding the second neighbor exchanges from $\mathcal{H'}$ (i.e. $J_{2}^\prime=J_{2}^{\prime\prime}=0$), the lower bound is $E_\text{LB}=-1.596933(7)$, reached for the $J_{1}^{\prime}=(1-J_{1}^{\prime\prime})/2=0.100148(3)$ parameter value [see Fig.~\ref{fig:lower_bound_plots}(b)]. For a cluster with uniform first neighbor exchanges $J'_1 =J''_1=1/3$ [green vertical line in Fig.~\ref{fig:lower_bound_plots}(b)] $E'_{\text{GS}} =-5.262256(5)$ and the estimated lower bound is considerably worse, $E_\text{LB}=-1.754085(5) J_1$. 
The energy is maximal at the cusp of the ground state energy plot, indicating a level crossing. Indeed, the ground state wave function has different parity under reflections.  

\begin{figure}[t]
	\centering
	\includegraphics[width=0.5\columnwidth]{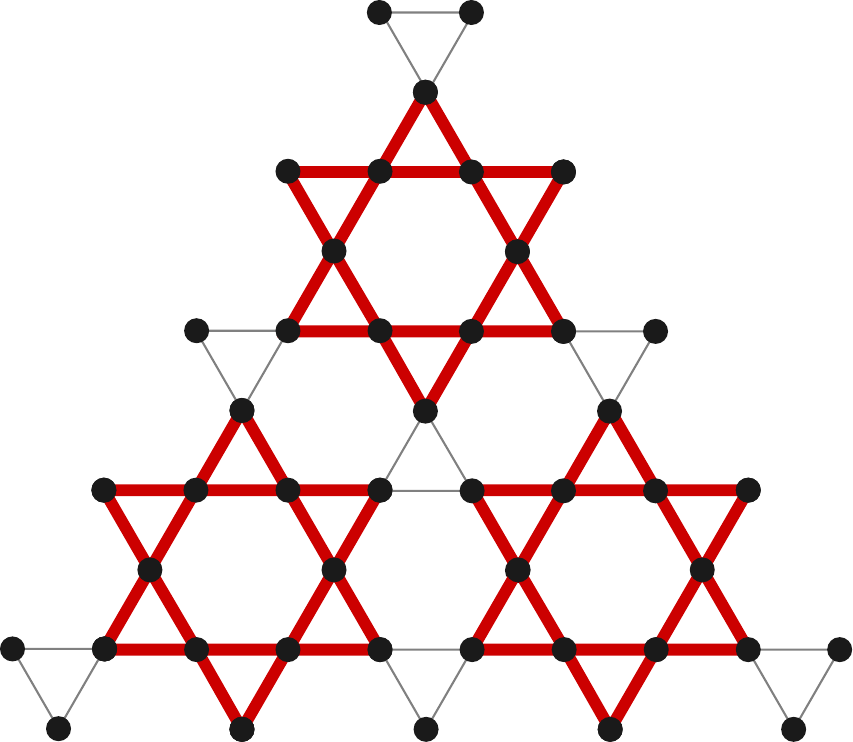}
	\caption{The tiling of the kagome lattice with the 12-site David stars for calculating the upper bound. The exchanges on the red thick bonds are $J_1$, and on the thin black bonds are $0$. The energy of the 12-site unit cell is then the energy of the decoupled David stars plus the energy of the thin bonds connecting the clusters. 
The energy of each thin bond is $\langle  \mathcal{P}_{i,j} \rangle = J_1/6$ from Eq.~(\ref{eq:TTP}) since the ground
    state of a David star is a singlet, and the expectation value $\langle \mathbf{T}_{i} \cdot \mathbf{T}_{j} \rangle$ vanishes on the thin bonds connecting the singlets.
    \label{fig:upper_bound}
	}
\end{figure}

We can also use the 12-site cluster with $J'_1 =J''_1=1$ to get a variational estimate (upper bound) for the energy \cite{WittmannStolzePRB1993}. The variational wave function is a product of the 12-site wave functions covering the lattice with David stars (see Fig.~\ref{fig:upper_bound}), the energy per site is simply the energy of the 12-site cluster plus the expectation values of the $\mathcal{P}$ exchange between the David stars.
 Since the ground state is a singlet, from Eq.~(\ref{eq:TTP}) we get the $\langle \mathcal{P}_{\text{1st}} \rangle = J_1/6$  for the latter. The energy of the 12-site unit cell is then the energy of the decoupled David-star ($-3\times 5.262256(5) =-15.78677(2)$, see the paragraph above)  plus the energy $J_1$ coming from the bonds connecting the clusters (there are six bonds in the unit cell, each having $J_1/6$), so the upper bound per site is $E_{\text{UB}} = [-15.78677(2)  + 1 ]J_1/12 =-1.232230(8) J_1$. Putting the lower and the upper bound together, the 12-site cluster provides 
\begin{equation}
 -0.778579 \leq \langle \mathcal{P}_{\text{1st}} \rangle \leq-0.616115.
  \label{eq:P1_bound}
\end{equation}
 The VMC energy of the $\pi$-flux state, $\langle \mathcal{P}_{\text{1st}} \rangle_{\text{VMC}} =-0.6885$, is roughly in the middle of this interval.
Alternatively, for the nearest neighbor correlations, we get
\begin{equation}
 -0.472623 \leq \langle \mathbf{T}_1\cdot \mathbf{T}_2 \rangle \leq -0.391391.
 \label{eq:T1T2_bound}
\end{equation}


\newcommand{\su}[3]{
  \ifthenelse{\equal{#2}{0}}{\ifthenelse{\equal{#1}{0}}{}{\mult{#1}}\overset{\bf #3}{\underset{\phantom{.}}{\bullet}}}{\ifthenelse{\equal{#1}{0}}{}{\mult{#1}}\underset{\phantom{.}}{\overset{\bf #3}{\Yboxdim9pt\yng(#2)}}}
}

\newcommand{\mult}[1]{{ \fcolorbox{gray!70}{gray!70}{\textcolor{white}{\footnotesize \bf #1}}}\;}

\section{The instability of the Ferromagnetic phase}
\label{sec:ferro}

In this section, we calculate the stability of the ferromagnetic state in the SU(6) Heisenberg model described by the Hamiltonian ~(\ref{eq:extended_Heisenberg_Hamiltonian}).
The ferromagnet is a lattice-symmetric state, which spontaneously breaks the internal SU(6) symmetry,
where both the two-site and the three-site permutations act trivially with eigenvalues $ + 1$.  Hence, the ground state energy per site is
\begin{equation}
 e_{\text{FM}} = 2 J_1 + 2 J_2  + 4 K /3.
 \end{equation}
  It is represented by the Young tableau having $N_s$ boxes in a single row.

\subsection{Single-magnon instability}

\begin{figure}[tb]
\includegraphics[width=.95\columnwidth]{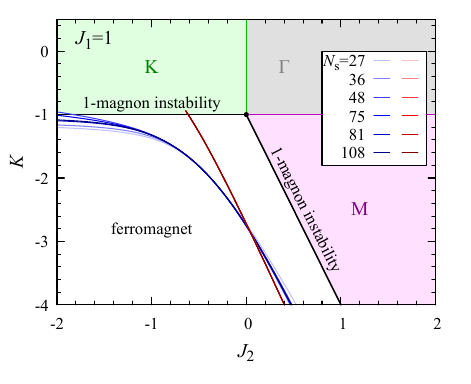}
\caption{
For negative values of the exchange parameters (lower left corner), the ground state of the model is the fully polarized ferromagnetic state (e.g., $|AA \dots A \rangle$) with a finite energy gap to one- and multi-magnon excitations. 
Along the black straight lines, the energy gap of a single magnon excitation vanishes. The energy of the one-magnon excitation is lower than that of the fully polarized state for $J_1 + K + 3 J_2 >0 $ or $J_1 + K >0$, making the fully polarized state unstable. 
The different colored regions indicate the minima of the magnon dispersion in the momentum space: $
\mathrm{K}$ (green area), $\mathrm{M}$ (magenta), and $\Gamma$ (grey area) points. 
Magnons attract each other for ferromagnetic $J_2$ or $K$, forming bound states. The bound states of two magnons --belonging to the irreducible representation labeled by the $(N_s-3,0,1,0,0)$ Young diagrams--  become energetically lower than the ferromagnetic state along the red and blue lines for system sizes ranging from 27 to 108. The red and blue colors indicate bound states with distinct point group symmetries.
 \label{fig:FM_instab}}
 \end{figure}

To determine the phase boundary of the ferromagnetic state, we first calculate the dispersion of a single magnon (two-magnon instabilities are discussed in the next section). By this, we mean configurations with identical flavors (say A) on all sites except one (say B), i.e., we flip an A spin into B. The Hamiltonian then hops the $B$ flavor to neighboring sites, endowing it with a dispersion determined by the eigenvalues of the three-by-three matrix
\begin{align}
&\mathcal{H}_{1m} = 2(J_1 + K)
\left(
\begin{array}{ccc}
-2  
 &  \cos q_{12} 
 & \cos q_{31} 
 \\
 \cos q_{12}
 &-2  
 &   \cos q_{23} 
 \\
   \cos q_{31} 
 &  \cos q_{23} 
 &-2
 \\
\end{array}
\right)
\nonumber\\
 &+ 
2 J_2
\left(
\begin{array}{ccc}
-2
 &  \cos(q_{23} \!-\! q_{31})
 &  \cos(q_{12} \!-\! q_{23}) 
 \\
 \cos(q_{23} \!-\! q_{31})
 &-2
 & \cos(q_{31} \!-\! q_{12})
 \\
 \cos(q_{12} \!-\! q_{23})
 & \cos(q_{31} \!-\! q_{12})
 &-2
 \\
\end{array}
\right)
\label{eq:magnon}
\end{align}
in reciprocal space, where the energy is measured from the energy of the ferromagnetic state and $q_{ij}=\mathbf{q}\cdot\boldsymbol{\delta}_{ij}$, where $\mathbf{q}=(q_x,q_y)$ is the momentum of the magnon and 
\begin{align}
 \boldsymbol{\delta}_{12} &= \frac{\mathbf{a}_2}{2}- \frac{\mathbf{a}_1}{2} = \left(-\frac{1}{4},\frac{\sqrt{3}}{4}\right), \nonumber\\
 \boldsymbol{\delta}_{31} &=-\frac{\mathbf{a}_2}{2} =\left(-\frac{1}{4},-\frac{\sqrt{3}}{4}\right), \\
 \boldsymbol{\delta}_{23} &= \frac{\mathbf{a}_1}{2} = \left(\frac{1}{2},0 \right) \nonumber 
\end{align} 
are lattice unit vectors.
 The Hamiltonian (\ref{eq:magnon}) describes three magnon branches measured from the energy of the ferromagnetic state. Since the $J_1$ and $K$ appear as a combination $J_1 + K$ in the matrix above, the dispersion depends only on two free parameters, $J_1 + K$ and $J_2$.
It turns out that the lowest magnon band is flat for two cases: (i) the kagome lattice with nearest neighbor exchanges when $J_2=0$ and $J_1 + K>0$ [the green line in Fig.~(\ref{fig:FM_instab})]; (ii) $J_1 + K=0$ and $J_2>0$ [the magenta line in Fig.~(\ref{fig:FM_instab})] that describes three intercalating decoupled kagome lattices.
 The extremal values of the dispersions appear at the symmetric points in the Brillouin zone. The one-magnon energies are 
$\varepsilon_M =-6 (J_1 + K)- 2 J_2$,
$-2 (J_1 + K)- 6 J_2$, and
$-4 (J_1 + K)- 4 J_2$
at the $\mathbf{q}_{\mathrm{M}}=(0,2 \pi/\sqrt{3})$ and the symmetry-related "M" points,
$\varepsilon_K =-6 (J_1 + K)$ and the 
two-fold degenerate $-3 (J_1 + K)- 6 J_2$
at the $\mathbf{q}_{\text{K}}=(\pm 4 \pi/3, 0)$ "K" points, and 
$\varepsilon_\Gamma =-6 (J_1 + K)- 6 J_2$ (two-fold degenerate)
and
$0$ at the $\Gamma$ point ($\mathbf{q}=0$).

When the one-magnon dispersion becomes negative, we do not need to invest energy to reverse a spin, and the ferromagnetic state is no longer the ground state. Examining the extremal values at the high symmetry points, we find that the dispersion becomes negative at the $\mathrm{M}$ point when  $J_1 + K =-3 J_2 >0$ and at the $\text{K}$ point when $ J_1 + K = 0$  and $J_2 \leq 0$. These instabilities are shown as black lines in Fig.~\ref{fig:FM_instab}, and the ferromagnet is stable against a single spin flip in the white region given by $J_1 + K + 3 J_2<0 $ and $ J_1 + K < 0$.

\begin{figure}[hb]
\includegraphics[width=.95\columnwidth]{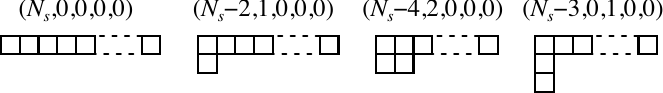}
\caption{
The Young diagrams appearing in the $[N_A,N_B,N_C,N_D,N_E,N_F] = [N_s-2,1,1,0,0,0]$ sector of the two magnon calculations for $N_s$ sites, labeled by their Dynkin indices. $N_A$ is the number of sites having $A$ flavors, and so on, so that $N_A + N_B + N_C + N_D + N_E + N_F=N_s$. The number of boxes is also $N_s$.  \label{fig:young_diagrams_2mag}
}
\end{figure}

\subsection{Instability due to two-magnon bound states}
In addition, we also examined the stability of the ferromagnetic state against the formation of bound pairs of magnons. To this end, we 
considered a two-magnon excitation of the ferromagnetic state, where two spins are flipped to different flavors. The two-magnon calculation reveals whether the interaction between the magnons is attractive and leads to a bound state, in which case the boundary of the ferromagnetic phase shrinks further. 
We followed the calculations in Ref.~\onlinecite{XU_SU3_kagome_PRB_2023} for the SU(3) Heisenberg model and Refs.~\onlinecite{Wortis1963, Hanus1963} for the SU(2) case. 

 Starting from the highest weight state of the fully symmetrical irreducible representation, $|AA\dots A\rangle$, as a vacuum, the two-magnon wave function is 
\begin{equation}
\Psi = \sum_{i,j\in \Lambda} c_{i,j} |A\dots AB_iA \dots AC_jA \dots A\rangle\;.
\end{equation}
The dimension of the Hilbert space spanned by $|A\dots AB_iA \dots AC_jA \dots A\rangle$ basis is $N_s (N_s-1)$, as $i = 1,\dots,N_s$ and $j = 1,\dots,N_s$, but the $B$ and $C$ cannot occupy the same site ($i\neq j$). Within these states, we will find $N_s-1$ states that belong to the $\mathbf{d}=(N_s-1)(N_s + 1)$ dimensional Young diagram with $(N_s-2,1,0,0,0)$ Dynkin label, which are just rotated single-magnon states, and a state with zero energy which belongs to the $\mathbf{d}=(N_s + 1)(N_s + 2)/2$ dimensional fully symmetrical irreducible representation of the ferromagnetic state. But also new irreducible representations appear, denoted by the $(N_s-4,2,0,0,)$ and $(N_s-3,0,1,0,0)$  Young diagrams (see Fig.~\ref{fig:young_diagrams_2mag}).

We diagonalized the Hamiltonian matrix for up to 108-site clusters numerically (in principle, being a two-body problem, it can be solved analytically, but the expressions would be pretty cumbersome). Using the Casimir operator, we separated the energies of the $(N_s-4,2,0,0,0)$ and  $(N_s-3,0,1,0,0)$ irreducible representations. The $(N_s-4,2,0,0,0)$ state also appears in the SU(2) case since the configurations with two $B$ type spins (instead of $B$ and $C$) are members of the multiplet. However, the $(N_s-3,0,1,0,0)$ is unique to the SU(N) models with $N\geq3$. We found that there are bound pairs in the $(N_s-3,0,1,0,0)$ irreducible representation in the region we investigated in Fig.~\ref{fig:FM_instab}.  The number of bound pairs is $2N_s/3$, which is also the number of triangles-- for substantial ring exchange $K$ with a negative sign, the $B$ and $C$ spins are localized on a triangle. In the momentum space, two bound pairs form two bands for each momentum. The lowest energy bound state is at the $\Gamma$ point in the Brillouin zone, but it can have different point group symmetry. The solid red and blue curves in Fig.~\ref{fig:FM_instab} show when these bound states start to have lower energy from the ferromagnet, leading to the instability of the ferromagnet. 
For the $J_2=0$ value, the bound state energy becomes negative for $K/J_1>-2.7532$ \cite{XU_SU3_kagome_PRB_2023}. We did not examine whether the interaction between the bound pairs is repulsive or attractive (in which case the ferromagnet region would shrink further).
 
\section{Flat bands}
\label{sec:flatBands}

\begin{figure}[hb]
\includegraphics[width=.95\columnwidth]{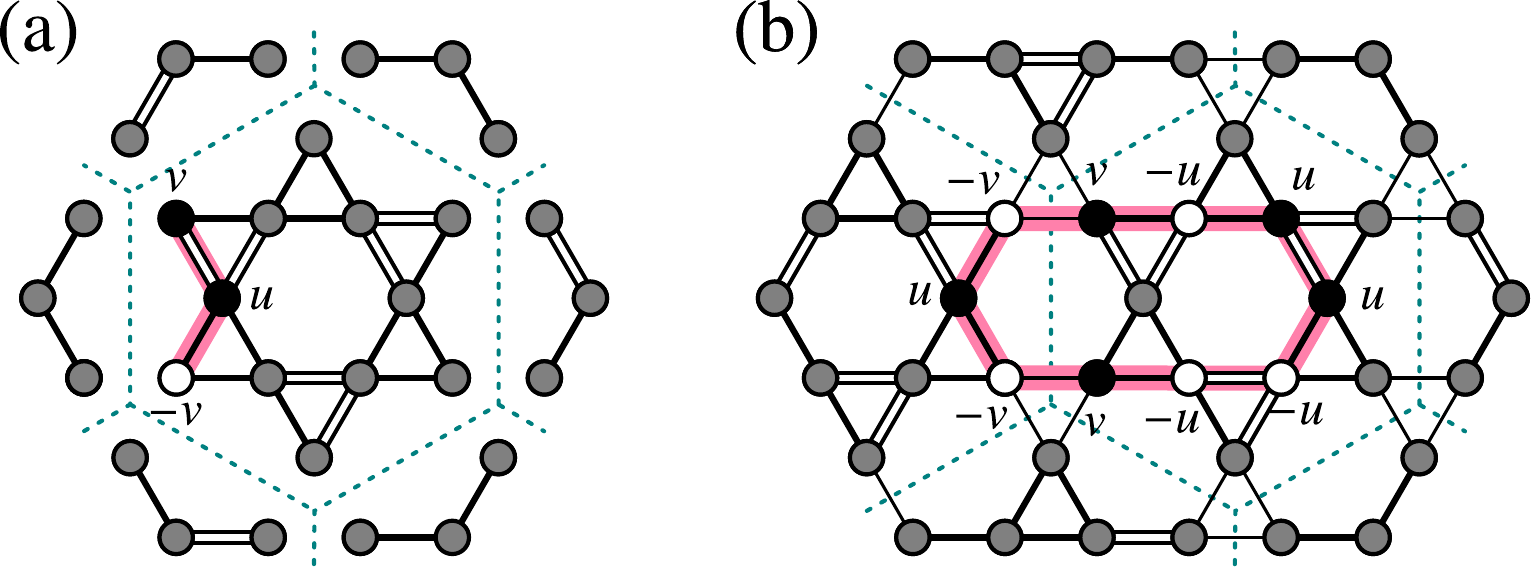}
\caption{
\label{fig:localized_states}
(a) The wave function of a localized fermion on a decoupled, $t_\triangle=0$, David star at the singular point. It extends over the three sites connected by thick red lines. Each David star can support six linearly independent localized states. 
(b) A localized state for coupled David stars  ($t_\triangle \neq 0$). It extends over two hexagons. 
}
\end{figure}

%

Here, we discuss the localized states leading to flat bands in the spectrum of the mean-field Hamiltonian for the David star ansatz. First, let us consider disconnected David stars when $t_\triangle=0$.
The equation for the banana-shaped localized state in Fig.~\ref{fig:localized_states}(a) reads
\begin{subequations}
\label{eq:local_state}
\begin{align}
t_{\DavidStarOut} v - t_{\hexagon} u &=0 \label{eq:local_state_a}\\
\begin{pmatrix}
0 &  2 t_{\DavidStarOut} \\
t_{\DavidStarOut}  & 0
\end{pmatrix}
\cdot
\begin{pmatrix}
  u   \\
  v  
\end{pmatrix}
&= \varepsilon
\begin{pmatrix}
  u   \\
  v  
\end{pmatrix}
\label{eq:local_state_b}
\end{align}
\end{subequations}   
where $u$ and $v$ are the amplitudes in the wave function, 
Eq.~(\ref{eq:local_state_a}) ensures the locality of the state, and (\ref{eq:local_state_b}) that it is an eigenstate. The two solutions are 
\begin{equation} 
\label{eq:local_state_sol}
 \varepsilon = 2 t_{\hexagon}, \quad
 \sqrt{2}  t_{\hexagon} = \pm t_{\DavidStarOut},  \quad
  u = \pm \sqrt{2} v .
\end{equation}
The solution with negative signs determines the singular point discussed in Sec.~\ref{sec:specialPoints}: Each David star has six banana-shaped localized orbitals, giving rise to a six-fold degenerate flat band in the momentum space. 

For the coupled David stars we can linearly combine the banana-shaped states into a state shown in Fig.~\ref{fig:localized_states}(b) that encompass two hexagons and zero flux, where the alternating amplitudes $u$ and $-u$ ensures the locality at the isosceles triangles, and Eqs.~(\ref{eq:local_state}) modify to include the effect of a finite $t_\triangle$ hopping, 
\begin{subequations}
\begin{align}
t_{\DavidStarOut} v - t_{\hexagon} u &=0 \;,\label{eq:local_state_A}\\
\begin{pmatrix}
0 &  2 t_{\DavidStarOut} \\
t_{\DavidStarOut}  & t_\triangle 
\end{pmatrix}
\cdot
\begin{pmatrix}
  u   \\
  v  
\end{pmatrix}
&= \varepsilon
\begin{pmatrix}
  u   \\
  v  
\end{pmatrix} .
\label{eq:local_state_B}
\end{align}
\end{subequations}   
This set of equations can be solved if the hopping amplitudes satisfy  $t_{\DavidStarOut}^2 - 2t_{\hexagon}^2  + t_{\hexagon}t_{\triangle} = 0$ [i.e., Eq.~(\ref{eq:greenline})] and we get 
\begin{equation}
\varepsilon = 2 t_{\hexagon}\;,
\quad
 \frac{t_{\hexagon}}{t_{\DavidStarOut}} = \frac{v}{u} \;,
  \quad\text{and}\quad 
 \frac{t_{\triangle}}{t_{\DavidStarOut}} = \frac{ 2  v^2  - u^2}{u v} .
\end{equation}
For $u=\pm v \sqrt{2}$, we recover Eq.~(\ref{eq:local_state_sol}) for disconnected David stars. Changing the ratio  $u/v$, we get the solid green line in Fig.~\ref{fig:Global_stability_of_David_star} that emerges from the singular point. It also describes the flat bands for the $t_{\hexagon}=t_{\DavidStarOut}=t_{\triangle}$ DSL mean-field Hamiltonian at higher energies by setting $u=v$. Unlike the flat bands for the $0$-flux kagome lattice \cite{Bergman_flatbands_2008}, these flat bands are non-topological (they form a set of $N_{\hexagon} = N_s/3$ linearly independent states, where $N_{\hexagon}$ is the number of hexagons).

\bibliography{su6} 
 
\end{document}